\documentclass[reprint,amsmath,aps,prd,nofootinbib,showkeys,superscriptaddress,onecolumn]{revtex4-2}
\pdfoutput=1
\usepackage{lmodern,anyfontsize}
\usepackage{amsfonts,array,bm,braket,cancel,comment,ctable,enumitem,float,footnote,%
graphicx,hhline,ifthen,keyval,latexsym,multirow,silence,slashed,xfrac,amsmath}
\usepackage[utf8]{inputenc}

\newcommand{\infinity}{\infty}
\newcommand{\be}{\begin{equation}}
\newcommand{\ee}{\end{equation}}
\newcommand{\bea}{\begin{eqnarray}}
\newcommand{\eea}{\end{eqnarray}}
\newcommand{\nn}{\nonumber}
\newcommand{\rmi}[1]{{\text{\scriptsize{#1}}}}
\newcommand{\e}{\text{e}}
\newcommand{\labsec}[1]{\label{sec:{#1}}}
\newcommand{\secn}[1]{(\ref{sec:{#1}})}
\newcommand{\hsec}[1]{section \secn{#1}}
\newcommand{\labeq}[1]{\label{eq:{#1}}}
\newcommand{\eqn}[1]{(\ref{eq:{#1}})}
\newcommand{\eq}[1]{eq.~\eqn{#1}}
\newcommand{\labfig}[1]{\label{fig:{#1}}}
\newcommand{\fign}[1]{(\ref{fig:{#1}})}
\newcommand{\vect}[1]{\bm{#1}}
\newcommand{\inte}[3]{{\int_{#2}^{#3}\kern-0.75em\text{{\small$%
\mathop{\text{{\small${\rm d}#1$}}}$}}\ }}
\newcommand{\pard}[2]{\frac{\partial{#1}}{\partial{#2}}}
\DeclareMathOperator{\Tr}{Tr}
\newcommand\Z{{\mathbb{Z}}}
\newcommand\N{{\mathbb{N}}}
\newcommand\MSbar{{\overline{\text{MS}}}}
\newcommand{\lmsb}{\bar{\Lambda}}
\newcommand{\Li}{\text{Li}}
\newcommand{\dbar}{d\hspace*{-0.08em}\bar{}\hspace*{0.1em}}
\newcommand{\half}{\frac{1}{2}}
\newcommand{\shalf}{\sfrac{1}{2}}
\newcommand{\partdiffat}[3]{\left.\frac{\partial #1}{\partial #2}\right\vert_{#3}}
\newcommand{\HS}[1]{\hspace{-#1cm}}
\newcommand{\Krondelt}[1]{{\delta_{\text{\tiny $#1$}}}}
\newcommand\hspa[1]{\hspace*{#1cm}}
\newcommand{\Rea}[1]{\mathrm{Re}\left(#1\right)}
\newcommand{\prefacmsb}{\left(\frac{\lmsb^{2} e^{\gamma_\rmi{\tiny E}}}{4\pi}\right)^{2-\frac{D}{2}}}
\newcommand{\minte}[2]{{\int\kern-0.50em\text{{\small$\mathop{\frac{\text{{\small${\rm d^{#2}}\vect{#1}$}}}%
{\text{{\small$(2\pi)^{#2}$}}}}$}}\ }}
\newcommand{\SPEprefacmsb}[2]{\left(\frac{e^{\gamma_\rmi{\tiny E}\epsilon}(\lmsb/T)^{2\epsilon}}%
{2^{#1+2\epsilon}\pi^{#2}}\right)}

\begin{document}

\title{Geometrically Confined Thermal Field Theory: Finite Size Corrections and Phase Transitions}

\author{Sylvain~Mogliacci}
\email{sylvain.mogliacci@gmail.com}
\affiliation{Department of Physics, University of Cape Town, Rondebosch 7701, South Africa}

\author{Isobel~Kolb\'e}
\email{ikolbe@uw.edu}
\affiliation{Department of Physics, University of Cape Town, Rondebosch 7701, South Africa}
\affiliation{Institute for Nuclear Theory, University of Washington, Seattle, Washington, 98195-1550}

\author{W.~A.~Horowitz}
\email{wa.horowitz@uct.ac.za}
\affiliation{Department of Physics, University of Cape Town, Rondebosch 7701, South Africa}


\begin{abstract}
	Motivated by evidence for quark-gluon plasma signatures in small systems, we study a simple model of a massless, noninteracting scalar field confined with Dirichlet boundary conditions.  We use this system to investigate the finite size corrections to thermal field-theoretically derived quantities compared to the usual Stefan-Boltzmann limit of an ideal gas not confined in any direction. Two equivalent expressions with different numerical convergence properties are found for the free energy in $D$ rectilinear spacetime dimensions with $c\le D-1$ spatial dimensions of finite extent.  We find that the First Law of Thermodynamics generalizes such that the pressure depends on direction.  For systems with finite dimension(s) but infinite volumes, such as a field constrained between two parallel plates or a rectangular tube, the relative fluctuations in energy are zero, and hence the canonical and microcanonical ensembles are equivalent. We present precise numerical results for the free energy, total internal energy, pressure, entropy, and heat capacity of our field between parallel plates, in a tube, and in finite volume boxes of various sizes in 4 spacetime dimensions.  For temperatures and system sizes relevant for heavy ion phenomenology, we find large deviations from the Stefan-Boltzmann limit for these quantities, especially for the pressure.  Our main result is the discovery that an isolated system of fields constrained between parallel plates reveals a divergent isoenergetic compressibility at a critical length $L_c\sim1/T$.  This divergence constitutes a novel phase transition, which, unlike the usual temperature-driven phase transition, is driven solely by the size of the system.
\end{abstract}{%



\maketitle

\section{Introduction}\labsec{Intro}

\subsection{Motivation and goals}\labsec{MotivationGoals}

 Multiple experimental signals from the Relativistic Heavy Ion Collider (RHIC)~\cite{Ackermann:2002ad,Adcox:2003zm} at Brookhaven National Laboratory (BNL) in Upton, New York and the Large Hadron Collider (LHC)~\cite{Aamodt:2008zz,Chatrchyan:2008aa,Aad:2008zzm} at CERN in Geneva confirm the creation of the quark-gluon plasma (QGP)~\cite{Adcox:2004mh,Adams:2005dq,Aamodt:2010jd,Chatrchyan:2011sx,Aad:2010bu}; remarkably, these colliders recreate for the first time conditions similar to those existing in the early universe. The future is bright with collider experiments such as the Nuclotron-based Ion Collider fAcility~\cite{Kekelidze:2012zzb} at the Joint Institute for Nuclear Research (JINR) in Dubna and the Facility for Antiproton and Ion Research (FAIR)~\cite{Barone:2005pu,Erni:2008uqa} at GSI in Darmstadt that will soon further explore the physics of heavy ion collisions (HIC). This abundance of data has led the HIC field to an era of high precision measurements, which demands a commensurate level of theoretical precision. 

Thermodynamics is one of the main avenues of theoretical investigation into the dynamics of the QGP, one that has a connection to the experimental measurements at RHIC and LHC through a number of observables such as identified particle spectra \cite{Bernhard:2016tnd}.  Understanding the thermodynamics of the QGP created in HIC has been, and still is, the subject of intensive theoretical research for the past three decades. Generally, there are three broad classes of theoretical tools used to explore QGP thermodynamics: weakly coupled thermal field theory \cite{Blaizot:2001vr,Andersen:2012wr,Mogliacci:2013mca,Andersen:2015eoa}, Monte Carlo methods \cite{Boyd:1996bx,Karsch:2000ps,Aoki:2006br,Bazavov:2009zn,Borsanyi:2010bp,Ratti:2011au,Bazavov:2011nk,Kitazawa:2016dsl,Borsanyi:2016ksw,Taniguchi:2016ofw,Allton:2002zi,Bazavov:2013dta,Borsanyi:2013hza,Schmidt:2017bjt}, and AdS/CFT \cite{CasalderreySolana:2011us}. Despite the long history of investigation into the thermodynamics of the QGP, an important aspect appears to have been overlooked by the HIC community: the effect of finite-sized rather than infinitely-sized systems.  The importance of finite-size corrections to the thermodynamics of QGP is of especial significance now as many signals of QGP creation in large ion-ion systems appear to depend on measured particle multiplicity rather than colliding system size.  Some examples include collective behavior~\cite{Khachatryan:2016txc,Aad:2015gqa,Sirunyan:2017uyl,Aaboud:2017blb}, strangeness enhancement~\cite{Abelev:2013haa,Khachatryan:2016yru,ALICE:2017jyt}, and quarkonium suppression~\cite{Aaboud:2017cif,Adare:2016psx}, while noticeable absence of appreciable jet quenching demands a thorough interrogation of our understanding of energy loss in small systems \cite{Kolbe:2015rvk}.  Further, a simple estimate of a quark or gluon mean free path of $\lambda\sim1-2$ fm indicates that even a large colliding Pb+Pb system of radius $r\sim6$ fm is not particularly close to the thermodynamic limit \cite{Romatschke:2016hle}.  

The present work is motivated by the pressing need for a better understanding of small droplets of QGP \cite{Kolbe:2019zvx}, including a quantification of the small system size corrections to the usual approximation of using dynamics derived in systems of infinite size. In terms of QGP phenomenology, our goals are modest.  As a first step, we concentrate on the finite size corrections to thermodynamic quantities such as the equation of state (EoS) computed in thermal field theory in the usual Stefan-Boltzmann limit of an ideal gas of infinite size in all directions. As a further simplification, we consider only noninteracting fields. Of the various types of boundary conditions one might use, we focus on Dirichlet boundary conditions (BC) for simplicity and, as we will argue below, because these are the most natural BC when considering a finite-sized QGP.  

Despite the simplicity of our model, we will find a suggestive phenomenology.  In our derivation of thermodynamic quantities, we use two independent methods that yield analytic formulae with different numerical convergence properties: one which converges exponentially fast when the dimensionless scale of the temperature times the system size is small, $T\times L\lesssim1$, and one that converges exponentially fast for $T\times L\gtrsim1$.  A careful application of these two formulae allow us to numerically investigate to arbitrary accuracy the various thermodynamic quantities such as the pressure, energy, entropy, and heat capacity for our model.  Of particular note for the heavy ion community, the introduction of Dirichlet instead of periodic BC leads to significant small system corrections to the usual infinite size results of the equation of state even out to fairly large system sizes $L\sim10/T$.

Our work naturally touches on the well-known Casimir effect \cite{Casimir:1948dh,Lifshitz:1956zz,Jaffe:2005vp,Graham:2003ib,Bordag:2009zzd}.  The original Casimir effect of a free scalar field between two slabs has been extended to systems with nontrivial geometries and to Fermi fields \cite{DePaola:1999im,Milton:2016tmr}.  For $T>0$, one encounters a number of interesting fundamental questions, many of which have only begun to be explored \cite{DeFrancia:1994us,Gross2001,Cheng_2002,Svaiter:2004ad,Ferrari2017}. Of particular interest to the current work are the consequences of a finite sized system on the thermodynamic limit \cite{Hill1962}, which in turn naturally raises questions surrounding the equivalence of statistical ensembles in such finite systems \cite{Gross2001,TOUCHETTE2004138,DUNKEL2006390,CAMPA200957}.  Although we will not address these questions directly, we will be careful to be vigilant of the uncertainties they introduce.  One very important problem in small systems that we will address is that of phase transitions; 
Phase transitions are notoriously difficult to rigorously identify.  This difficulty is only compounded in finite-sized systems \cite{Chomaz:2002sv,Pule2004,CAMPA200957,Bausch2018} with some authors claiming that phase transitions in systems of finite size are impossible.  
It is also important to note that, in finite-sized systems, one loses extensivity in thermodynamics \cite{Mogliacci:2018hif}, where by extensivity we mean that the entropy is positively homogeneous of degree 1, i.e.\ $S(\lambda E,\,\lambda X_1,\,\lambda X_2,\,\ldots) = \lambda S(E,\,X_1,\,X_2,\,\ldots)$.  This lack of extensivity then leads to questions about the use of the Tsallis distribution \cite{Tsallis:1987eu,Tsallis:1998ws} and the minimal set of assumptions required for a consistent thermodynamics \cite{Swendsen2017}.  We will also address the problem of thermal fluctuations in finite-sized systems, which has begun to be explored in the heavy-ion hydrodynamic community \cite{Akamatsu:2016llw}.  

Importantly, we also find a connection between our work and the study of phase transitions.  Phenomenologically, one finds substances for which a phase transition can be driven by a change of pressure at constant temperature, for example the liquid-gas phase transition for water above 0$^{\circ}$C.  We are aware of only one example of a phase transition driven by the size of the system~\cite{Bausch2018}, in which case the transition is due to the self interactions of the system.  What we will show is that for an isolated ideal gas constrained within parallel plates, the system resists compression until the separation of the plates is on the order of the thermal de Broglie wavelength; at this critical length $L_c\sim1/T$ the susceptibility diverges and the system collapses.  This divergence of the susceptibility indicates a second order phase transition driven by the size of the system, which is conjugate to the pressure on the system.  Unlike other works that draw a connection between Bose-Einstein condensation between parallel plates \cite{Biswas2007} and a first order phase transition in a finite volume box \cite{Park:2008sk}, the phase transition found here is second order and also exists for Fermionic fields.  

This paper is organized as follows:  In the remainder of the present section we will outline a model for a confined field theory by describing a simple thermal field-theoretical method.  In sec.\,\secn{FieldTheoCalculations} we apply these methods and compute the partition function and Helmholtz free energy for a single, massless scalar field confined between two parallel plates, in an infinite tube, and in a finite box.  In sec.\,\secn{FiniteSizeThermo} we outline a number of thermodynamical subtleties before exploring the thermodynamics of a geometrically confined scalar field in sec.\,\secn{ResultThermo}.  In sec.\,\secn{PhaseTransition} we develop and discuss our main result: a novel phase transition, driven by the system size.

\subsection{Geometric confinement for HIC}\labsec{GeoConf}

There have been a number of studies of finite size effects using periodic boundary conditions~\cite{Meyer:2009kn,Giusti:2011kt,Braun:2011iz,Bijnens:2013doa,Fister:2015eca,Parvan:2015mta,Fraga:2016oul,Juricic:2016tpt,Almasi:2016zqf}.  Consider, however, that a boundary-less manifold, e.g. a three-dimensional sphere, is entirely decoupled from the rest of the universe: there is no possibility for any signal or information to come through the QGP and reach the detectors. Therefore, spatial boundary conditions---other than periodic ones---should be considered for a more realistic approach to the finite size corrections of the EoS. 

Consider now a QGP system inside of which the quarks and gluons are color deconfined and propagate relatively freely while outside of this QGP system the quarks and gluons are color confined into hadrons. Inside the system the quark and gluon fields, then, are weakly coupled while these same quark and gluon fields are strongly coupled outside the system.  There is further some very small region of space over which the fields transform from weakly to strongly coupled. For our purposes here, we are most interested in the dynamics inside the QGP system.  If we approximate the small transition region from weak to strong coupling as a decoupling, we must impose a boundary condition that prevents an inside weakly coupled field from propagating outside of the geometric region defining the QGP system. We refer readers to~\cite{Bazavov:2007ny,Flachi:2012pf,Berg:2013jna} for related investigations that demonstrate the need for such more realistic boundaries. We also refer to~\cite{Bender:1976wb,Grossmann:1995,Kirsten:1999zx,Kirsten:1991mi,Chernodub:2017gwe} for somewhat related investigations, as well as to~\cite{Sharma:2018uma} regarding the importance of accounting for finite size effects in the different context of proton-proton collisions. While such a decoupling might seem irrelevant from a perturbative point of view, since this change of degrees of freedom across the two regions must involve nonperturbative physics, our system appears to be analogous to the Quantum Electrodynamics (QED) Casimir effect~\cite{Casimir:1948dh,Bordag:2009zzd}, which can be reproduced by imposing Dirichlet Boundary Conditions (DBCs). Such boundary conditions indeed follow the requirement for the fields to vanish at the boundary, and any two-point correlation function connecting the inside part to the boundary would identically vanish, thereby decoupling both sides of the boundary. Thus, DBCs for the QGP can avoid the propagation of QGP particles across the boundary, keeping the relatively free quarks and gluons geometrically confined inside the volume of the QGP system, while at the same time allowing for other fields, e.g.\ electromagnetic, to freely propagate out from the confined system.

We then introduce the notion of {\textit{geometric confinement}} for such a system, by implementing appropriate spatial compactifications (depending on the geometry to be characterized).  In each of the compactified directions we impose DBCs\footnote{Note that by a compactification we do not mean, e.g., the addition of the point at infinity to turn $\mathbb{R}$ into $S^1$.  Rather, we mean that we are considering directions that are compact, i.e.\ of finite extent (technically, closed and bounded).}.  This is done in much the same spirit as \cite{Ambjorn:1981xw} where a similar calculation is performed in order to study the Casimir energy of the vacuum. We will significantly extend that program here. It should be noted that such a boundary is not a material boundary, but rather a thin layer subdividing different regions of space with different degrees of freedom. Since we are only interested in the bulk physics away from the boundary (avoiding then possible technical complications~\cite{Graham:2002fw,Elizalde:2003jw}), we will not consider the microscopic nature of the boundary. As a result, the physical space is separated into two distinct regions: An inside part of nearly free quarks and gluons characterizing the quark-gluon plasma, and an outside part (of nearly free hadrons composed of strongly coupled quarks and gluons). For the sake of our investigations in this manuscript we ignore the details of color confinement and the microscopic nature of the boundary. Moreover, it should be noted that while DBCs may be implemented in more physically realistic geometries, for the sake of simplicity we choose to work here with simple rectangular geometries, that is planar pairwise parallel spatial boundaries forming a cuboid cavity.  In three spatial dimensions, we consider two infinite parallel plates, the infinite rectangular tube, and the finite volume box.

We stress that our concept of geometric confinement is in no way related to actual color confinement, for it barely reproduces only one consequence of it (the fact that quark and gluon fields should not propagate outside of the QGP); we do not address the fundamental mechanism of color confinement. In addition, our concept is different from the MIT bag model~\cite{Chodos:1974je} since our model is not meant to be relevant for strongly coupled fields. 

\subsection{A model for first investigations}\labsec{Model1stInvestigations}

Recalling that the EoS for a noninteracting gas of gluons is the same, up to a group theory prefactor, as the EoS given by a massless scalar field, for simplicity we choose to consider a single neutral noninteracting massless scalar field at finite temperature under geometric confinement.  Further, we will work in the canonical ensemble: we will compute the partition function using thermal field theoretic methods.  Even though the quark-gluon plasma created in heavy ion collisions is an isolated system, we choose to work in the canonical ensemble both for simplicity and because we may more readily connect our results to lattice QCD calculations \cite{Boyd:1996bx,Karsch:2000ps,Allton:2002zi,Aoki:2006br,Bazavov:2009zn,Borsanyi:2010bp,Ratti:2011au,Bazavov:2011nk,Bazavov:2013dta,Borsanyi:2013hza,Kitazawa:2016dsl,Borsanyi:2016ksw,Taniguchi:2016ofw,Schmidt:2017bjt}; and second, for simplicity.

In the following, we will keep the mass of the field nonzero for pure convenience during the calculations. At a later stage, we will, however, take the massless limit which is the present case of interest. We will work in $D$ spacetime dimensions of which an arbitrary number $c$ will be compactified spatial dimensions with DBCs.  We will leave the noncompactified dimensions, if any, in the usual infinite Euclidean form. Each such compactified spatial direction will then have a distinct compactification length $L_i$, $i=1,\ldots c\le D-1$.  The $L_i$'s do not necessarily have to be equal, which allows us to investigate systems of asymmetric sizes. Our starting point is therefore the free, Wick rotated Euclidean action
{\allowdisplaybreaks
\be
\mathcal{S}\Big[\phi\left(\tau,\left\{z_i\right\},\vect{x}\right)\Big]\equiv%
\int_{0}^{1/T}\!\!\mathop{{{\rm d}}\tau}\!\!%
\inte{z_1}{0}{L_1}\!...\inte{z_{c}}{0}{L_{c}}\!\!%
\!\!\int_{R}\!\mathop{{{\rm d}^{\text{\scriptsize{$D$}}-1-\text{\scriptsize{$c$}}}}\vect{x}}%
\bigg(\mathcal{L}\Big[\phi\left(\tau,\left\{z_i\right\},\vect{x}\right)\Big]\bigg)%
,\labeq{Action1}%
\ee}%
where beside the usual trace induced periodic boundary condition along the Euclideanized, Wick-rotated $\tau$ direction, and the new geometric confinement induced DBCs along the $z_i$ spatial directions, the $\vect{x}$ coordinates will be momentarily compactified around circles of radius $R$, i.e.\ with periodic boundary conditions.  At the end of the calculation we will take $R\rightarrow\infinity$. The free Lagrangian $\mathcal{L}\left[\phi\left(\tau,\left\{z_i\right\},\vect{x}\right)\right]$ then reads
\begin{multline}
\mathcal{L}\left[\phi\right]\equiv%
\frac{1}{2}\left(\pard{\phi\left(\tau,\left\{z_i\right\},\vect{x}\right)}{\tau}\right)^2%
+\frac{1}{2}\sum_{i=1}^{c}\left(\pard{\phi\left(\tau,\left\{z_i\right\},\vect{x}\right)}{z_i}\right)^2%
\\+\frac{1}{2}\bigg(\vect{\nabla_\vect{x}}\phi\left(\tau,\left\{z_i\right\},\vect{x}\right)\bigg)^2%
+\frac{m^{2}}{2}~\phi^2\left(\tau,\left\{z_i\right\},\vect{x}\right)%
,\labeq{LagrangeDensity1}%
\end{multline}
where we are working in $\hbar=c=k_B=1$ natural units.

\section{Partition function and free energies for parallel plates, a tube, and a box}\labsec{FieldTheoCalculations}
\subsection{Deriving the partition function}\labsec{PartitionFunc}

In this subsection, we derive the partition function $\mathcal{Z}(T,\,\left\{L_i\right\})$ for a single, neutral, noninteracting, massless scalar field that is geometrically confined within $c\leq D-1$ spatial dimensions.

Formally, the partition function in a theory with Hamiltonian operator $\mathcal{\hat{H}}$ and no globally conserved charges is obtained from the trace of the density matrix
{\allowdisplaybreaks
\be
\mathcal{Z}(T,\,V)\equiv\Tr\left[\hat{\rho}(T,\,V)\right]\equiv\Tr\left[\exp\left\{-\beta\int_{V}d^{D-1}x\,\mathcal{\hat{H}}\right\}\right]%
,\labeq{PrimaryZDef}%
\ee}
\HS{0.12}where $\beta=1/T$ is the inverse temperature, $V$ the spatial volume of the system, and the trace represents a summation over all possible physical states. We will employ the Matsubara imaginary time formalism~\cite{Matsubara:1955ws} in order to compute the partition function using path integral techniques~\cite{Feynman:1948ur}. For more details on this formalism as well as on thermal field theory, we refer readers to~\cite{Landsman:1986uw,Kraemmer:2003gd,Andersen:2004fp,Kapusta:2006pm,Bellac:2011kqa,Laine:2016hma}.

The usual procedure for expressing the partition function as a path integral leads to a path integral with a {\textit{periodic boundary condition}} on the temporal line which, up to an irrelevant constant, reads
{\allowdisplaybreaks
\be
\mathcal{Z}(T,\,V)\propto\left.\int_{\phi(0)}^{\phi(\beta)}%
\left[\mathcal{D}\phi\right]\exp\left\{-\int_{0}^{\beta}d\tau\int_{V}d^{D-1}x\,\mathcal{L}\right\}\right|_{\phi(\beta)=\phi(0)}%
.\labeq{ZInf}%
\ee}%
We now extend the above to a manifold with $c$ compactified spatial dimensions. The procedure will require DBCs for the compactified spatial dimensions, in addition to the periodic boundary condition required by the trace operation. The derivation of the analogue of eq.~\eqn{ZInf} with compactified spatial dimensions closely follows that of the spatially noncompactified case, which we call the Stefan-Boltzmann limit, the details of which can be seen in the aforementioned textbooks. For pedagogical reasons we will first compactify only one dimension ($c=1$), which gives the canonical ensemble description for a noninteracting scalar field at temperature $T$ constrained between two infinite parallel plates separated by a distance $L_1$. Extending the result to $c$ compactified dimensions will be straightforward and is presented in appendix~\secn{AppDetailsCalculations}.

Following the usual procedure, we start by decomposing our field into the relevant Fourier modes. The dimensions that will ultimately be left noncompactified are, as usual, momentarily compactified onto circles of identical sizes $R$ (which means we impose periodic boundary conditions), while the dimension we wish to permanently compactify is done so along a finite interval with length $L_1$. We choose a convenient normalization constant (see appendix~\secn{AppFourierDecomp} for more details) and express the Fourier decomposition of our field as
{\allowdisplaybreaks
\bea
\phi(\tau,z_1,\vect{x})&=&\sum_{n\in\Z}\sum_{\ell_1\in\N}\sum_{\vect{k}\in\Z^{D-2}}\sqrt{\frac{2\beta}{L_1 R^{D-2}}}~%
\exp\Big\{i\omega_n\tau-i\vect{\omega}_k\cdot\vect{x}\Big\}~\sin\Big\{\omega_{\ell_1}z_1\Big\}~\tilde{\phi}_{n,\ell_1}(\vect{\omega}_k)%
,\labeq{phiFourier}
\eea}%
where the components of $\vect{x}$ are the spatial components in the non permanently compactified dimensions, $z_1$ is the spatial component in the compactified dimension, and $\tilde{\phi}_{n,\ell_1}(\vect{\omega}_k)$ are the dimensionless Fourier modes. The Matsubara, the compactified spatial dimension, and the non permanently compactified spatial dimensions related frequencies are given by $\omega_n=2\pi n T$, $\omega_{\ell_1}=\pi\ell_1/L_1$ and $\vect{\omega}_k=2\pi \vect{n}_k/R$, respectively, and we also refer to~\cite{Kirsten:2010zp} for more details on the derivation of the modes given DBCs. The latter frequencies are to be replaced by a $D-2$ dimensional continuous momentum $\vect{k}$ (with corresponding momentum integrals), in the asymptotically large $R$ limits.

Setting $c=1$ and employing eq.~\eqn{LagrangeDensity1} for the Lagrangian, the partition function~\eqn{ZInf} becomes, after an integration by parts,
{\allowdisplaybreaks
\be
\mathcal{Z}\propto\int_{\phi_{\text{cond.}}}\!\!\!\!\!\!\!\!\!\left[\mathcal{D}\phi(\tau,\,z_1,\,\vect{x})\right]\exp\left\{-\int_0^{\beta}\!%
\inte{z_1}{0}{L_1}\!\!\int_{R}\!\mathop{{{\rm d}^{2}}\vect{x}}~~%
\half\phi(\tau,\,z_1,\,\vect{x})(\partial_{\mu}\partial^{\mu}+m^2)\phi(\tau,\,z_1,\,\vect{x})\right\}%
,~\labeq{ZM1}%
\ee}
\HS{0.12}where, among others, the periodic boundary condition from the trace is imposed through setting $\phi_{\text{cond.}}$ such that $\phi(\beta,\,z_1,\,\vect{x})=\phi(0,\,z_1,\,\vect{x})$, and where $\partial_{\mu}\partial^{\mu}\equiv\frac{\partial^2}{\partial\tau^2}+\frac{\partial^2}{\partial z_1^2}+\vect{\nabla_\vect{x}}^2$.

Substituting eq.~\eqn{phiFourier} into eq.~\eqn{ZM1}, and simplifying the argument of the exponential by performing the integrations (see appendix~\secn{AppFourierDecomp} for more details)~\cite{Kapusta:2006pm,Bellac:2011kqa,Laine:2016hma}, we obtain
{\allowdisplaybreaks
\be
\mathcal{Z}\propto\int_{\phi_{\text{cond.}}}\!\!\!\!\!\!\!\!\!\left[\mathcal{D}\phi(\tau,\,z_1,\,\vect{x})\right]\exp\left\{%
-\frac{\beta^2}{2}\sum_{n\in\Z}\sum_{\ell_1\in\N}\sum_{\vect{n}_k\in\Z^{D-2}}\left\lvert\tilde{\phi}_{n,\ell_1}%
(\vect{\omega}_k)\right\rvert^2\left(\omega_{n}^2+\omega_{\ell_1}^2+\vect{\omega}_k^2+m^2\right)\right\}%
.~\labeq{ZM1Bis}%
\ee}
\HS{0.12}As in the Stefan-Boltzmann case, we are faced with an issue of double counting when performing explicitly the path integral. This problem can be accounted for, in the noninteracting case, by separating the Fourier modes into real and imaginary parts
{\allowdisplaybreaks
\begin{align}
\tilde{\phi}_{n,{\ell_1}}(\vect{\omega}_k)&=a_{n,{\ell_1}}(\vect{\omega}_k)+ib_{n,{\ell_1}}(\vect{\omega}_k)\nn\\%
\!\!\Rightarrow\left\lvert\tilde{\phi}_{n,{\ell_1}}(\vect{\omega}_k)\right\rvert^2%
&=a^2_{n,{\ell_1}}(\vect{\omega}_k)+b^2_{n,{\ell_1}}(\vect{\omega}_k),%
\end{align}}
and since the field is required to be real, we obtain the following restrictions
{\allowdisplaybreaks
\begin{alignat}{2}
&a_{-n,{\ell_1}}(-\vect{\omega}_k)=a_{n,{\ell_1}}(\vect{\omega}_k)\quad\quad\text{and}\!\!\!\!\!\!\quad\quad%
&\quad b_{-n,{\ell_1}}(-\vect{\omega}_k)=-b_{n,{\ell_1}}(\vect{\omega}_k)\nn\\%
&\hspa{3.5}\Rightarrow b_{0,{\ell_1}}(\vect{0})=0,%
\end{alignat}}
\HS{0.12}from which one may choose a set of physically relevant independent $\phi$-variables over which to integrate. Following the standard procedure~\cite{Kapusta:2006pm,Bellac:2011kqa,Laine:2016hma}, one may then perform the infinite set of Gaussian integrals (dropping any $T$- and $L_i$-independent factors~\cite{Bernard:1974bq}) to obtain the partition function of a free scalar field in $D-1$ spatial dimensions with $c=1$ geometrically confined dimension. However, since thermodynamic quantities are straightforwardly related to the logarithm of the partition function, we will find the more useful quantity to be
{\allowdisplaybreaks
\be
\ln\mathcal{Z}^{(1)}(T,\,L_1)=\ln\left\{\prod_{n\in\Z}\prod_{\ell_1\in\N}\prod_{\vect{n}_k\in\Z^{D-2}}%
\left[\beta^2\left(\omega_{n}^2+\omega_{\ell_1}^2+\vect{\omega}_k^2+m^2\right)\right]^{-\shalf}\right\}%
.\labeq{ZM1final}
\ee}
\HS{0.12}Formally, eq.~\eqn{ZM1final} is our final result for the logarithm of the partition function of a single neutral noninteracting scalar field in between two parallel plates. Extending~\eqn{ZM1final} to arbitrary $c\leq D-1$ compactified spatial dimensions is relatively straightforward, and the corresponding logarithm of the partition function with $c$ geometrically confined dimensions is therefore given by
{\allowdisplaybreaks
\be
\ln\mathcal{Z}^{(c)}(T,\,\left\{L_i\right\})=-\half\sum_{n\in\Z}\sum_{\vect{\ell}\in\N^{c}}\sum_{\vect{n}_k\in\Z^{D-1-c}}%
\left[\ln\left\{\beta^2\left(\left(2\pi nT\right)^2+\sum_{i=1}^{c}\left[\left(\frac{\pi\ell_i}{L_i}\right)^2\right]%
+\vect{\omega}_k^2+m^2\right)\right\}\right]%
,\labeq{LnZc}
\ee}
\HS{0.12}where we still have to send $R$ to infinity.

In Appendix \secn{AppCanonicalComputation} and \secn{AppAlternativeComputation} we present two independent methods of evaluating \eq{LnZc}.  We have confirmed numerically that both methods yield the same results.  The two methods are mutually complementary as they naturally yield results with different numerical convergence properties.  More precisely, the usual method yields a result that is always better suited for small values of the dimensionless parameters ($T\times L_i$).  The alternative method yields a result better suited for high values of these dimensionless parameters and thus to recover the usual Stefan-Boltzmann limits. Moreover, the usual result explicitly isolates the $T$-independent contributions, and one can pass from one compactified dimension case to the next in an iterative manner; therefore the usual method explicitly yields the known zero temperature Casimir pressure. 

\subsection{Evaluating the free energy}\labsec{EvalFreeEnergy}

We now present our final results for the free energy in the massless limit using both eq.~\eqn{LnZcfinal2} and results from eq.~\eqn{MasterIAlpha}. Recall that the free energy density is given by $f\equiv F/V$,~$F$ being the total free energy, and therefore reads
{\allowdisplaybreaks
\be
f(T,\,\left\{L_i\right\})\equiv -\frac{T}{V}\ln\mathcal{Z}(T,\,\left\{L_i\right\})%
,\labeq{FreeEnergyPartitionFunction}
\ee}
\HS{0.12}where $V\equiv \prod_{i=1}^{D-1}~L_i$. 

\paragraph{Case I: Two infinite parallel plates\\*}

In the usual approach, and after considering $D-1=3-2\epsilon$, simplifying the summations with Bessel functions, expanding about $\epsilon=0$, and performing summations using analytic continuation of the Epstein-zeta functions, eq.~\eqn{LnZcfinal2} leads to the further refined expression of the free energy density, $f^{(c=1)}(T,L_1)$, of a system geometrically confined in between two infinite parallel plates separated by a distance $L_1$
{\allowdisplaybreaks
\be
f^{(1)}=-\frac{\pi^2}{1440L_1^4}%
-\frac{T^2}{2L_1^2}\sum_{\ell=1}^{\infty}\left[\ell~\Li_2\,\left(\e^{-\frac{\pi\ell}{TL_1}}\right)\right]%
-\frac{T^3}{2\pi L_1}\sum_{\ell=1}^{\infty}\left[\Li_3\,\left(\e^{-\frac{\pi\ell}{TL_1}}\right)\right]%
,\labeq{M1finalResultWAHIK}%
\ee}
\HS{0.12}where the $\Li_n$ are the usual polylogarithm functions. Although it is not obvious at first glance, eq.~\eqn{M1finalResultWAHIK} reduces to the correct infinite volume limit, as is shown in appendix~\secn{AppCanonicalComputation} (the same will be true for the tube and box cases, eq.~\eqn{M2finalResultWAHIK} and eq.~\eqn{M3finalResultWAHIK}, below).

We can also follow the alternative approach, as presented in appendix~\secn{AppAlternativeComputation}, to compute the same free energy density. Employing similar techniques and making use of eq.~\eqn{lnA} in order to analytically continue the Epstein-zeta functions without formally manipulating a divergence, we end up with a similar expression which we further resum using the contour integral representation of the polylogarithm function. We find
{\allowdisplaybreaks
\bea
f^{(1)}&=&-\frac{\pi^2T^4}{90}+\frac{\zeta(3) T^3}{4\pi L_1}%
-\frac{T^2}{8L_1^2}\sum_{\ell=1}^\infty\left[\frac{\text{csch}^2\left(2\pi TL_1\ell\right)}{\ell^2}\right]\nn\\%
&-&\frac{\zeta(3)T}{16\pi L_1^3}-\frac{T}{16\pi L_1^3}\sum_{\ell=1}^\infty\left[\frac{\coth\left(2\pi TL_1\ell\right)-1}{\ell^3}\right]%
,\labeq{M1finalResultSM}%
\eea}%
where in the last summation, we kept the $-1$ explicit---even though there exists a simple closed form for $\coth(x)-1$---since the less simple form improves the convergence properties of the expression.

As previously mentioned, we see that eq.~\eqn{M1finalResultWAHIK} converges exponentially fast for low values of the dimensionless variable $TL_1$, while eq.~\eqn{M1finalResultSM} converges exponentially fast for high values of this dimensionless variable. The latter even has enhanced convergence properties (due to the additional resummation which we performed), for nearly all $TL_1$ down to $TL_1\sim 10^{-6}$.

A similar resummation could have been performed on eq.~\eqn{M1finalResultWAHIK}, but we will refrain in doing so since the alternative result covers nearly all $TL_1$ values. 

\paragraph{Case II: Infinite rectangular tube\\*}

Using the usual approach together with the same type of procedure as in the parallel plates case, we may then set $c=2$ in eq.~\eqn{LnZcfinal2}, perform a few more refinements, and obtain the free energy density, $f^{(c=2)}(T,L_1,L_2)$, of a system geometrically confined within an infinite rectangular tube of section $L_1\times L_2$
{\allowdisplaybreaks
\bea
f^{(2)}&=&\frac{45\,L_1\,\zeta(3)-\pi^3L_2}{1440\pi L_2L_1^4}-\frac{1}{64\pi\,L_1^2\,L_2^3}\sum_{\ell=1}^{\infty}%
\left[\frac{L_1\left(1-\e^{-\frac{2\pi L_2}{L_1}\ell}\right)+2\pi L_2\ell}{\ell^3}\times\text{csch}^2\left(\frac{\pi L_2}{L_1}~\ell\right)\right]\nn\\%
&-&\frac{T}{L_1L_2}\sum_{\ell,\ell_1,\ell_2=1}^{\infty}\left[\frac{1}{\ell}\sqrt{\left(\frac{\ell_1}{L_1}\right)^2%
+\left(\frac{\ell_2}{L_2}\right)^2}~K_1\left(\frac{\pi\ell}{T}\sqrt{\left(\frac{\ell_1}{L_1}\right)^2+\left(\frac{\ell_2}{L_2}\right)^2}\right)\right]%
,\labeq{M2finalResultWAHIK}%
\eea}%
where $K_n$ are the usual modified Bessel function of the second kind.  Although eq.~\eqn{M2finalResultWAHIK} does not appear to be manifestly invariant under exchange of $ L_i \leftrightarrow L_j $, it is formally symmetric since it is the direct result of eq.~\eqn{LnZcfinal2}, which is itself manifestly symmetric.  We have also directly checked that eq.~\eqn{M2finalResultWAHIK} is numerically invariant under exchange of $ L_i $.  These comments hold also for eq.~\eqn{M3finalResultWAHIK}.

We may also use our alternative approach, in order to compute the very same free energy density, doing so we obtain
{\allowdisplaybreaks
\bea
f^{(2)}&=&-\frac{\pi^2 T^4}{90}+\frac{\zeta\left(3\right)T^3(L_1+L_2)}{4\pi L_1L_2}%
-\frac{\pi T^2}{24L_1 L_2}+\frac{\pi T(L_1+L_2)}{96L_1^2 L_2^2}-\frac{\zeta\left(3\right)T(L_1^3+L_2^3)}{32\pi L_1^3 L_2^3}\nn\\%
&-&\frac{T^2}{4L_1^2}\sum_{\ell=1}^\infty\bigg[\ell~\text{Li}_{2}\left(e^{-4\pi TL_1\ell}\right)\bigg]%
-\frac{T^2}{4L_2^2}\sum_{\ell=1}^\infty\bigg[\ell~\text{Li}_{2}\left(e^{-4\pi TL_2\ell}\right)\bigg]\nn\\%
&-&\frac{T}{16\pi L_1^3}\sum_{\ell=1}^\infty\bigg[\text{Li}_{3}\left(e^{-4\pi TL_1\ell}\right)\bigg]%
-\frac{T}{16\pi L_2^3}\sum_{\ell=1}^\infty\bigg[\text{Li}_{3}\left(e^{-4\pi TL_2\ell}\right)\bigg]\nn\\%
&+&\frac{T^2}{L_1L_2}\sum_{n,\ell=1}^\infty\left[\frac{n}{\ell}~K_{1}\bigg(4\pi TL_1 n\ell\bigg)%
+\frac{n}{\ell}~K_{1}\bigg(4\pi TL_2 n\ell\bigg)\right]\\%
&-&\frac{T}{8L_1 L_2^2}\sum_{\ell_1=1}^\infty~\sum_{(n,\ell_2)\in\Z^{2}\setminus\{\vect{0}\}}%
\left[\frac{\sqrt{\ell_2^2+(2TL_2)^2n^2}}{\ell_1}~K_{1}\left(\frac{2\pi L_1}{L_2}~\ell_1~\sqrt{\ell_2^2+(2TL_2)^2n^2}\right)\right]\nn\\%
&-&\frac{T}{8L_2 L_1^2}\sum_{\ell_2=1}^\infty~\sum_{(n,\ell_1)\in\Z^{2}\setminus\{\vect{0}\}}%
\left[\frac{\sqrt{\ell_1^2+(2TL_1)^2n^2}}{\ell_2}~K_{1}\left(\frac{2\pi L_2}{L_1}~\ell_2~\sqrt{\ell_1^2+(2TL_1)^2n^2}\right)\right]\nn%
.\labeq{M2finalResultSM}%
\eea}

\paragraph{Case III: Finite volume box\\*}

Using, again, the same type of procedure as for the two previously cases, we may set $c=3$ in eq.~\eqn{LnZcfinal2}, perform a few more refinements, and then obtain within the usual approach, the free energy density, $f^{(c=3)}(T,L_1,L_2,L_3)$, of a system geometrically confined in a finite volume box
{\allowdisplaybreaks
\bea
f^{(3)}
	&=&-\frac{\pi^2}{1440L_1^4}
	-\frac{1}{64\pi\,L_1\,L_2^3}\sum_{\ell=1}^{\infty}%
\left[\frac{1}{\ell^3}\left(1+2\pi\ell\frac{L_2}{L1}-\e^{-2\pi\ell\frac{L_2}{L1}}\right)
\times\text{csch}^2\left(\pi\ell\frac{L_2}{L1}\right)\right]\nn\\%
	&-&\frac{\pi}{96L_1^2L_2L_3}+\frac{(L_2+L_3)~\zeta(3)}{32L_1^3L_2L_3}%
+\frac{1}{4L_1^2L_2L_3}\sum_{\ell_1,\ell_2=1}^\infty\left[\frac{\ell_1}{\ell_2}~K_{1}\left(2\pi\ell_1\ell_2\frac{L_2}{L_1}\right)\right]\nn\\%
	&-&\frac{1}{2 L_1L_2L_3}\sum_{\ell_1,\ell_2,\ell_3=1}^{\infty}\left[\frac{1}{\ell_3}\sqrt{\left(\frac{\ell_1}{L_1}\right)^2%
+\left(\frac{\ell_2}{L_2}\right)^2}~%
K_1\left(2\pi L_3\ell_3~\sqrt{\left(\frac{\ell_1}{L_1}\right)^2+\left(\frac{\ell_2}{L_2}\right)^2}\right)\right]\nn\\%
	&-&\frac{\sqrt{2}}{L_1L_2L_3}\sum_{\ell,\ell_1,\ell_2,\ell_3=1}^\infty
		\left[\sqrt{\frac{T}{\ell}}\left(\frac{\ell_1^2}{L_1^2}+\frac{\ell_2^2}{L_2^2}+\frac{\ell_3^2}{L_3^2}\right)^{\frac{1}{4}}%
K_{\half}\left(\frac{\ell\pi}{T}\sqrt{\frac{\ell_1^2}{L_1^2}+\frac{\ell_2^2}{L_2^2}+\frac{\ell_3^2}{L_3^2}}\right)\right]%
.\labeq{M3finalResultWAHIK}%
\eea}

Introducing a new notation, in order to shorten our next representation, namely the set $\sigma$ of permutations of $L_i$,\footnote{Containing six elements. For instance, if the third element of $\sigma$ is $\sigma_3=(L_3,\,L_1,\,L_2)$, then $\sigma_3(2)=L_1$.} as well as $\Omega=\{L_1,\,L_2,\,L_3\}$, we may also use our alternative approach to compute the same free energy density,
{\allowdisplaybreaks
\bea
f^{(3)}&=&-\frac{\pi^2 T^4}{90}+\frac{T~\log\Big(8T^3L_1L_2L_3\Big)}{24L_1L_2L_3}%
+\frac{\zeta\left(3\right)T^3(L_1L_2+L_1L_3+L_2L_3)}{4\pi L_1L_2L_3}-\frac{\pi T^2(L_1+L_2+L_3)}{24L_1L_2L_3}\nn\\%
&-&\frac{\zeta\left(3\right)T(L_1^3L_2^3+L_1^3L_3^3+L_2^3L_3^3)}{48\pi L_1^3L_2^3L_3^3}%
+\frac{\pi T(L_1^2L_2+L_1L_2^2+L_1^2L_3+L_1L_3^2+L_2^2L_3+L_2L_3^2)}{144L_1^2L_2^2L_3^2}\nn\\%
&+&\frac{T}{4L_1L_2L_3}\sum_{L_i\in\Omega}\sum_{\ell\in\N}\Bigg[\log\bigg(1-\e^{-4\pi TL_i \ell}\bigg)%
-\frac{T^2\ell}{6L_i^2}~\Li_2\bigg(\e^{-4\pi TL_i \ell}\bigg)-\frac{T}{24\pi L_i^3}~\Li_3\bigg(\e^{-4\pi TL_i \ell}\bigg)\Bigg]\nn\\%
&+&\sum_{\sigma_i\in\sigma}\left[\frac{T^2}{4L_1L_2L_3}\sum_{(n,\ell)\in\N^2}%
\Bigg[\frac{\sigma_i(1)+\sigma_i(2)\ell}{n}~K_1\left(4\pi T\sigma_i(3) n\ell \right)\Bigg]\right.\nn\\%
&&\hspa{0.75}-\frac{T}{16L_1L_2L_3}\sum_{(n,\ell)\in\Z^{2}\setminus\{\vect{0}\}}%
\log\left(1-\e^{-\frac{2\pi\sigma_i(1)}{\sigma_i(2)}\sqrt{\ell^2+(2\pi T\sigma_i(2))^2n^2}}\right)\\%
&&\hspa{0.75}-\frac{T}{24L_1^2L_2^2L_3^2}\sum_{\ell_1\in\N}\sum_{(n,\ell_2)\in\Z^{2}\setminus\{\vect{0}\}}\Bigg[%
\sigma_i(1)(\sigma_i(2))^2\frac{\sqrt{\ell_2^2+(2nT\sigma_i(3))^2}}{\ell_1^2}\times\nn\\%
&&\hspa{5.375}\times K_1\left(\frac{2\pi \sigma_i(1)}{\sigma_i(3)}\ell_1\sqrt{\ell_2^2+(2T\sigma_i(3))^2n^2}\right)\Bigg]\nn\\%
&&\hspa{0.75}+\left.\frac{T}{48L_1L_2L_3}\sum_{(n,\ell_1,\ell_2)\in\Z^{3}\setminus\{\vect{0}\}}%
\log\left(1-\e^{-4\pi T\sigma_i(1)\sqrt{n^2+\frac{\ell_1^2}{(2T\sigma_i(2))^2}+\frac{\ell_2^2}{(2T\sigma_i(3))^2}}}\right)\right]\nn%
.\labeq{M3finalResultSM}%
\eea}

\section{Thermodynamic expressions, first and third laws of thermodynamics, statistical fluctuations, and nonextensivity}\labsec{FiniteSizeThermo}

In this section we 1) examine how the First Law of Thermodynamics is altered by the compactified directions, in particular showing how the pressure is no longer a scalar but depends on direction, 2) provide some formal manipulations to arrive at the formulae for the usual thermodynamic quantities such as the EoS that we ultimately evaluate numerically in \hsec{ResultThermo}, 3) comment on the effects of geometric confinement on the Third Law of Thermodynamics, 4) quantify the thermal fluctuations of our geometrically confined system, and 5) show explicitly that a thermal system between parallel plates is not extensive.

\subsection{Modification of the first law}\labsec{ModifOfFirstLaw}

Recall that the fundamental object that is computed when using the canonical ensemble is the partition function, $\mathcal{Z}(T,\,\left\{L_i\right\})$. The partition function is indeed fundamental in the sense that it allows one to compute any thermodynamic potential, and therefore any thermodynamic quantity of interest.  Recall also that the natural potential to use in the canonical ensemble is the Helmholtz free energy $F(T,\,\left\{L_i\right\})$, as previously derived in section~\secn{FieldTheoCalculations}.

Besides being related to the partition function, the free energy is also defined as the Legendre transform~\cite{Zemansky:1937} of the total energy $E(S,\,\left\{L_i\right\})$ with respect to the total entropy $S$
{\allowdisplaybreaks
\be
F(T,\,\left\{L_i\right\})\equiv E(S,\,\left\{L_i\right\})-TS%
,\labeq{LegendreTransformDef}%
\ee}%
where
{\allowdisplaybreaks
\be
T\equiv\partdiffat{E(S,\,\left\{L_i\right\})}{S}{\{L_{i}\}}%
\labeq{Tcomp}%
\ee}%
provides a generalization of the definition for the usual Stefan-Boltzmann thermodynamic temperature~\cite{Callen:1985tit,LeBellac:2004}. Recall also that the Legendre transform in eq.~\eqn{LegendreTransformDef} is well defined only if $E$ is a convex function of $S$, at constant lengths $L_i$, i.e.\ $E^{''}(S,\,\ldots)>0$. We will see below that $E$ is convex for our geometrically confined systems. Furthermore, the total energy as a function of the total entropy $S$ and the lengths $L_i$, i.e.\ the entropy as a function of only extensive parameters,\footnote{The concept of extensivity being meaningful only in the large size, thermodynamic limit~\cite{TOUCHETTE200284}.} is a thermodynamically complete function: $S(E,\,L_i)$ contains the full thermodynamic information.  Other relations such as, e.g., the pressure as a function of $T$ and $L_i$, $p(T,\,L_i)$, are {\textit{equations of state}}, which may only contain partial information about the thermodynamics of the system. See, e.g.,~\cite{Callen:1985tit,LeBellac:2004} for more details on these general concepts.

From both the variable dependence of the total energy and eq.~\eqn{Tcomp}, we can readily write
{\allowdisplaybreaks
\be
dE=TdS+\sum_{i=1}^3\left[\left(\partdiffat{E}{L_i}{S,\{L_{k\neq i}\}}\right)dL_i\right].\labeq{dEForm1}%
\ee}%
Let us then comment on the subsequent modification of the first law of thermodynamics, recalling that for infinitesimal changes the most general form~\cite{Callen:1985tit,LeBellac:2004} of the First Law is
{\allowdisplaybreaks
\be
dE=\dbar Q+\dbar W,\labeq{General1stLaw}%
\ee}%
where $\dbar Q$ and $\dbar W$ are respectively the infinitesimal transfers of heat and work. Considering quasistatic processes, without any loss of generality, and since the only work which can be done is the one coming from a displacement of the boundaries in any of the three directions, we obtain a generalization of the First Law of Thermodynamics to include the possibility of asymmetric pressures:
{\allowdisplaybreaks
\be
dE=TdS-V\sum_{i=1}^3\left[\frac{P_i}{L_i}~dL_i\right]%
,\labeq{NewFirstLaw}%
\ee}%
where the $P_i$ are the pressures along each ($i$) of the three directions and are defined as
{\allowdisplaybreaks
\be
P_j\equiv-\frac{L_j}{V}\partdiffat{E(S,\,\left\{L_i\right\})}{L_j}{S,{\{L_{k\neq j}\}}}%
.\labeq{Picomp}%
\ee}%

Eq.\ \eqn{Picomp} shows explicitly that for asymmetric systems, the pressure may be anisotropic.  The above manipulations also allow us to make the temperature and pressure functions of the entropy, $S$, and the system side lengths, $L_i$: $T=T(S,\,\left\{L_i\right\})$ and $P_i=P_i(S,\,\left\{L_i\right\})$.  Note that in the thermodynamic limit, in which all lengths $L_i$ become asymptotically large, the second term of eq.~\eqn{NewFirstLaw} reduces to the usual $-PdV$. Thus, in the thermodynamic limit, we recover that the pressure $P$ is thermodynamically complete, an obvious consequence of the relation $PV=-F$, which holds in the thermodynamic limit. 

\subsection{Thermodynamic expressions}\labsec{ThermoExpressions}

Let us now give some details concerning the practical computation of various thermodynamic quantities.  We start from the free energy as derived in section~\secn{FieldTheoCalculations}.  

Starting from eq.~\eqn{LegendreTransformDef}, we already have that
{\allowdisplaybreaks
\be
S(T,\,\left\{L_i\right\})\equiv -\partdiffat{F(T,\,\left\{L_i\right\})}{T}{\left\{L_i\right\}}%
,\labeq{TotEntropy}%
\ee}
\HS{0.12}and thus
{\allowdisplaybreaks
\be
E(T,\,\left\{L_i\right\})\equiv F(T,\,\left\{L_i\right\})-T~\partdiffat{F(T,\,\left\{L_i\right\})}{T}{\left\{L_i\right\}}%
.\labeq{TotEnergy}%
\ee}
\HS{0.12}In addition, given the proof in section~\secn{AppLengthDerivatives}, we know that eq.~\eqn{Picomp} is equivalent to
{\allowdisplaybreaks
\be
P_j\equiv-\frac{L_j}{V}\partdiffat{F(T,\,\left\{L_i\right\})}{L_j}{T,{\{L_{k\neq j}\}}}%
.\labeq{PicompBis}%
\ee}
\HS{0.12}Furthermore, using eqs.~\eqn{TotEnergy} and~\eqn{PicompBis}, we obtain the trace of the energy-momentum tensor
{\allowdisplaybreaks
\be
T^{\mu}_{\mu}(T,\,\left\{L_i\right\})\equiv \varepsilon(T,\,\left\{L_i\right\})-\sum_{j=1}^3P_j(T,\,\left\{L_i\right\})%
,\labeq{Tmumucomp}%
\ee}
\HS{0.12}where $\varepsilon\equiv E/V$ is the energy density. Not surprisingly, given that our noninteracting model is indeed conformal, this quantity clearly vanishes. Notice further that it is indeed a generalization of the Stefan-Boltzmann limit result, $T^{\mu}_{\mu}=\varepsilon-3P$, and encodes the aforementioned system anisotropy.

Coming back now to eq.~\eqn{General1stLaw}, and using the explicit form for the total amount of infinitesimal work that can be done from eq.~\eqn{NewFirstLaw}, we obtain a heat function $Q$ which admits an exact differential, when all the lengths $L_i$ are kept fixed\footnote{Hence when the volume is fixed too, even if the latter constraint does not imply the former.}. More precisely, we have
{\allowdisplaybreaks
\be
\dbar Q=\left.dE\right|_{\{L_i\}}%
,\labeq{HeatFunction1}%
\ee}%
which leads to the following definition of the heat capacity at constant sizes (hence volume):
{\allowdisplaybreaks
\be
C_\rmi{v}(T,\,\left\{L_i\right\})\equiv\partdiffat{E(T,\,\left\{L_i\right\})}{T}{\{L_j\}}.%
\labeq{HeatCapacity}%
\ee}%
We keep the usual $V$ index on our heat capacity to emphasize the connection with the usual heat capacity at fixed volume.  It is worth noting that given eqs.~\eqn{TotEntropy} and~\eqn{TotEnergy}, the heat capacity can be re-written as
{\allowdisplaybreaks
\be
C_\rmi{v}(T,\,\left\{L_i\right\})= T\partdiffat{S(T,\,\left\{L_i\right\})}{T}{\{L_j\}}%
.\labeq{HeatCapacityBis}%
\ee}%
We will see in section~\secn{ResultThermo} that our finite-size heat capacity is always positive.

The positivity of the finite size heat capacity means the connection between the pressure, temperature, and entropy is straightforward:  the relation $T=T(S,\,\left\{L_i\right\})$ is  invertible and hence equivalent to $S=S(T,\,\left\{L_i\right\})$.  Thus we may equivalently use $P_i(T,\,\left\{L_i\right\})$ in place of $P_i(S,\,\left\{L_i\right\})$.  Further, this statement together with eq.~\eqn{HeatCapacityBis} imply that $E^{''}(S,\,\ldots)=T/C_\rmi{v}>0$, thus making the Legendre transform connecting the free energy with the total energy well defined.

\subsection{On the third law}\labsec{ZerothAndThirdLaws}

The Third Law of Thermodynamics states that the entropy reduces to a constant value (usually 0) as the temperature of a system goes to 0~\cite{Callen:1985tit,LeBellac:2004}.  Some implementations of the Lifschitz theory of the Casimir effect in QED claim that in their system the Third Law of Thermodynamics is violated~\cite{Klimchitskaya:2009cw}.  We first recall that in our system the spatial compactification(s) required for establishing a geometric confinement lead to a zero temperature Casimir-type geometrical contribution to the free energy $F(T,\,\left\{L_i\right\})$. However, given the definition of the entropy~\eqn{TotEntropy}, we expect that this temperature independent contribution vanishes from the expression of the entropy. The numerical value of the entropy can then be assessed by appropriately rescaling $S$ by the dimensionless combination $T^3V$. 
Numerical evaluation of this quantity (see section~\secn{ResultThermo}) clearly shows that $\tilde{s}\rightarrow 0$ as the temperature vanishes. We refer to~\cite{Hoye:2002at} for another study of the Third Law in the context of QED Casimir systems.


\subsection{Fluctuations of the energy}\labsec{FluctuEnergy}

Recall that in the micro-canonical ensemble the total energy is fixed. However, in the canonical ensemble the system is in contact with a heat bath of infinite heat capacity and all energies of the system are accessible through a probability distribution.  For the canonical ensemble, then, there is a mean total energy together with a standard deviation (the square root of the statistical variance $\overline{\Delta E^2}=\overline{E^2}-\overline{E}^2$).  It is precisely the variance of the energy of the system that we take to provide us with insight into the energy fluctuations of our canonical ensemble system.  

One may compute the mean total energy $\overline{E}\equiv E$ in the usual way with partition functions
{\allowdisplaybreaks
\be
E\equiv\big\langle\hat{H}\big\rangle_{\e^{-\beta H}}%
=\frac{1}{\mathcal{Z}}\pard{\mathcal{Z}}{(-\beta)},%
\ee}%
where the angular brackets denote the usual ensemble average
{\allowdisplaybreaks
\be
\left\langle\ldots\right\rangle_{\e^{-\beta H}}\equiv\frac{1}{\mathcal{Z}}\Tr\left[\ldots~\exp\left\{-\beta\int_{V}d^{D-1}x\,\mathcal{\hat{H}}\right\}\right].%
\ee}%
The standard deviation, $\sigma_\rmi{E}=\sqrt{\overline{\Delta E^2}}$, is then 
{\allowdisplaybreaks
\be
\sigma_\rmi{E}\equiv\sqrt{\big\langle\hat{H}^2\big\rangle_{\e^{-\beta H}}-\big\langle\hat{H}\big\rangle_{\e^{-\beta H}}^2}%
=\sqrt{\pard{E}{(-\beta)}}.%
\ee}%
Therefore, after a little manipulation, the standard deviation may be simply written as
{\allowdisplaybreaks
\be
\sigma_\rmi{E}=T\sqrt{C_\rmi{v}}%
,\labeq{StandardDeviation}
\ee}%
where we recall that the heat capacity $C_\rmi{v}$ is defined in eq.~\eqn{HeatCapacity}.

Recall that both the heat capacity and the total energy of a system scale with the volume $V$ of the system.  Therefore as the system volume increases, the relative size of the energy fluctuations decreases, $\sigma_E/E\sim1/V^{1/2}\rightarrow0$.  Systems that are not compactified in all directions, such as the case of two infinite parallel plates or a rectangular tube of infinite length, have infinite volume and therefore experience \emph{no} fluctuations in their total energy.  Crucially, then, for these systems with no fluctuations in total system energy, we conclude that the canonical ensemble must exactly reproduce the results of the microcanonical ensemble.  We will exploit this equivalence of ensembles later to demonstrate a phase transition at a critical length (instead of temperature) for isolated systems of finite extent in some (but not all directions) in section~\secn{PhaseTransition}.

In figure~\fign{FluctuEBox111&113&133}, we plot the mean total energy and its standard deviation as a function of temperature $T$ (in units of $1/L$) for the case of a massless, noninteracting scalar field geometrically confined in a finite-sized box.  The upper left panel shows the results for a symmetric box, with side ratios 1:1:1; the upper right panel shows the same for an asymmetric box of side ratios 1:1:3, a finite volume symmetric tube; the lower panel shows the same for an asymmetric box of side ratios 1:3:3, a set of two finite area parallel plates.  One can see that as $T\times L$ grows large, the system approaches the thermodynamic limit, which we denote by ``SB,'' for Stefan-Boltzmann.  Even out to relatively large $T\times L\sim20$, energy fluctuations are on the order of 10\%.  In the limit that $T\times L\rightarrow0$ the energy density becomes negative, which is the usual case for Casimir-like systems\footnote{The negative energy density implies that at some $T\times L$ the energy density is 0.  Hence it is not very enlightening to plot the relative fluctuations in energy $\sigma_E/E$ (as opposed to $\sigma_E/E_{\mathrm{SB}}$ for small $T\times L$).}.

\begin{figure}[!htbp]\centering
\begin{tabular}{c}
\begin{minipage}[c]{0.5\linewidth}
\includegraphics[scale=0.225]{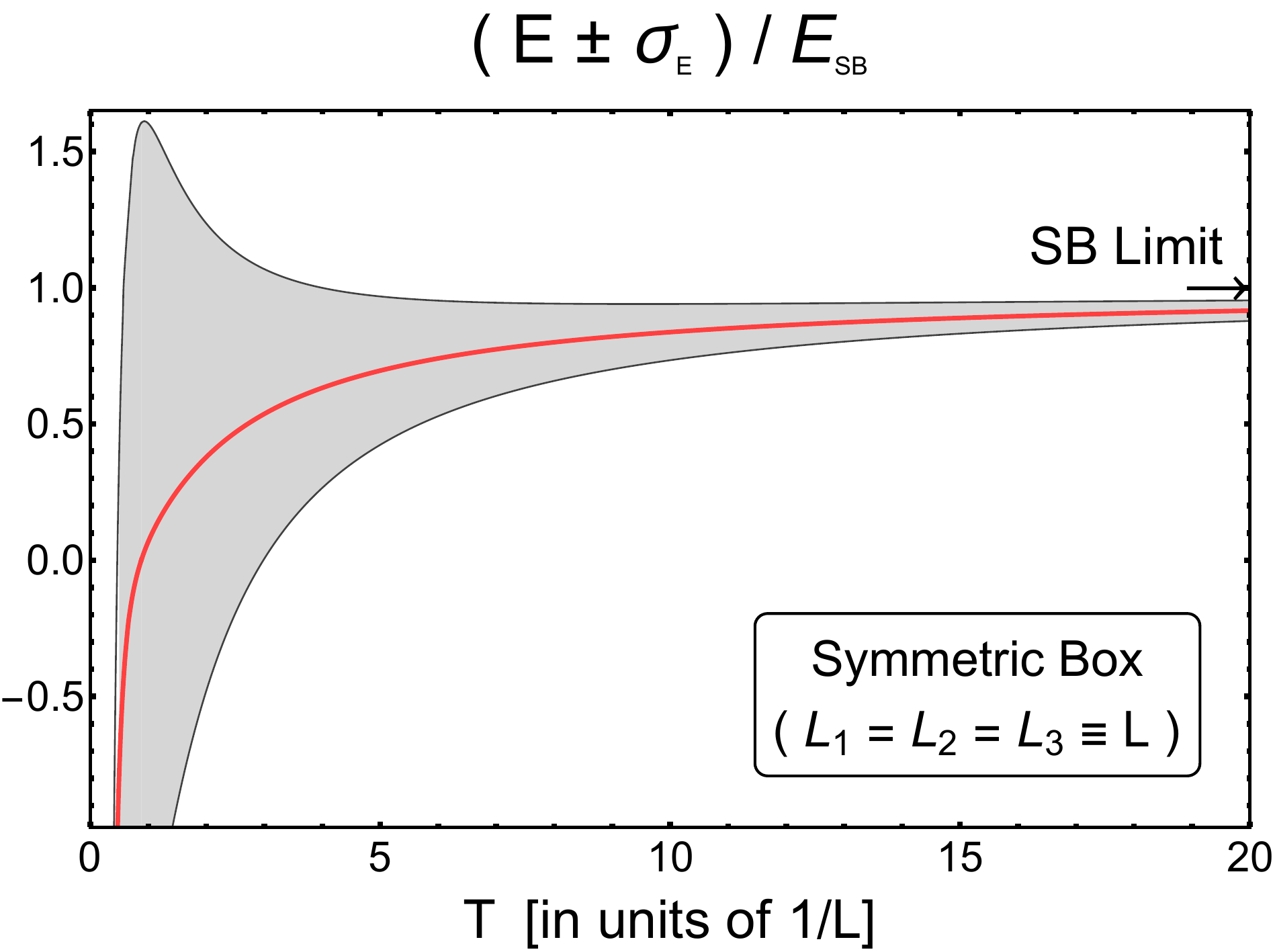}
\end{minipage}\hfil\begin{minipage}[c]{0.5\linewidth}
\includegraphics[scale=0.225]{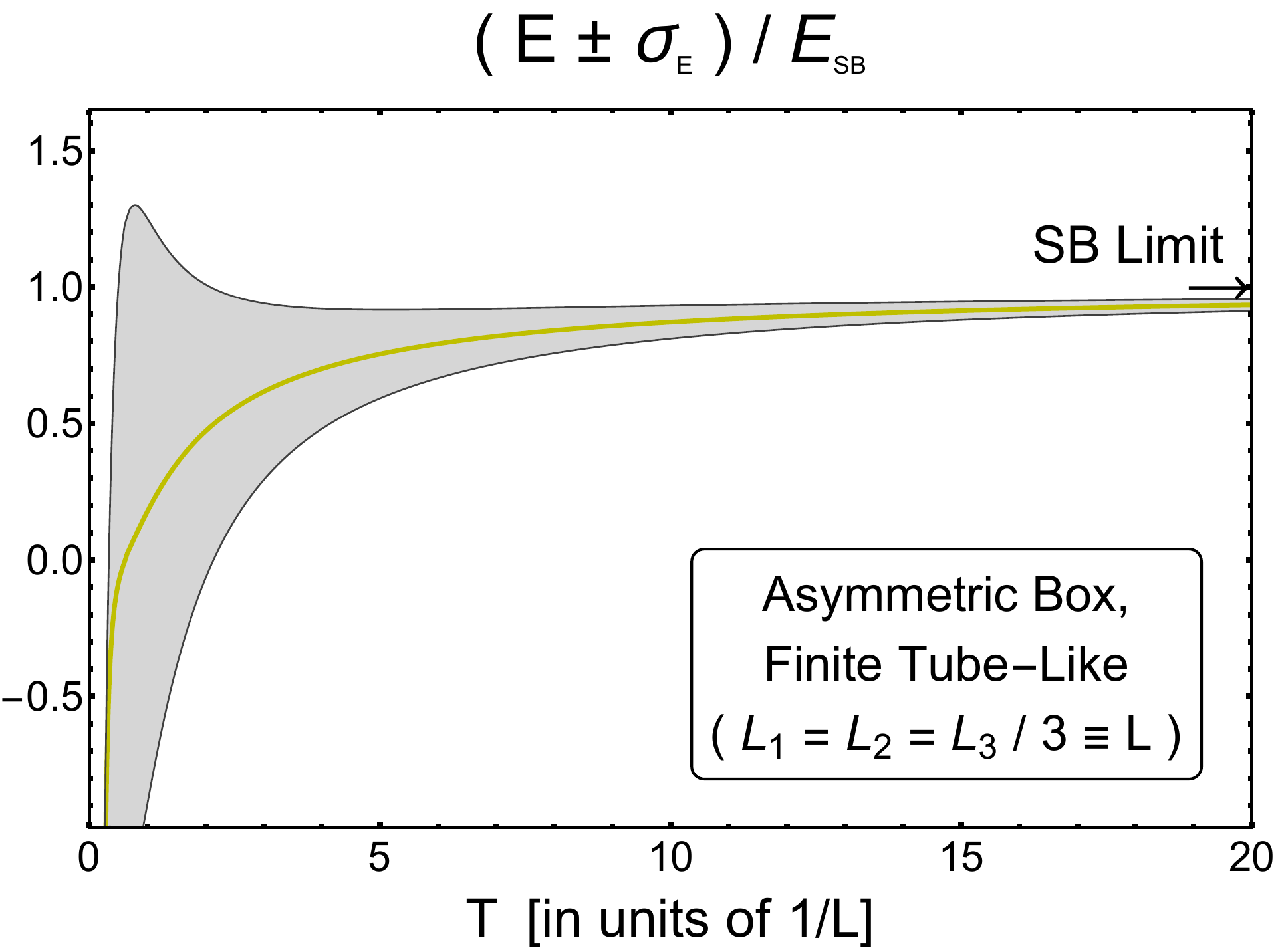}
\end{minipage} \\
\rule{0pt}{40ex}
\includegraphics[scale=0.26125]{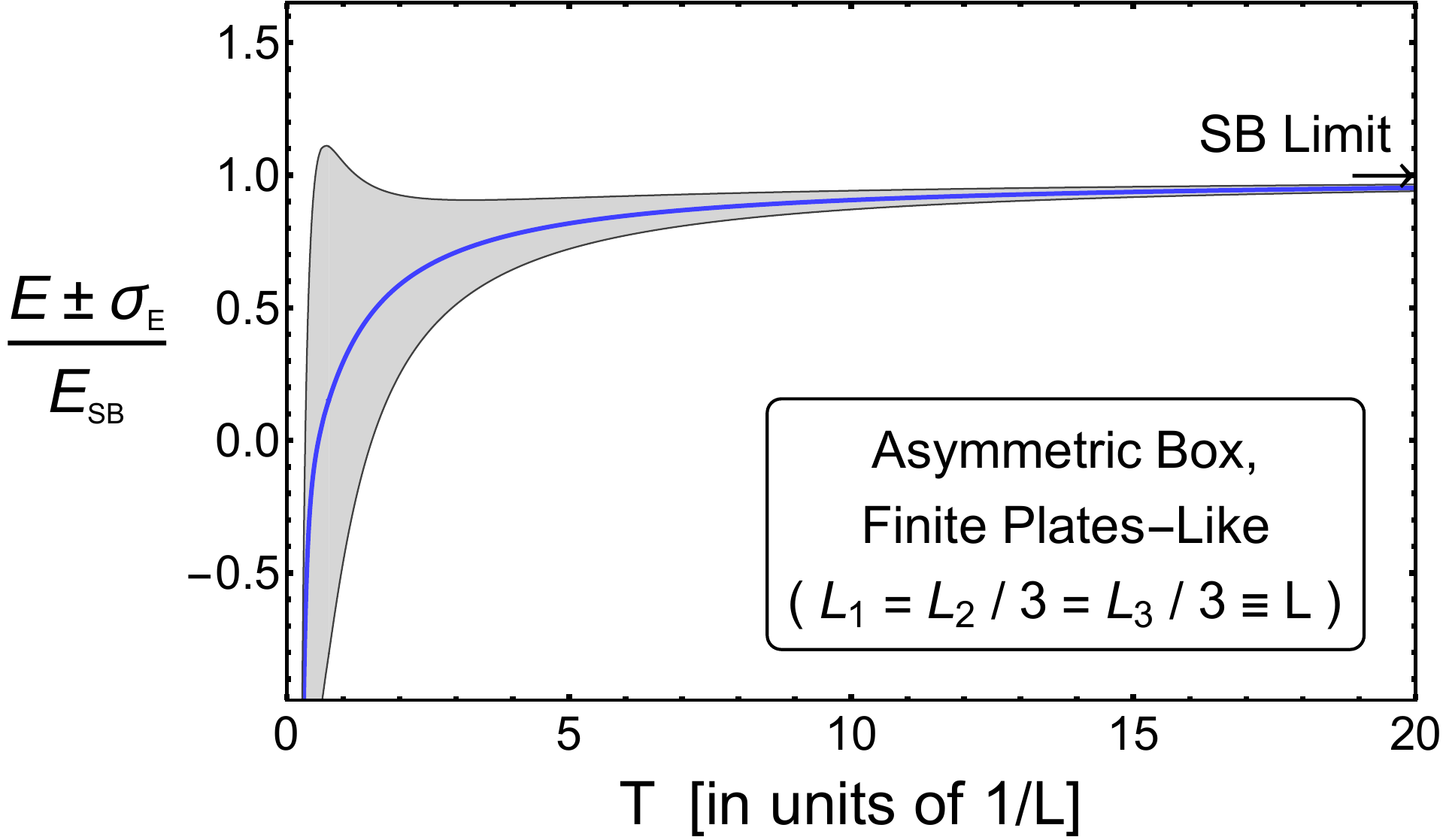}
\end{tabular}
\caption{The mean total energy (thick colored lines) with corresponding standard deviation bands describing the fluctuations (whose edges are the thin black lines), in different cases of finite volume symmetric and asymmetric boxes.  The results have been rescaled to the usual Stefan-Boltzmann limit and are presented as a function of the temperature $T$ in units of $1/L$, where $L$ measures the length of the compactified direction(s). The upper left panel shows the result for a symmetric box of side ratios 1:1:1, while the upper right panel shows that for an asymmetric box of side ratios 1:1:3, and the lower panel for an asymmetric box of side ratios 1:3:3.}
\labfig{FluctuEBox111&113&133}
\end{figure}

\subsection{Nonextensivity of finite size systems}
We show in figure~(\ref{fig:nonextensive}) the deviation from extensivity for the parallel plates case.  The plot shows the difference between the entropy for a massless, noninteracting scalar field between two parallel plates a distance $2L$ apart, divided by the Stefan-Boltzmann limit for the entropy in such a case, $S(2V)/S_{SB}(2V)$, and twice the entropy for plates a distance $L$ apart divided by the Stefan-Boltzmann limit in such a case, $2S(V)/2S_{SB}(V)$ as a function of temperature $T$, measured in units of inverse length $1/L$.  Since the entropy in the Stefan-Boltzmann limit is extensive, the two denominators are the same:
\begin{align}
	\label{eq:extensive}
	\frac{S(2V)}{S_{SB}(2V)} - \frac{2S(V)}{2S_{SB}(V)} = \frac{S(2V)-2S(V)}{2S_{SB}(V)}.
\end{align}
For an extensive system, Eq.\ (\ref{eq:extensive}) is identically 0.  A system with all lengths infinite is extensive.  Hence we may understand why the system approaches extensivity as $T$ increases in the figure as follows.  As $T$ increases, the thermal de Broglie wavelength decreases, and the effective size of the system becomes larger.  The nonextensivity goes to zero as $T\rightarrow0$ trivially as $S(V)/S_{SB}(V)$ goes to zero for small $T$, which we show in detail below in the left panel of figure~\fign{SAndCvPlanes}.

\begin{figure}[!htbp]
\centering
\includegraphics[width=3in]{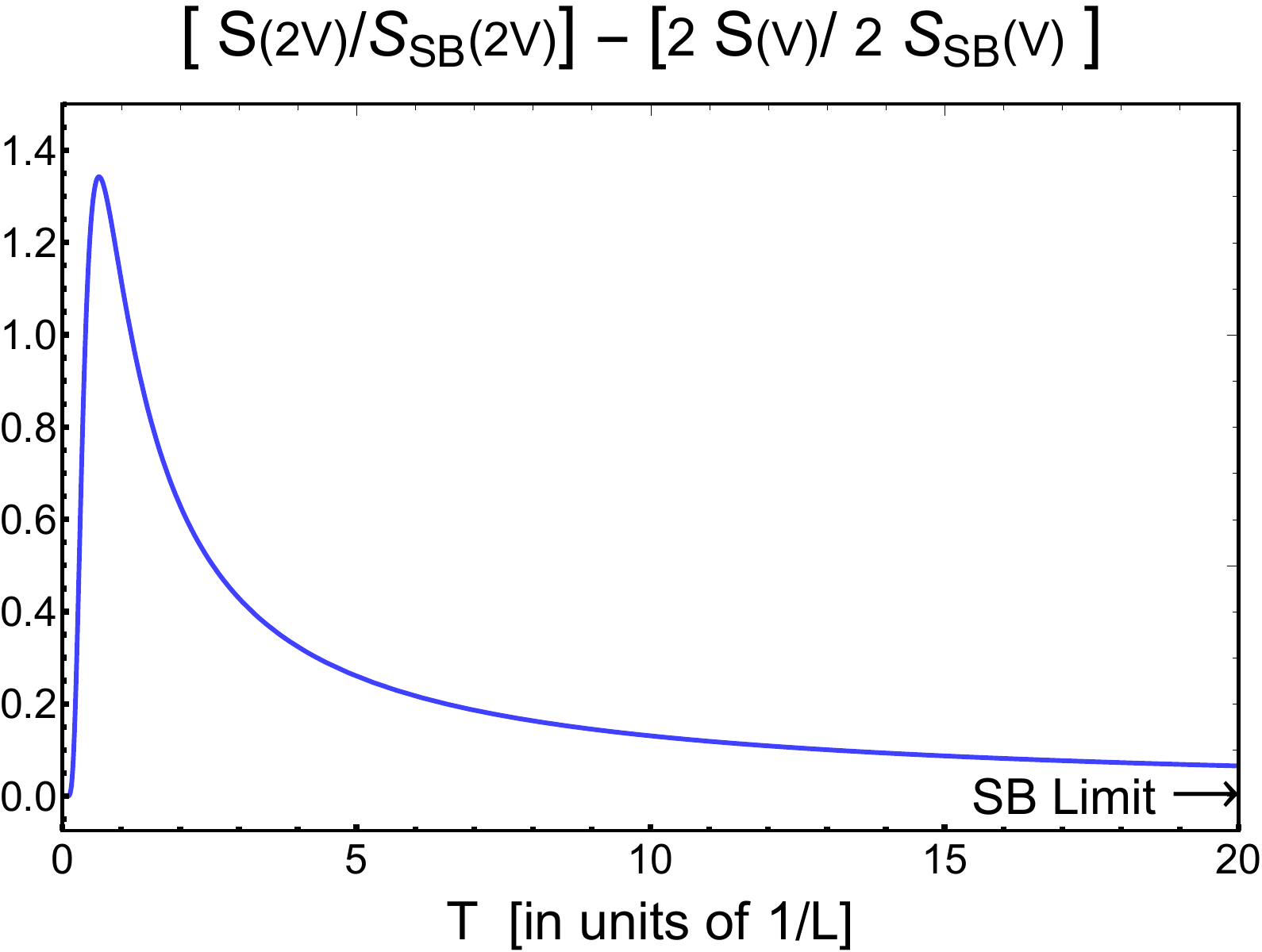}
\caption{The difference in entropy for a massless, noninteracting scalar field between parallel plates a distance $2L$ apart and twice the entropy for a system with plates a distance $L$ apart at a temperature $T$ measured in units of $1/L$ scaled by the relevant Stefan-Boltzmann limit. A fully extensive entropy would give zero.  The two denominators are equivalent since the Stefan-Boltzmann result is extensive.}
\label{fig:nonextensive}
\end{figure}

\section{Thermodynamic properties of a geometrically confined scalar field}\labsec{ResultThermo}
In this section, we present a number of thermodynamic quantities, ranging from the free energy to the specific heat capacity at constant lengths, relevant to various geometrically confined systems (infinite parallel plates, infinite tube, and finite volume box). The results are all computed from the canonical ensemble, which is to say for systems in contact with an infinite heat bath held at constant temperature $T$, for a massless, noninteracting scalar field.  Each of the plots has been rescaled by the Stefan-Boltzmann result. Recall that in the Stefan-Boltzmann limit of zero coupling and infinite size in all directions, various thermodynamic expressions we wish to evaluate in the finite size case are $F_\rmi{SB}\equiv -\pi^2T^4V/90$, $p_\rmi{SB}\equiv \pi^2T^4/90$, $E_\rmi{SB}\equiv \pi^2T^4V/30$, $S_\rmi{SB}\equiv 2\pi^2T^3V/45$, $C_\rmi{v SB}\equiv 2\pi^2T^3V/15$.

In figure~\fign{FAndEPlanes}, we show the free and total internal energies in the case of two infinite parallel plates as a function of the temperature $T$ in units of $1/L$, where $L$ is the distance between plates. Recall that for a plasma of temperature $\sim400$ MeV and width of $\sim2$ fm, relevant for a high multiplicity $pp$ or $pA$ collision at RHIC or LHC, $T\times L\sim4$.  For a mid-central $AA$ collision resulting in a plasma of temperature $T\sim400$ MeV and width $\sim10$ fm, $T\times L\sim20$. One can see in the figure that both results tend towards the thermodynamic limit as $T\times L\rightarrow\infinity$. However, both the energy and the free energy are $\gtrsim5\%$ different from their Stefan-Boltzmann limits even at the relatively large value of $T\times L\sim 20$. The total energy of a system geometrically confined in between two infinite parallel plates separated by a distance $L$, and in contact with a thermal bath at temperature $T$, is thus noticeably affected by its finite size. 

\begin{figure}[htbp]\centering
\includegraphics[scale=0.225]{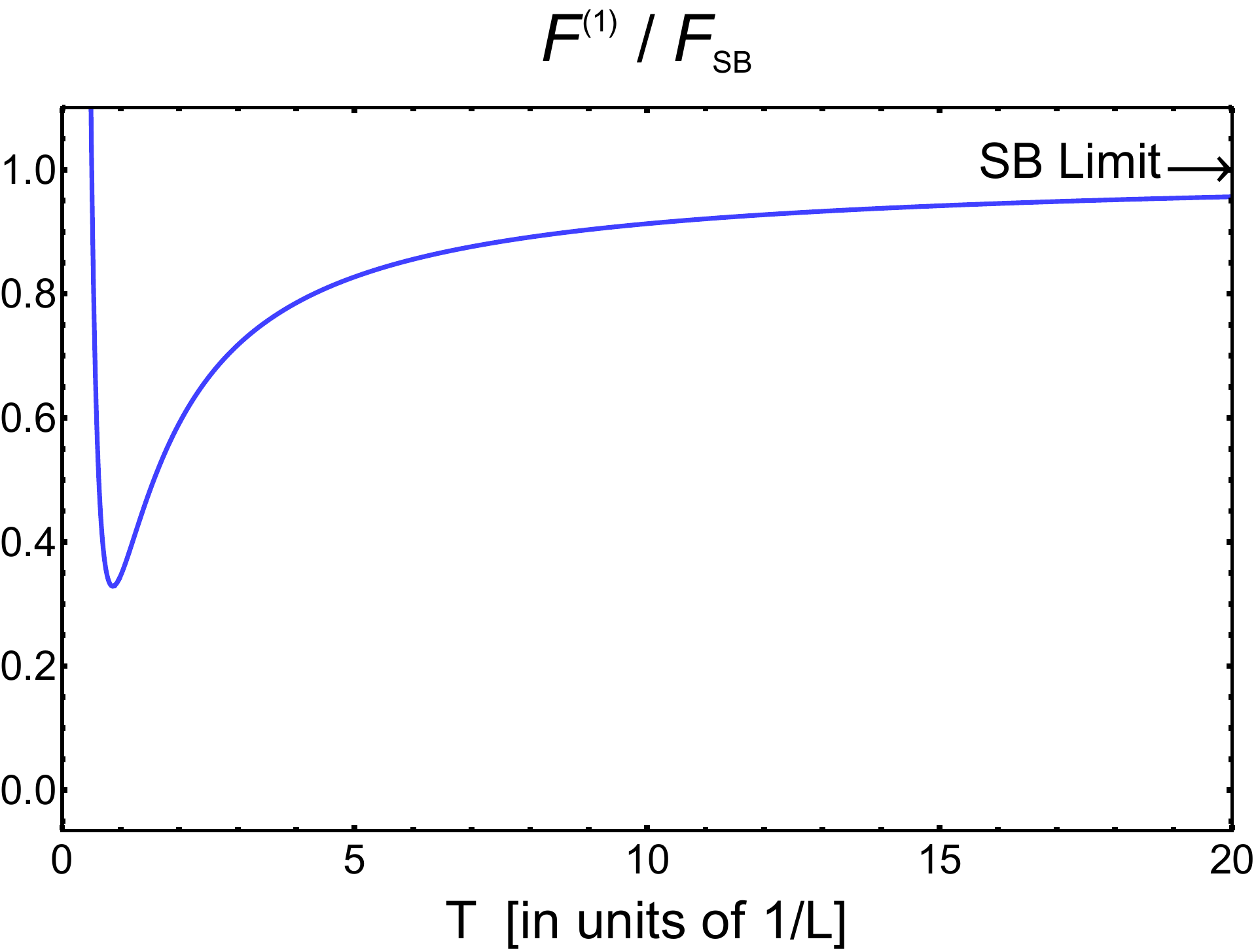}~\includegraphics[scale=0.225]{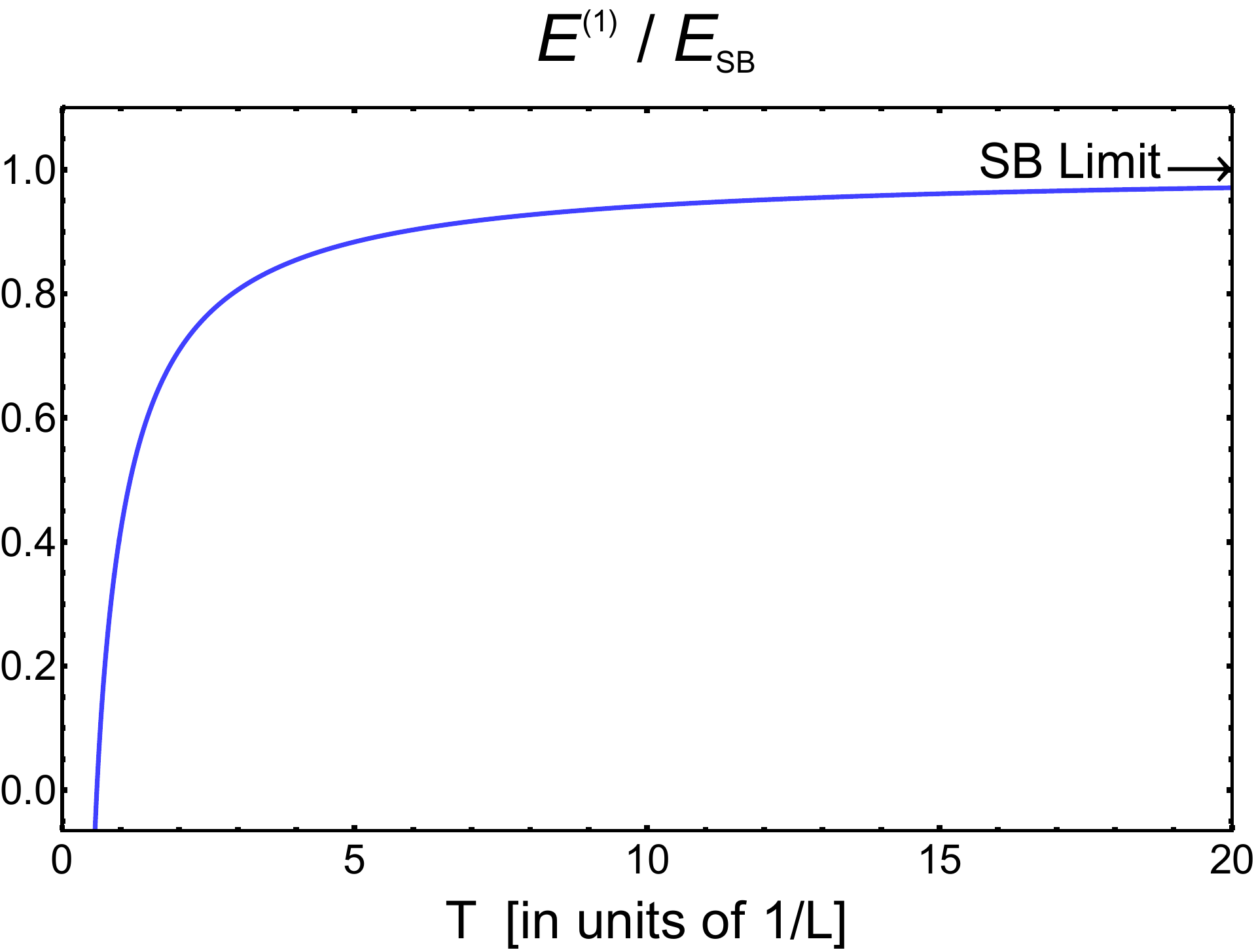}
\caption{The free energy (left) and the total energy (right) for a massless, noninteracting scalar field between two infinite parallel plates and rescaled to their Stefan-Boltzmann limits as a function of temperature $T$ in units of $1/L$, where $L$ is the distance between the sides of the system that are of finite length.}
\labfig{FAndEPlanes}
\end{figure}
%
\begin{figure}[htbp]\centering
\includegraphics[scale=0.23]{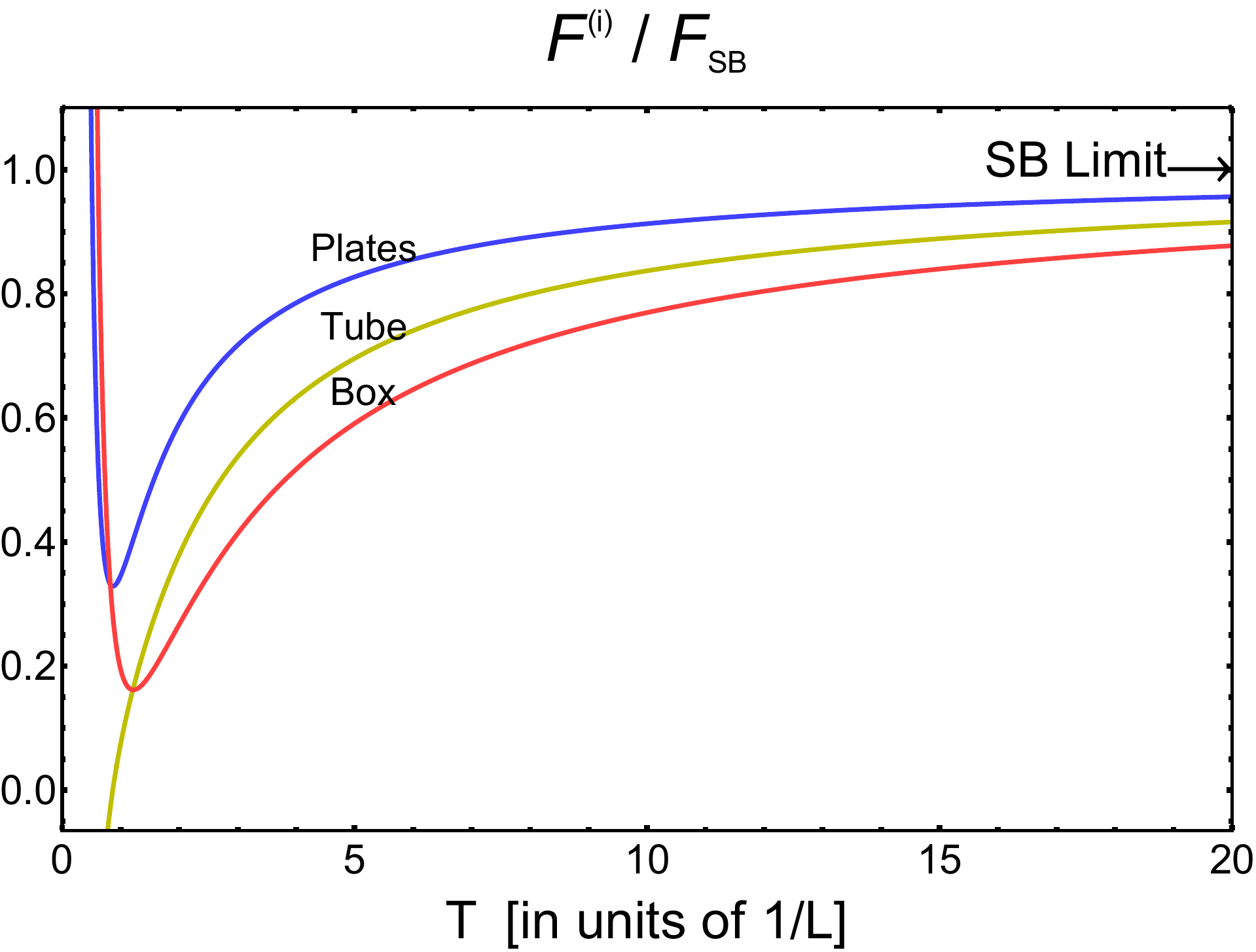}~\includegraphics[scale=0.23]{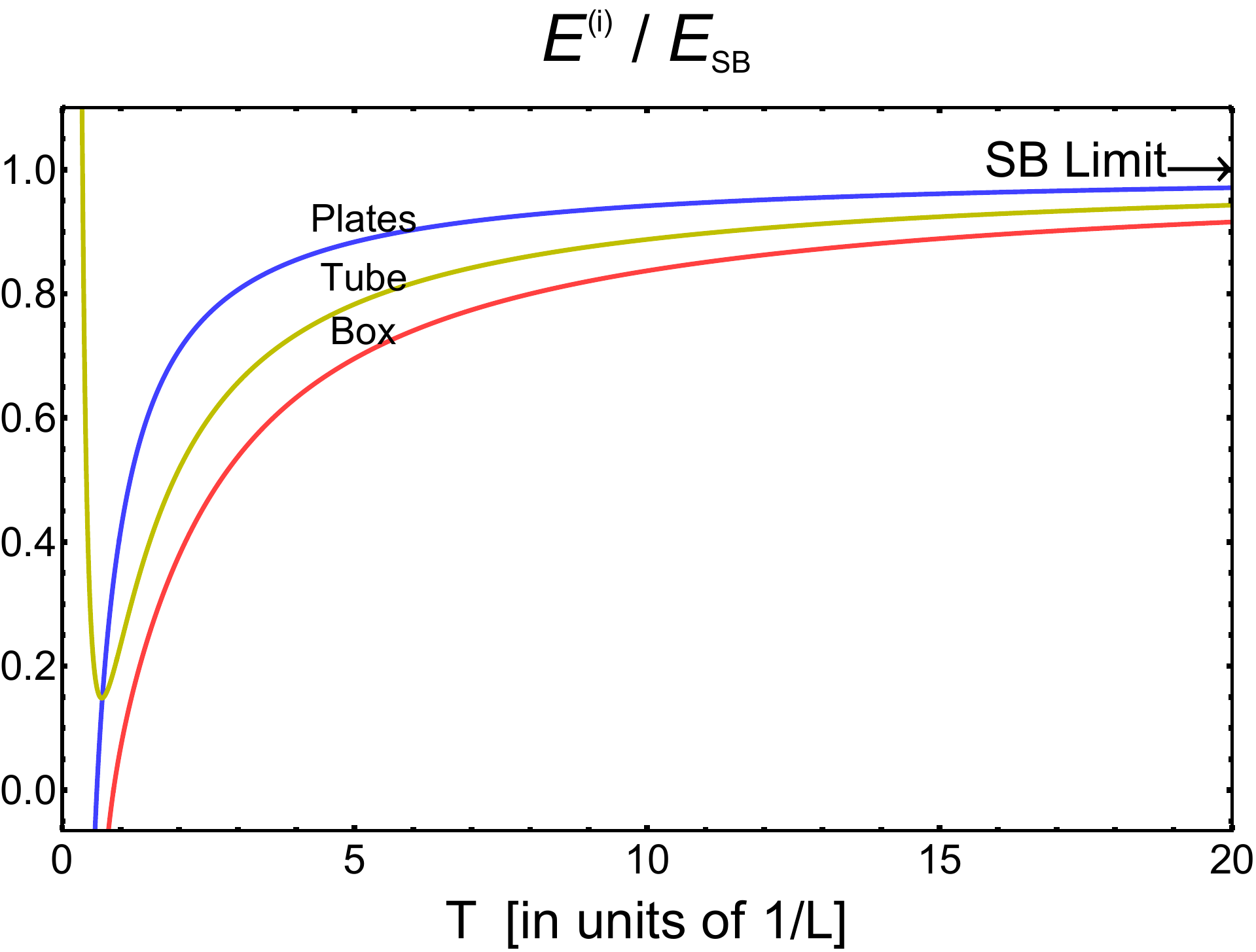}
\caption{The free energy (left) and the total energy (right) for a massless, noninteracting scalar field between two infinite parallel plates (blue lines), in an infinite symmetric tube (yellow lines), and in a finite volume symmetric box (red lines).  The results have been rescaled by the Stefan-Boltzmann limits and are shown as a function of temperature $T$ in units of $1/L$, where $L$ is the distance between the sides of the system that are of finite length. }
\labfig{FAndEPlanesTubeBox}
\end{figure}

In figure~\fign{FAndEPlanesTubeBox}, we display the free and total energies in the three cases of infinite parallel plates (blue lines), infinite tube (yellow lines), and finite volume box (red lines). All results tend towards the Stefan-Boltzmann limits as $T\times L\rightarrow\infinity$. We see again the large finite size corrections to the Stefan-Boltzmann limits, even for $T\times L\sim 20$, with the size of the corrections increasing with increasing number of compactified directions. Thus the total and free energies of a system geometrically confined are noticeably affected by its finite sized directions. We note that the peculiar behavior of the tube case, that its total energy reaches a $T=0$ limit which is positive unlike the plates and box cases, is not a surprise. It is indeed due to the dimensionality of the space-time~\cite{Caruso:1991ge}.

We now turn towards the different pressures.  In figure~\fign{P1AndP2Or3Planes} we plot the perpendicular (left, labeled ``1'') and parallel (right, labeled ``2,3'') pressures as a function of the temperature $T$ in units of length $1/L$ for a massless, noninteracting scalar gas held between two infinite parallel plates that are a perpendicular distance $L$ apart. The plots have a number of interesting properties. First, the pressure in the perpendicular direction ($p_1$, left panel), i.e.\ between the plates, is different from the pressure in the parallel direction ($p_{2}$ or $p_{3}$, right panel), i.e.\ along the plates.  There is an intrinsic anisotropy due to the compactification. Furthermore, the compactification of one direction appears to have a greater effect on the pressure in the noncompactified parallel direction than on the pressure in the compactified perpendicular direction: the parallel pressure is nonmonotonic, unlike in the perpendicular case, and approaches the Stefan-Boltzmann limit much slower than the perpendicular pressure. 

Note that the pressure for our geometrically confined system dimensionally reduces in the correct way.  For our infinite parallel plates case in $4D$ spacetime 
\begin{align}
	\label{eq:dimensionalreduction}
	p_\parallel(\beta,\,L) \equiv \frac{\partial(T\ln Z)}{\partial V_\parallel} = \frac{1}{2\pi\beta^3}\sum_{\ell=1}^\infinity\Bigg[ \frac{\pi\beta\ell}{L}\Li_2\big( e^{-\pi\beta\ell/L} \big) + \Li_3\big( e^{-\pi\beta\ell/L} \big) \Bigg] + vacuum,
\end{align}
where $vacuum$ depends only on $L$.  One can see that in the limit $T\times L \ll 1$ eq.~\ref{eq:dimensionalreduction} reproduces the pressure of a massive, noninteracting scalar theory in 3 spacetime dimensions in the Stefan-Boltzmann limit
\begin{align}
	p_{2D}(\beta,\,m) = \frac{1}{2\pi\beta^3}\Big( m\beta\Li_2\big( e^{-m\beta} \big) + \Li_3\big( e^{-m\beta/L} \big) \Big),
\end{align}
for $m\equiv\pi/L$, where we have dropped the infinite, temperature independent zero point pressure from the last expression.  One sees then that the inverse of the compactified length $L$ in $4D$ acts as an effective mass in the $3D$ theory.

\begin{figure}[!tbp]\centering
\includegraphics[scale=0.23]{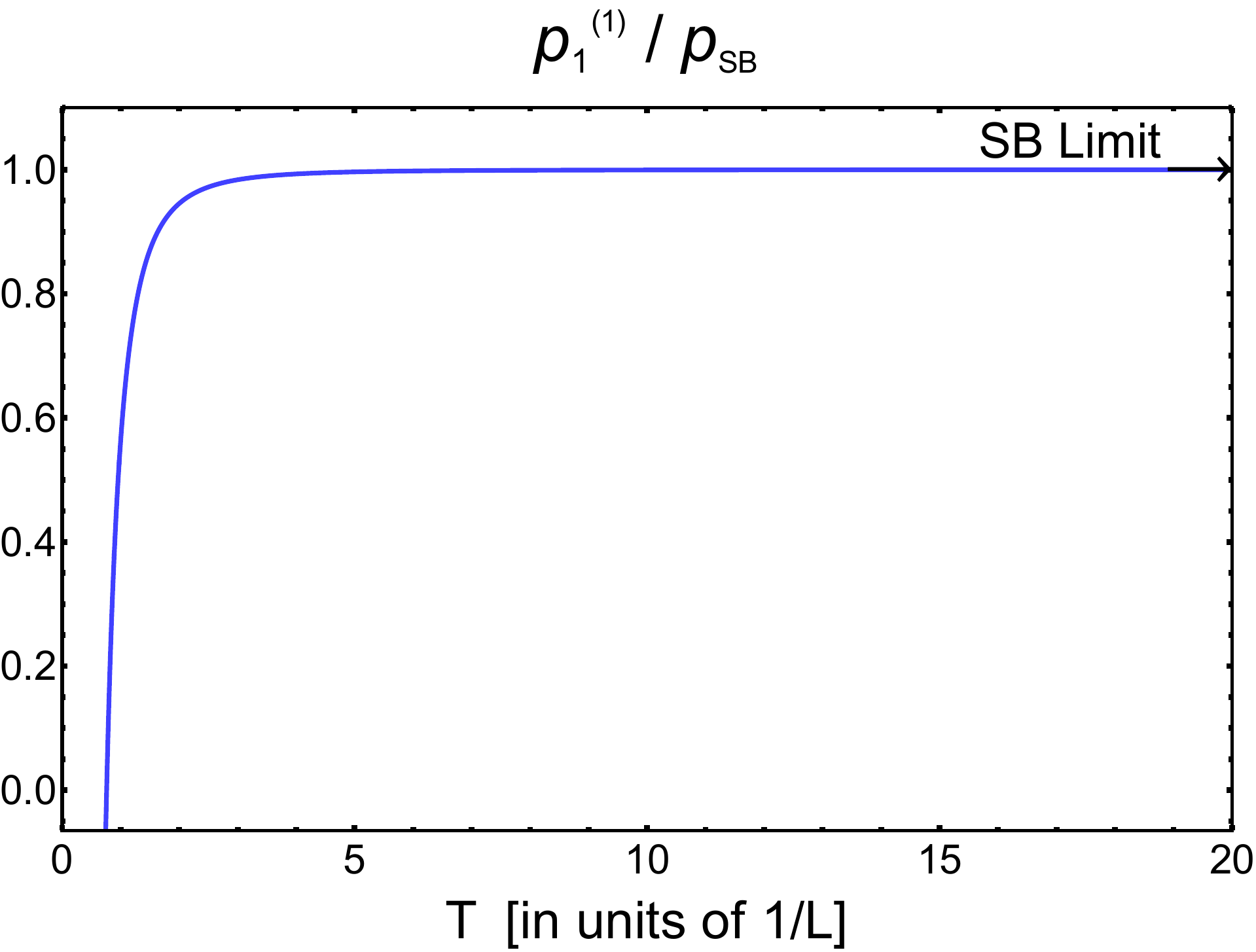}~\includegraphics[scale=0.23]{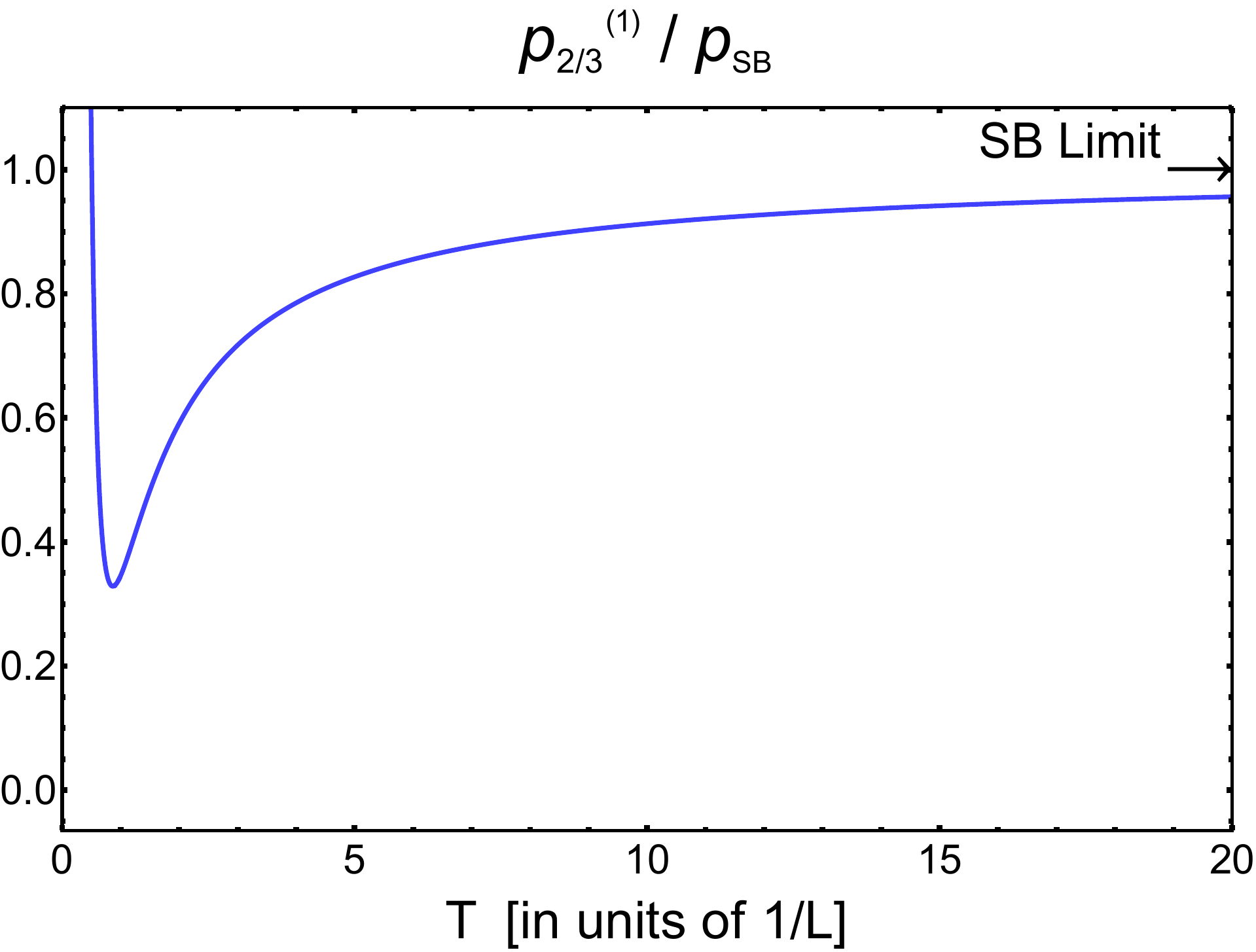}
\caption{The pressures, perpendicular (left, $p_1$) and parallel (right, $p_2$ or $p_3$), for a massless, noninteracting scalar field restricted between two infinite parallel plates as a function of temperature $T$ in units of the perpendicular distance $L$ between the two plates. Both quantities are relative to the usual Stefan-Boltzmann pressure.}
\labfig{P1AndP2Or3Planes}
\end{figure}

We show in figure~\fign{P1PlanesTubeBox} the pressure in one of the compactified directions for the two infinite parallel plates, the infinite (symmetric) tube and the finite volume (symmetric) box cases, respectively represented by the blue, yellow, and red lines.  The pressures are rescaled by their Stefan-Boltzmann limits and are plotted as a function of the temperature $T$ in units of $1/L$, where $L$ is the length of the compactified direction.

\begin{figure}[!btp]\centering
\includegraphics[scale=0.315]{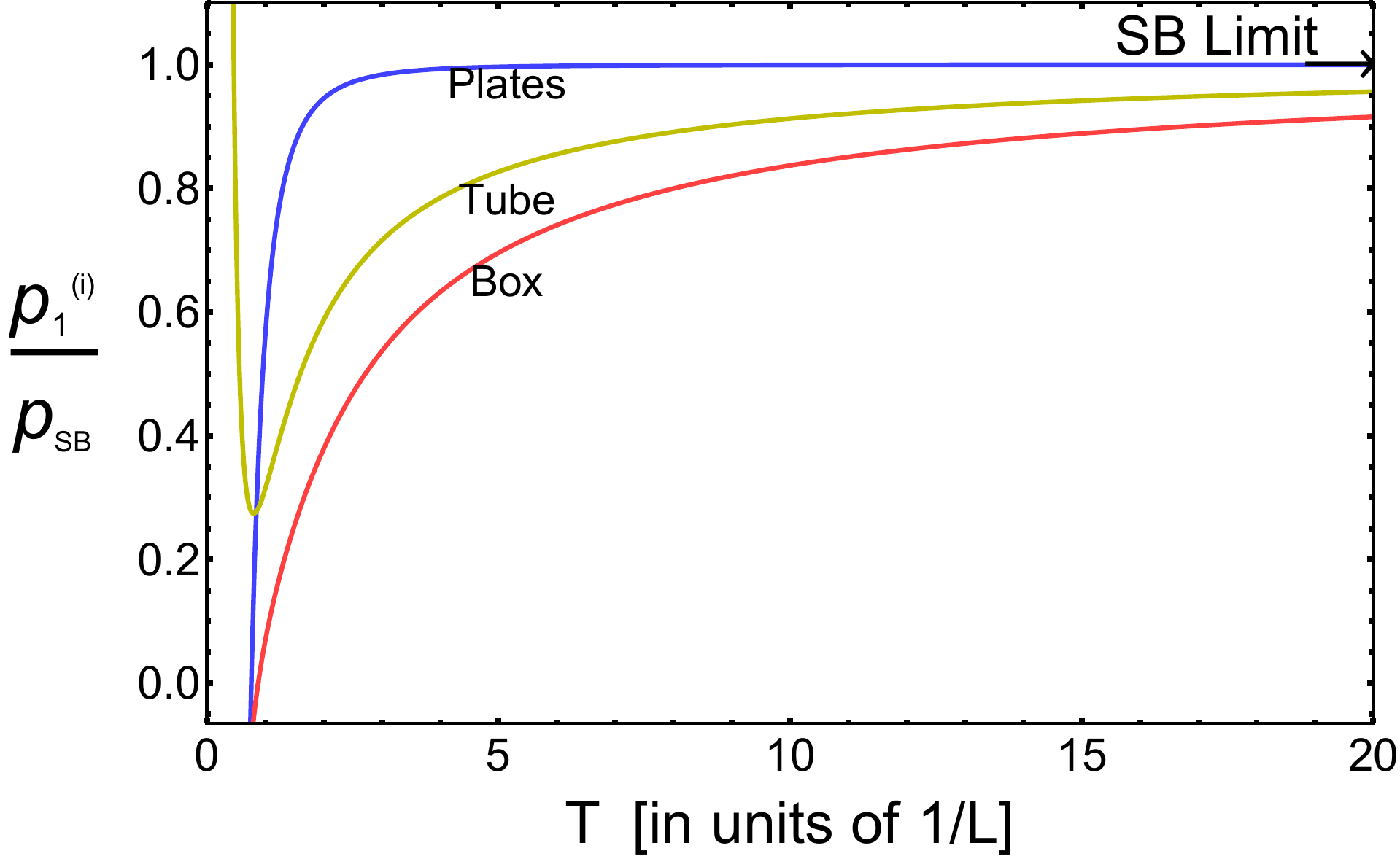}
\caption{Longitudinal pressure of a massless, noninteracting scalar field in between two infinite parallel plates (blue), in an infinite symmetric tube (yellow), and in a finite volume symmetric box (red). All results are rescaled to the Stefan-Boltzmann limit. The pressures are plotted as a function of the system temperature $T$ in units of $1/L$, where $L$ measures the length of the compactified direction(s).}
\labfig{P1PlanesTubeBox}
\end{figure}

Notice that the effect of further compactifications on the pressure is drastic.  In the finite box case, for $T\times L\sim4$, which is relevant for a $pp$ collision resulting in a $\sim400$ MeV QGP, the pressure sees a $\sim40\%$ correction.  Even for $T\times L\sim20$, there are $\sim10\%$ corrections to the pressure of the finite volume box.  Note that while it might appear that the pressures diverge at low $T$, this apparent divergence is an artifact of plotting the ratio of our finite sized results with the Stefan-Boltzmann limit; recall that the Stefan-Boltzmann case scales as $T^4$.  One may use the un-rescaled expression eq.~\eqn{M1finalResultWAHIK} to investigate the asymptotic $T=0$ behavior, in which case one recovers the usual Casimir pressure in the longitudinal direction in each compactification case. (This comment also applies to the other pressures, and the free and total energies.)

We plot in figure~\fign{SAndCvPlanes} the entropy (left) and specific heat (right) for the infinite plates case as a function of the temperature $T$ in units of $1/L$, where $L$ measures the length of the finite direction(s) of the systems, and scaled by the Stefan-Boltzmann limit.  The inset in the left plot shows that the entropy of our finite sized system indeed goes to zero as the temperature vanishes, as dictated by the Third Law of Thermodynamics, and as opposed to some implementations of Lifschitz's theory for the QED Casimir effect~\cite{Klimchitskaya:2009cw}.  The inset in the right plot shows that the specific heat for our finite sized system always remains positive.

We plot in figure~\fign{SAndCvPlanesTubeBox} the entropy (left) and specific heat (right) for the infinite plates (blue), infinite tube (yellow), and finite box (red) cases as a function of the temperature $T$ in units of $1/L$, where $L$ measures the length of the finite direction(s) of the systems, and scaled by the Stefan-Boltzmann limit.  Although we do not provide insets in these plots, the entropy for all our cases again goes to 0 as $T\rightarrow 0$, in accordance with the Third Law of Thermodynamics and the specific heat remains strictly positive.  Notice again that the size of the deviations from the Stefan-Boltzmann limit increase as the number of compactified dimensions increases.  In particular the deviation from the Stefan-Boltzmann limit is significant ($\sim 10 - 15\%$) for the finite box case even out to $T\times L\sim 20$. 

\begin{figure}[!tbp]\centering
\includegraphics[scale=0.23]{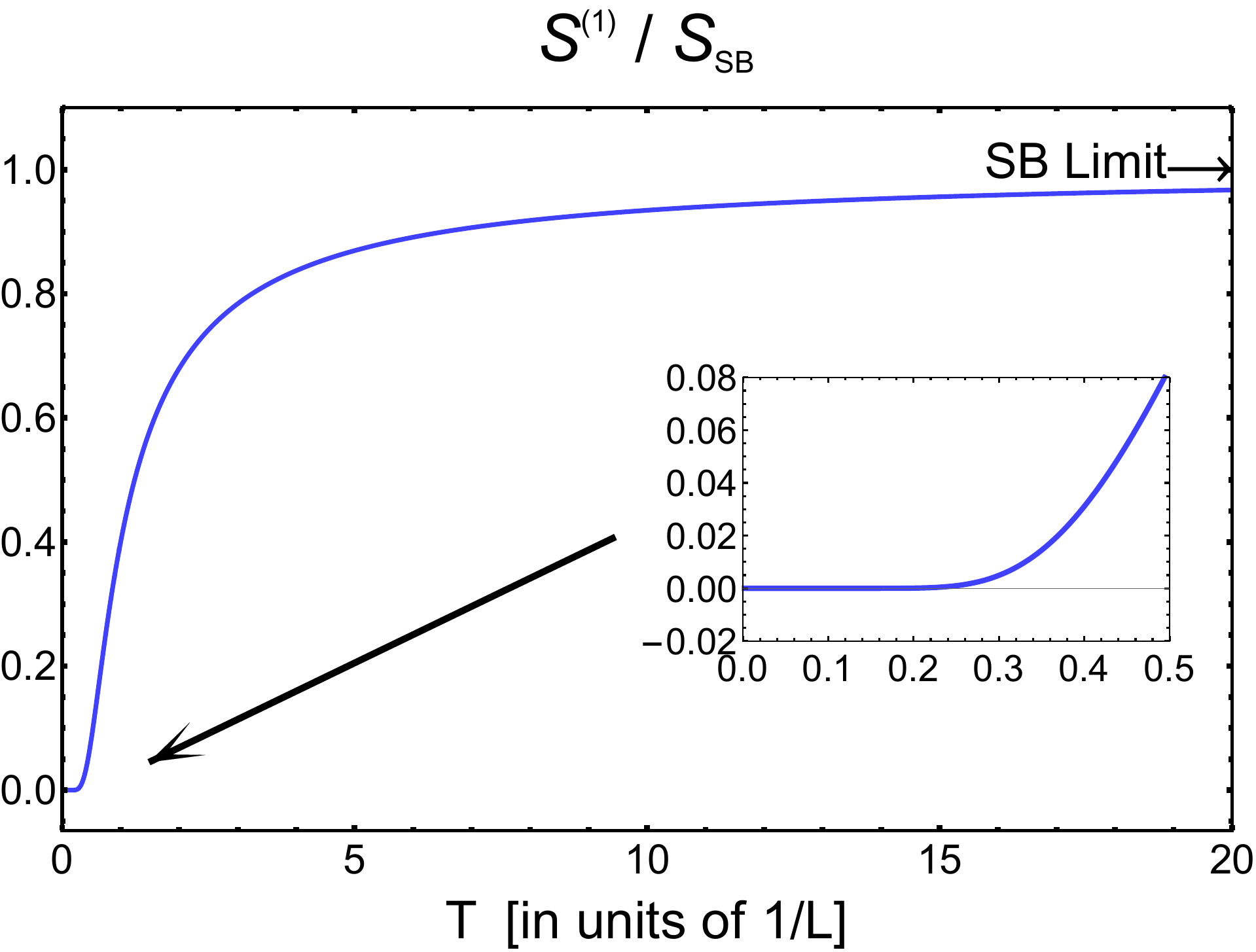}~\includegraphics[scale=0.23]{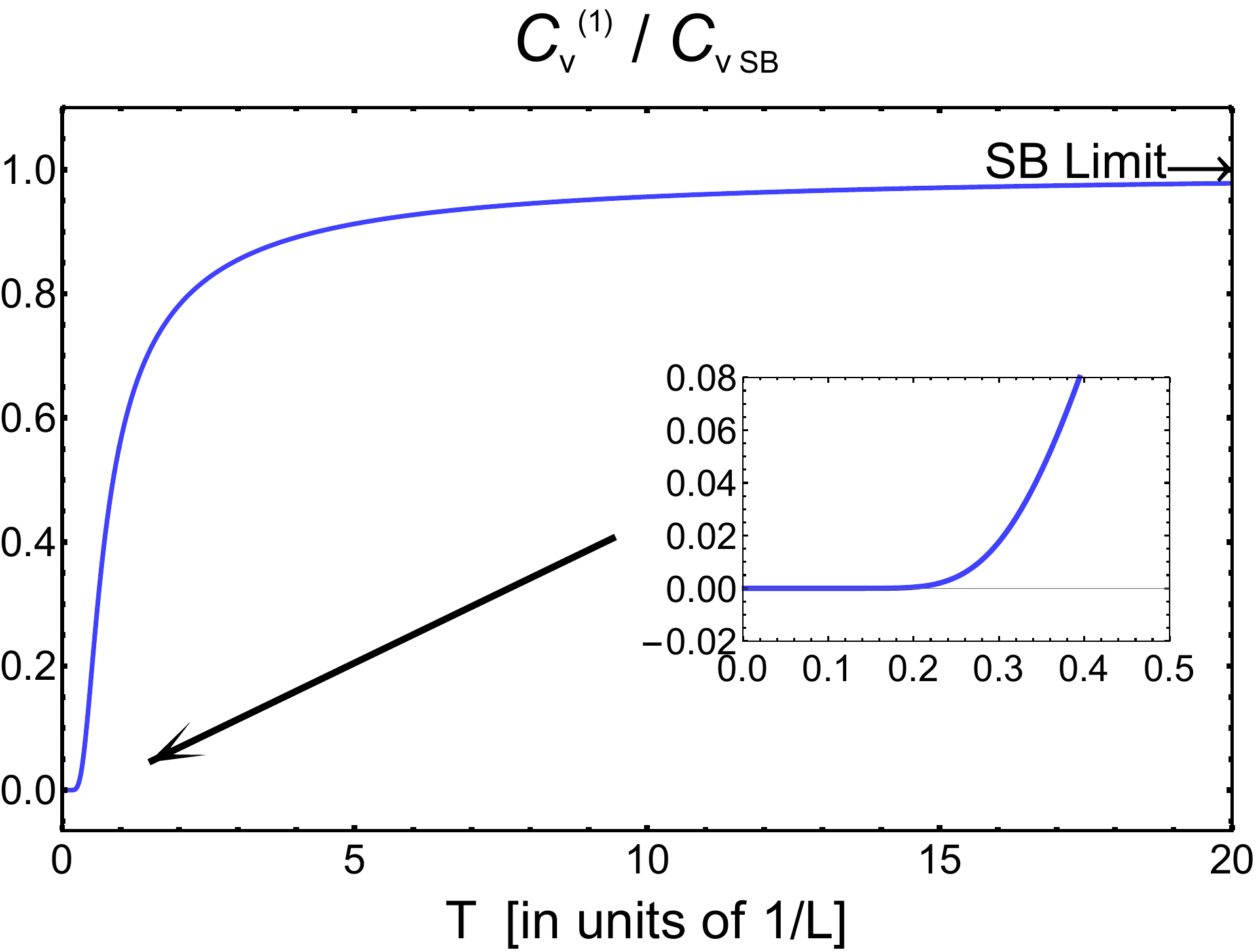}
\caption{The total entropy (left) and specific heat at constant lengths (right) for a massless, noninteracting scalar field between two infinite parallel plates separated by a distance $L$ as a function of temperature $T$ measured in units of $1/L$. Both quantities are rescaled to their Stefan-Boltzmann limits. The insets show the small temperature limits of the quantities.}
\labfig{SAndCvPlanes}
\end{figure}

\begin{figure}[!tbp]\centering
\includegraphics[scale=0.23]{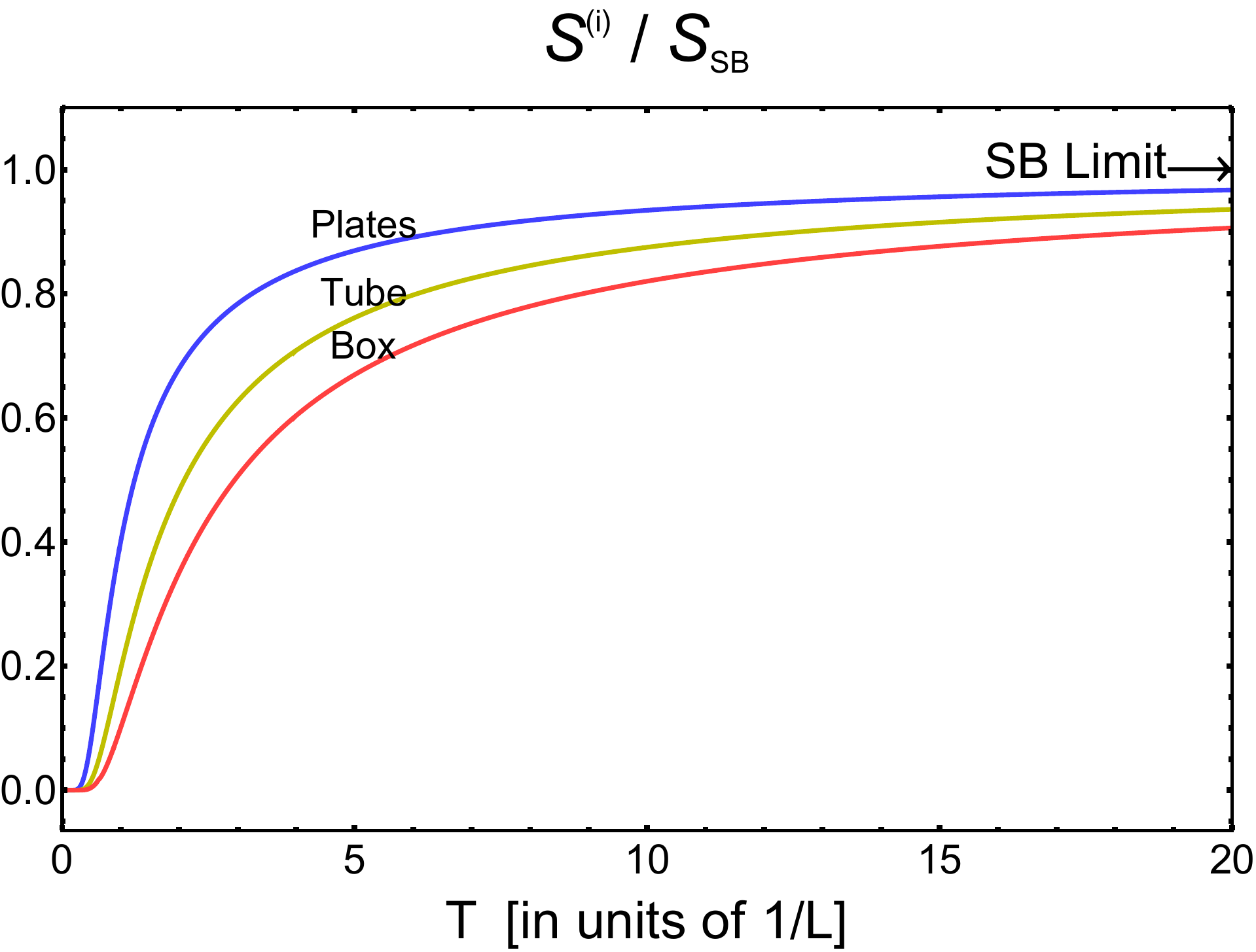}~\includegraphics[scale=0.23]{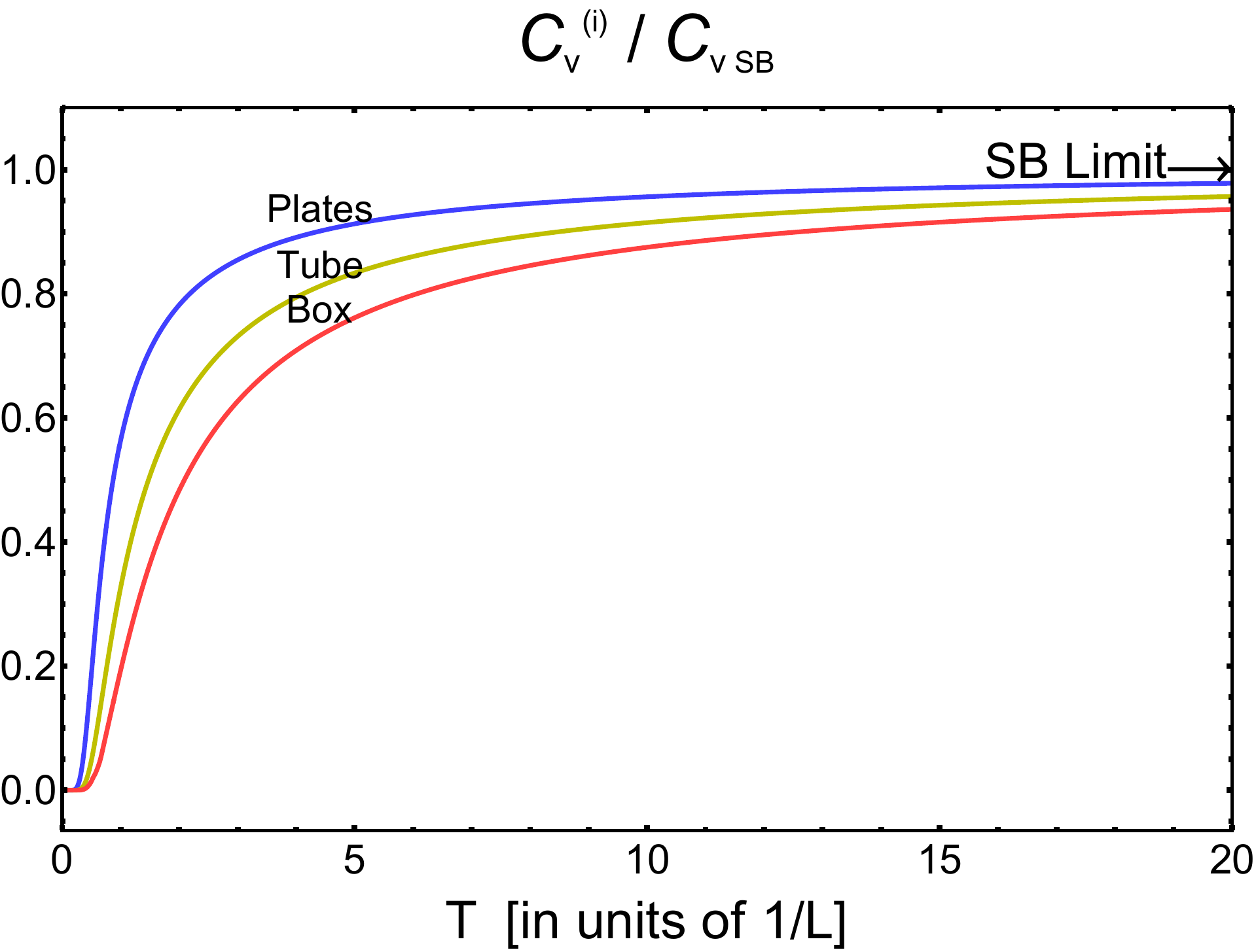}
\caption{(Left) The entropy and (right) the specific heat for a massless, noninteracting scalar field for the cases of two infinite parallel plates (blue lines), the infinite symmetric tube (yellow lines), and the finite volume symmetric box (red lines).  The quantities are plotted as a function of the temperature $T$ in units of $1/L$, where $L$ is the compactification length for the system, and the quantities are rescaled to their Stefan-Boltzmann limits. }
\labfig{SAndCvPlanesTubeBox}
\end{figure}
\paragraph*{On boundary conditions:}
We would like to better understand the importance of the type of boundary condition imposed on our systems on the size of the finite size corrections to the Stefan-Boltzmann limit.  Intuitively, one might expect that periodic boundary conditions cause the least difference since the system has a finite size but is boundaryless.  We show in figure~\fign{FPlanesDiriVsOthers} the free energy of a massless, noninteracting scalar field between two infinite parallel plates rescaled to its Stefan-Boltzmann limit as a function of the temperature $T$ in units of $1/L$, where $L$ is the distance between the plates.  The plot shows the results for Dirichlet (solid blue), Neumann (dotted brown), and periodic (dashed purple) boundary conditions.  One can clearly see from the figure that the system with periodic boundary conditions reaches the Stefan-Boltzmann limit at much smaller $T\times L$ than either the Neumann or Dirichlet cases.  One should perhaps then hesitate to conclude that small finite sized corrections computed for a system with periodic BCs~\cite{Meyer:2009kn,Giusti:2011kt,Braun:2011iz,Bijnens:2013doa,Fister:2015eca,Parvan:2015mta,Fraga:2016oul,Juricic:2016tpt,Almasi:2016zqf} will remain small for heavy ion phenomenology.  

\begin{figure}[!btp]\centering
\includegraphics[scale=0.3]{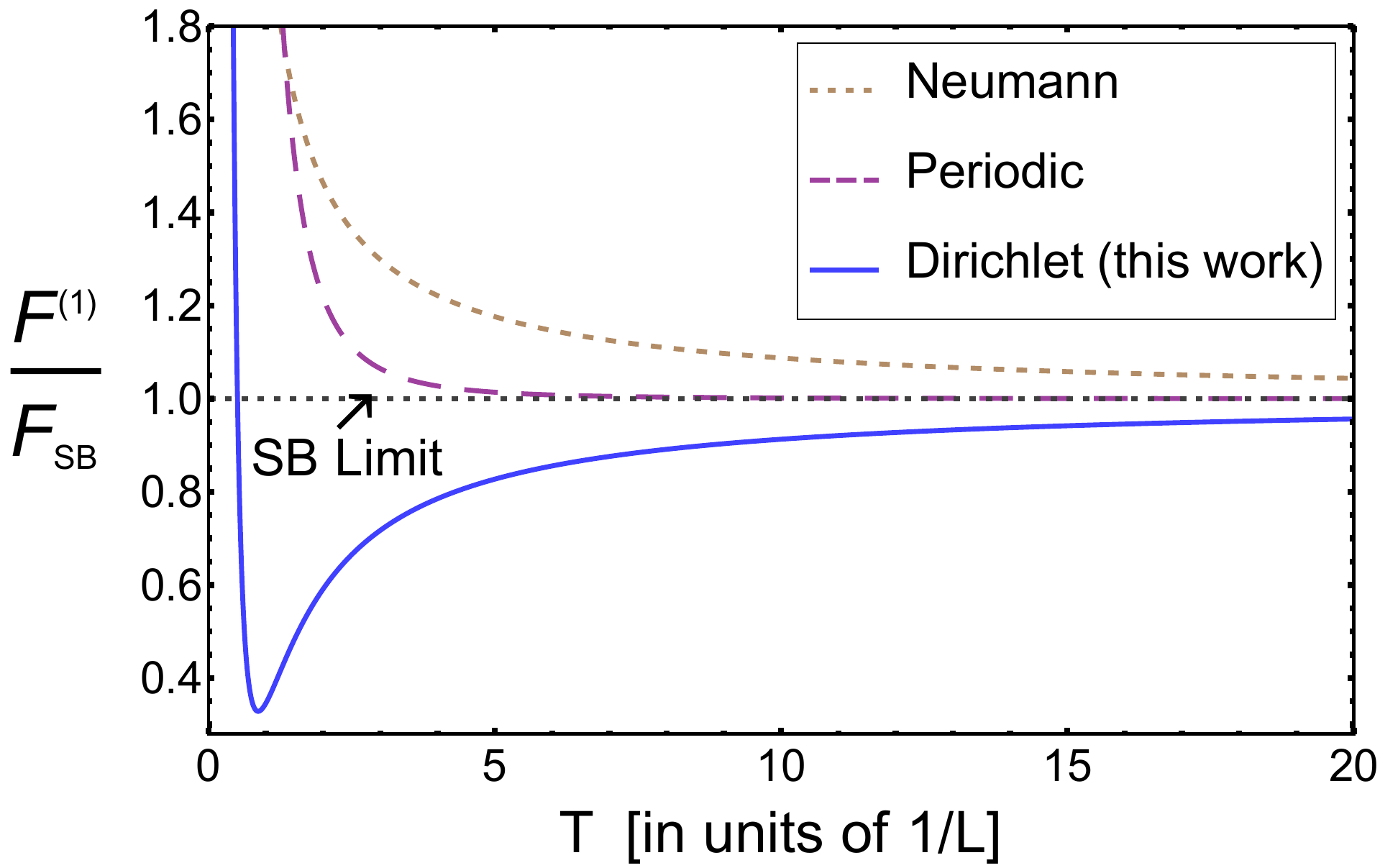}
\caption{The free energy of a massless, noninteracting scalar field in between two infinite parallel plates rescaled to its Stefan-Boltzmann limit for Dirichlet (blue), Neumann (dotted brown), and periodic (dashed purple) boundary conditions.}
\labfig{FPlanesDiriVsOthers}
\end{figure}

Let us now come back to our geometric confinement boundary conditions. Since our noninteracting scalar field theory is conformal, the trace of the energy momentum tensor should vanish identically.  We performed a nontrivial check of our numerics and confirmed that our results do respect $T^{\mu}_{\mu}\equiv\varepsilon-\left(p_1+p_2+p_3\right)=0$. 

\section{A novel geometric phase transition}\labsec{PhaseTransition}
In this section we consider how our results depend on the physical setup of our system.  In the previous section we considered our system to be in contact with an infinite thermal heat bath.  Since the equations are numerically tractable, we now focus on the infinite parallel plates case and consider the possibility of the system existing in isolation: instead of a system at constant temperature $T$ we rather consider a system with constant energy $E$. All results shown in this section are derived from the partition function calculated in the canonical ensemble.  We are justified in this approach as argued in section~\secn{FluctuEnergy}: the ratio of the energy fluctuations away from the mean energy divided by the mean energy computed in the canonical ensemble for the Stefan-Boltzmann parallel plates case is zero; hence we will have ensemble equivalence between the canonical and microcanonical ensembles.  Thus the canonical ensemble provides exact results for all thermodynamic quantities for a set of parallel plates kept in isolation\footnote{In principle it should be possible to derive, via double Legendre transform of the canonical ensemble free energy, the fundamental thermodynamic relations of the microcanonical ensemble.  Such an exercise is highly non-trivial and is not attempted in the present work.}, i.e.\ with fixed energy $E$.

As we explore the different physics associated with a noninteracting scalar field between two infinite parallel plates, it is worth keeping in mind two related thermodynamic concepts.  First, a system is thermodynamically stable if, when squeezed and all other parameters are held fixed, the pressure increases.  I.e.\ our parallel plates system separated by a length $L$ is thermodynamically stable so long as
\begin{align}
	\frac{\partial p}{\partial L} < 0
	\label{stabilitycondition}
\end{align}
for fixed $E$ and plate area $V_2$, which is to say that the pressure in the system decreases with increasing system size; see, e.g., \cite{Park:2008sk}.  Second, a system undergoes a second order phase transition should a susceptibility diverge; see, e.g., \cite{Altland:2006si}.  Recall that the susceptibility is the derivative of an extensive parameter with respect to its conjugate intensive parameter.  The susceptibility is then nothing more than the multiplicative inverse of the second derivative of the entropy with respect to some parameter.  For our case of two parallel plates separated by a distance $L$, the susceptibility is the length-scaled negative of the compressibility:
\begin{align}
	\chi\equiv \frac{\partial L}{\partial p} = -L\kappa.
	\label{phasetransitioncondition}
\end{align}
Comparing the condition for stability, eq.~(\ref{stabilitycondition}), to the definition of susceptibility above, one can see that a second order phase transition occurs when the system goes from being thermodynamically stable to unstable.  Further, the order parameter associated with the compressibility is the size of the system $L$.

We can further understand physically what happens at a phase transition by examining the relationship between the entropy and an intensive variable such as the pressure.  There are three common, and equivalent, definitions of pressure:
\begin{align}
	p\equiv\frac{1}{\beta}\left.\frac{\partial S}{\partial V}\right|_E \equiv -\left.\frac{\partial E}{\partial V}\right|_S \equiv \frac{1}{\beta}\left.\frac{\partial \ln Z}{\partial V}\right|_\beta.
	\label{pressuredefinitions}
\end{align}
The first expression is the quantity that must be equal for two systems in thermodynamic equilibrium separated by a moveable wall.  The second expression is the generalized force conjugate to the volume.  The third is the generalized thermodynamic intensive variable conjugate to the extensive volume variable.  In the parallel plates case of current interest, the derivatives with respect to volume $V$ become derivatives with respect to separation length $L$.  One can then see that a phase transition occurs precisely when the second derivative of the entropy changes sign; i.e.\ when the entropy goes from a convex function of $L$ to a concave function of $L$.

While all three expressions of the pressure in eq.~(\ref{pressuredefinitions}) are equivalent, the different expressions are naturally a function of different variables: $p(E,\,L)$, $p(S,\,L)$, and $p(\beta,\,L)$, respectively, where we have already switched over to using the plate separation distance $L$ instead of the volume $V$ for our system.  
Since we have shown that one may freely move from the canonical to the microcanonical ensembles, one may freely switch between different definitions of pressure by using relations between the various independent variables.  For example, equipped with an expression that relates the energy to the entropy and volume, one can equivalently use the first definition of pressure in the same way as the second definition with $p\big(E(S,L),L\big) = p(S,L)$.

\begin{figure}
	\centering
	\includegraphics[width=2.75in]{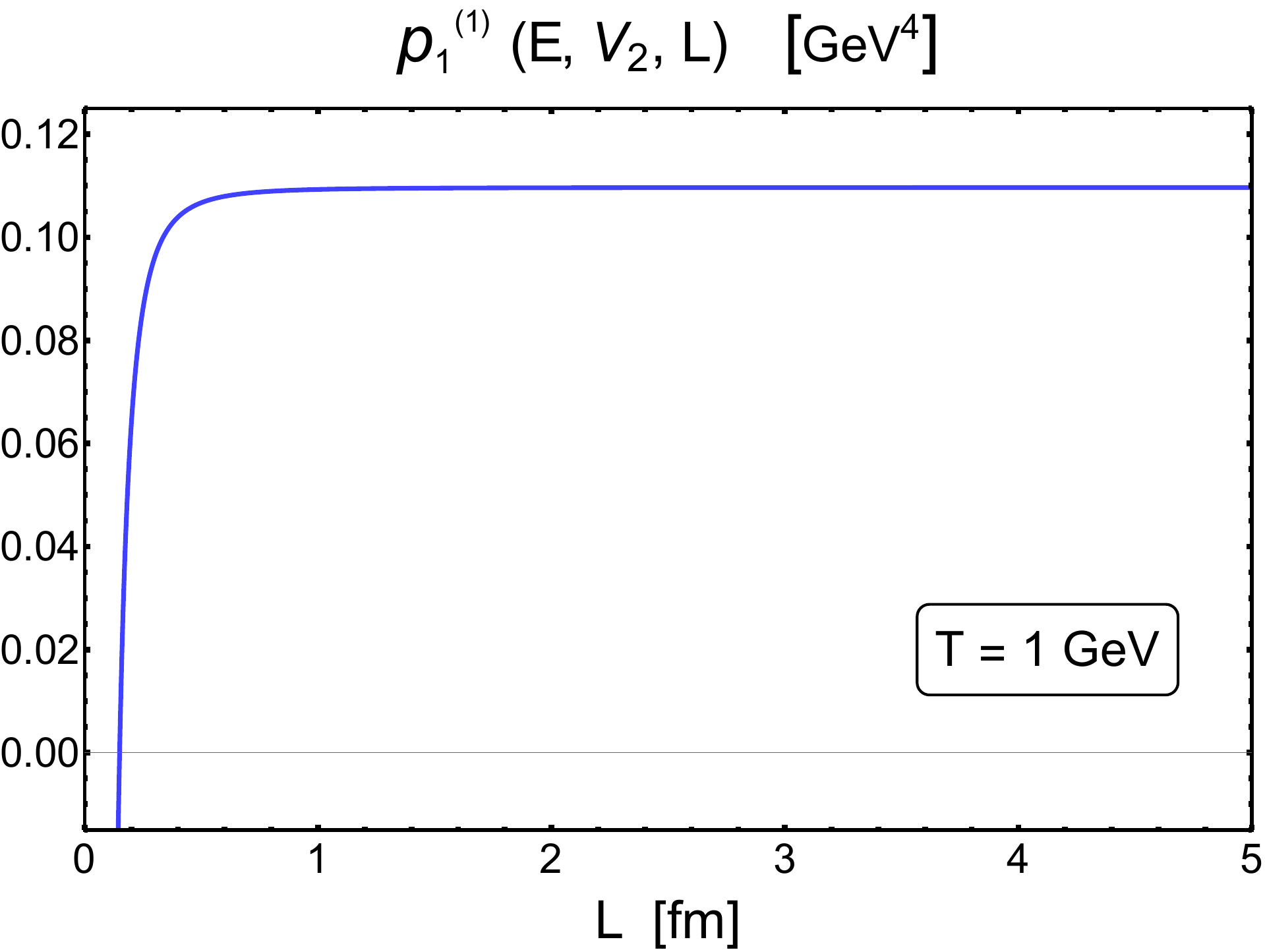} \hspace{.2in}
	\includegraphics[width=2.75in]{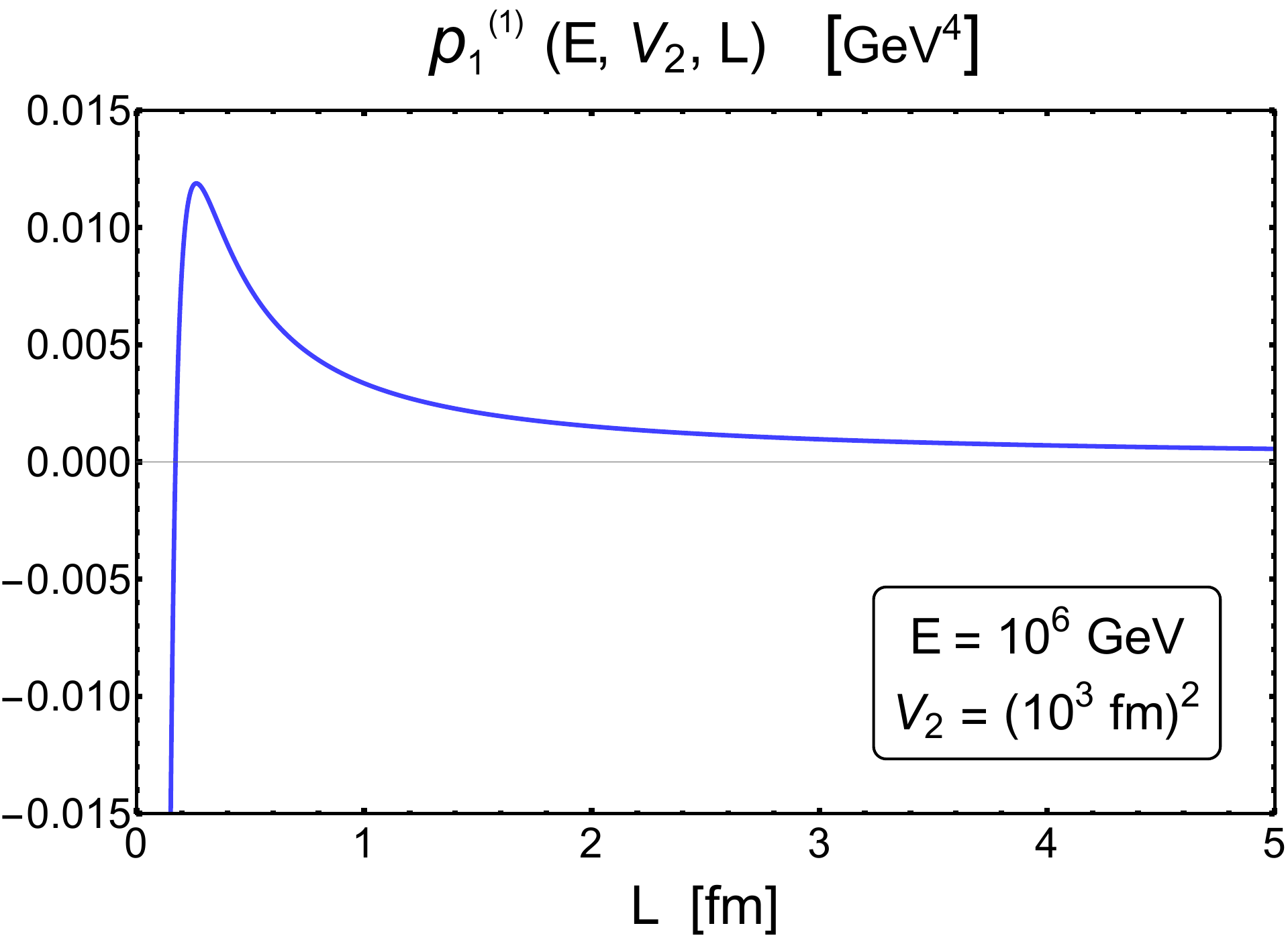}
	\caption{
		\label{pofLplot}
		(Left) Pressure $p$ as a function of $L$ for a massless, noninteracting scalar field between two parallel plates separated by a distance $L$ in contact with a thermal heat bath at a temperature $T = 1$ GeV.  (Right) Same but for an isolated system at constant energy $E = 10^6$ GeV and with parallel plates area $V_2=10^6$ fm$^2$.
	}
\end{figure}

In figure~(\ref{pofLplot}) we compare the pressure of a massless, noninteracting scalar field between parallel plates of area $V_2$ as a function of the plate separation length $L$ for fixed temperature $T$ (left) and for fixed energy $E$ (right); i.e.\ the left plot shows the pressure as a function of $L$ for a system in contact with a heat bath whereas the right plot shows the pressure as a function of $L$ for an isolated system.  We computed $p(L)$ from the partition function \eq{M1finalResultWAHIK} by numerically inverting $E(T,\,V_2,\,L)$ to find $p\big(T(E,\,V_2,\,L),\,V_2,\,L\big)$.  
It is perhaps not so easy to see in the figure, but for the system in contact with a heat bath (left), the pressure always decreases for decreasing $L$; i.e.\ $\partial p/\partial L > 0$ and the system is always unstable: the system always wants to collapse.  The isolated system, however, resists collapse as the system size is decreased---i.e.\ $\partial p/\partial L<0$ and the system is thermodynamically stable---so long as the system starts off large enough, which is to say the length $L$ is greater than some critical length $L_c$.  As soon as $L\le L_c$, the system is unstable and will collapse, shrinking until the plates are no longer separated at all.  We show explicitly the susceptibilities as a function of plate separation $L$ for these two systems in figure~(\ref{kappaofLplot}).  As claimed, the isothermal compressibility is purely negative while the isoenergetic compressibility clearly exhibits a divergence. We would like to emphasize that the two systems considered here, one at constant $ T $ and one at constant $ E $, study two different physical situations, each characteristic of a different ensemble, but we have not explicitly computed the micro-canonical ensemble result. We have merely studied the physics of the microcanonical ensemble from the canonical ensemble by invoking ensemble equivalence, where we noted earlier in this section that ensemble equivalence is ensured by the lack of energy fluctuations in the canonical ensemble for the parallel plates setup.  Computing the micro-canonical result directly requires the execution of a double Legendre transform, which is non-trivial and not attempted here. 

\begin{figure}
	\centering
	\includegraphics[width=2.75in]{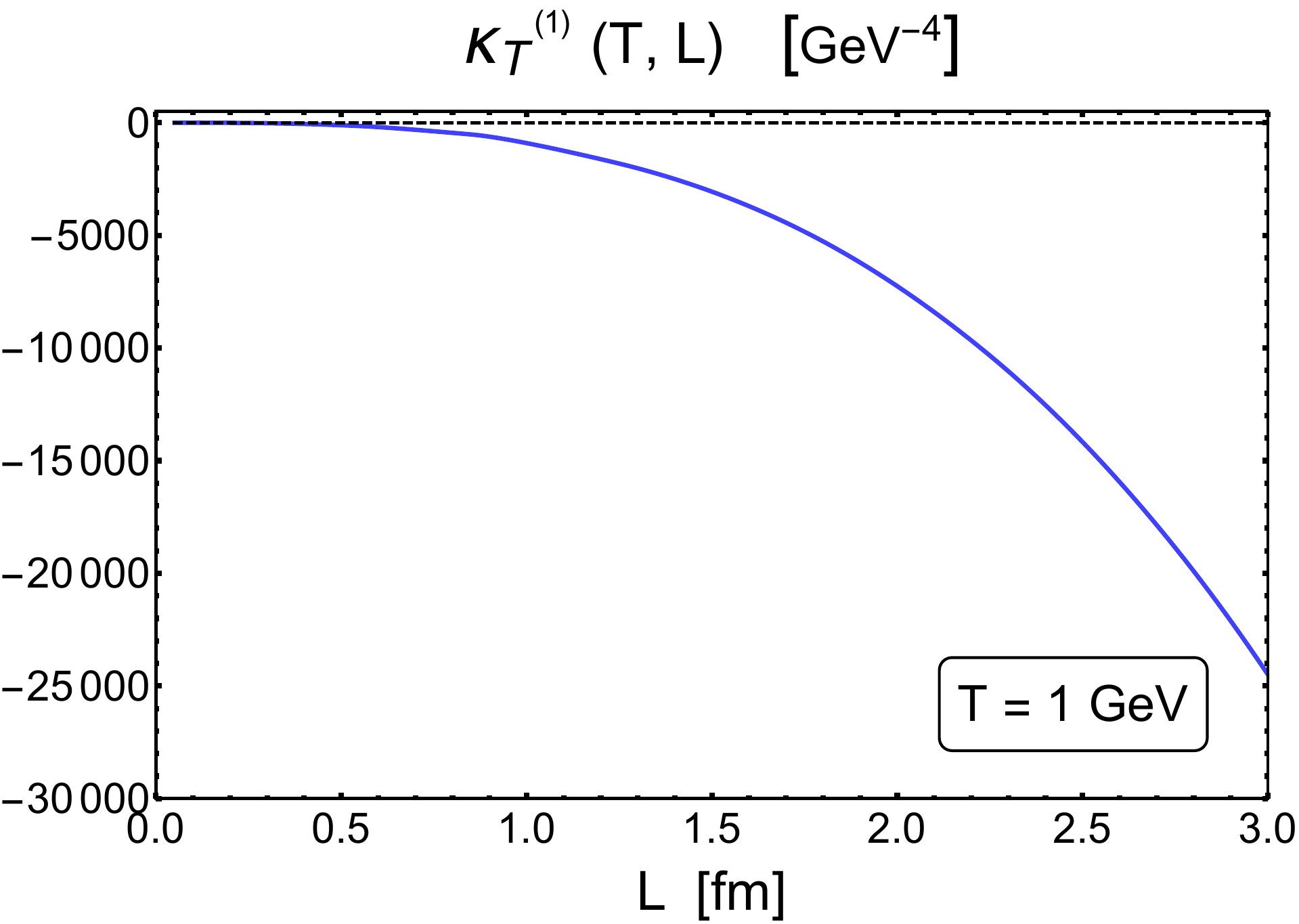} \hspace{.2in}
	\includegraphics[width=2.75in]{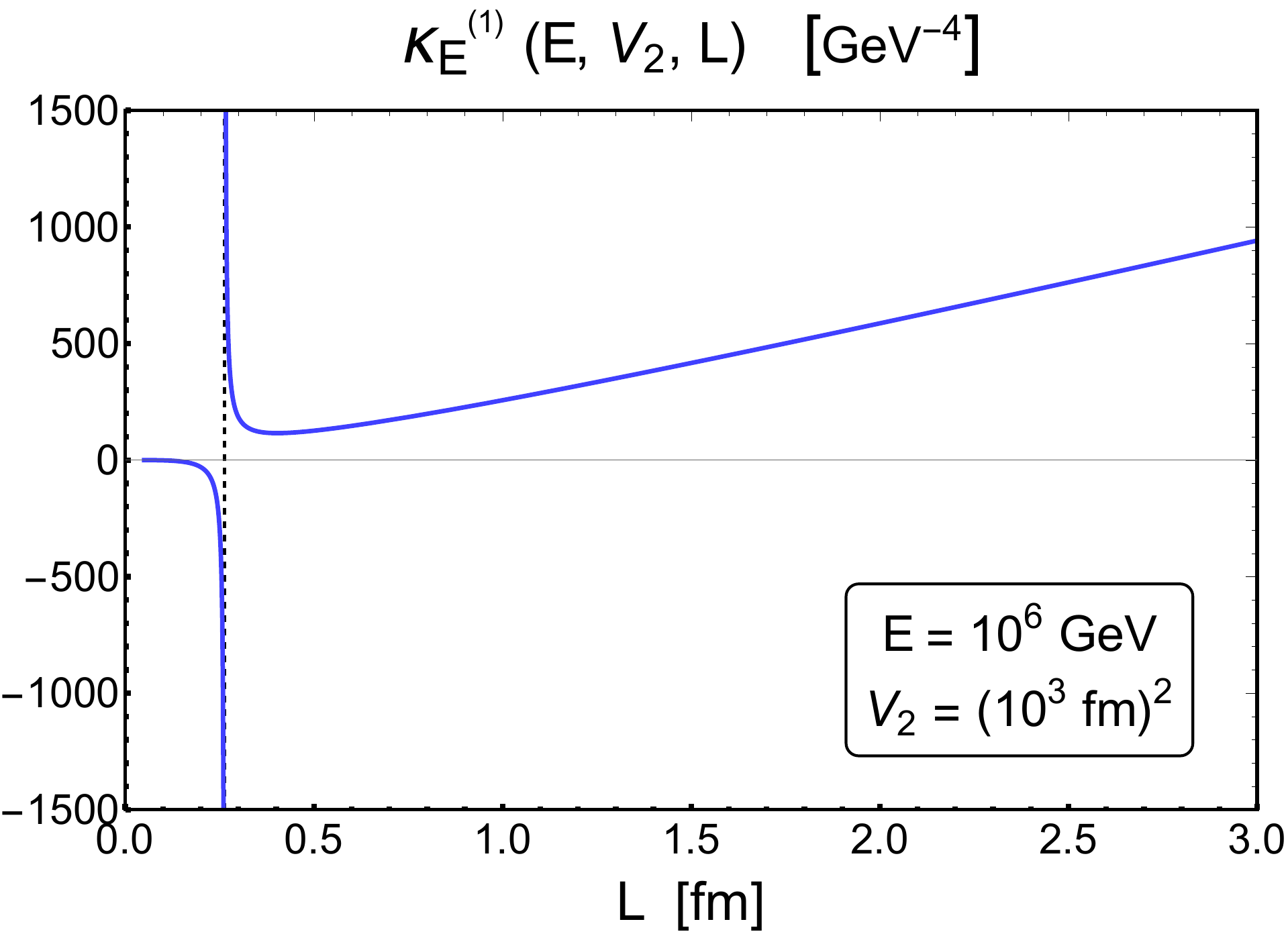}
	\caption{
		\label{kappaofLplot}
		(Left) Isothermal compressibility $\kappa_T = (1/L)\partial L/\partial p|_T$ as a function of plate separation $L$ for a noninteracting, scalar field between two parallel plates a distance $L$ apart and held at a temperature $T = 1$ GeV.  (Right) Isoenergetic compressibility $\kappa_E = (1/L)\partial L/\partial p|_E$ for the same scalar field system with constant energy $E = 10^6$ GeV and with parallel plates area $V_2=10^6$ fm$^2$.
	}
\end{figure}

We thus conclude that our massless, noninteracting scalar field theory constrained between two parallel plates exhibits a phase transition
 at a critical \emph{length} $L_c$.  We believe this is the first example of the explicit derivation of a phase transition induced by changing the size of the system, as opposed to the usual means of inducing a phase transition by changing the temperature of the system.  This is interesting because, in the derivation shown above, the phase transition is due to changing an extensive variable, as opposed to the usual description of a phase transition due to changing an intensive variable.  To be clear: we envision the (somewhat artificial) construct of a pair of (approximately infinitely large) parallel plates of \emph{fixed} separation length $L$.  The space between the plates is filled with a noninteracting, massless scalar field, and one measures the pressure on the plates.  Since the separation length is fixed, the system cannot actually collapse.  However, one is tempted to interpret the right plot of figure~(\ref{pofLplot}) as follows;  Consider a system of two (approximately infinitely large) parallel plates filled with a noninteracting, massless scalar field.  One plate is fixed, but the other plate is freely allowed to move and is exposed to a constant external pressure $p$; at equilibrium, the parallel plates settle down to an average separation distance $\langle L\rangle$ set by $p$.  As one slowly dials up the external pressure $p$, $L$ decreases but the system continues to find a new, smaller $\langle L\rangle$ at which the plates are in equilibrium with the external pressure $p$\footnote{To really match in principle the right plot of figure~(\protect\ref{pofLplot}) one would have to keep the total energy constant by continuing to remove the energy put into the system by the work of the external pressure squeezing the plates together.}.  At a certain large enough critical pressure $p_c$, the scalar field inside the plates can no longer resist the external pressure, and the system collapses.  One would very much like to confirm this extrapolated interpretation from the canonical ensemble; to do so would require going to the higher ensemble in which the system is in contact with a hypothetical thermal pressure bath.  It is perhaps not so easy to see from figure~(\ref{pofLplot}) or figure~(\ref{kappaofLplot}), but we will show below that the phase transition occurs at a length of order the thermal de Broglie wavelength, $L_c\sim1/T$.  Since the phase transition seen from the canonical ensemble calculation occurs at a length of order the thermal de Broglie wavelength, it is possible that fluctuations in the separation distance $L$ from the mean distance $\langle L\rangle$ that one would necessarily observe in a system exposed to a thermal pressure bath spoil the observation of a phase transition.  A quantitative derivation of the properties of a massless, noninteracting scalar field between two parallel plates of variable separation distance and exposed to a thermal pressure bath via a higher ensemble is left for future work.

Since we observe this phase transition for a massless, noninteracting scalar field theory in a small enough geometric confinement, a natural question to ask is: is this transition one of Bose-Einstein condensation?  One can see an indication of an answer in the negative from figure~(\ref{entropyhessianplot}).  On the left, we plot the entropy $S$ as a function of the area of the parallel plates, $V_2$, and the separation of the plates, $L$, for our isolated scalar field theory system with constant total energy $E$.  For small enough $L$ one can explicitly see how the entropy transitions from a convex to a concave function.  We further quantify this phase transition by examining the most positive eigenvalue of the Hessian of $S(V_2,\,L)$, i.e.\ with $E$ fixed.  For a concave function, all eigenvalues of the Hessian are negative.  The plot on the right of figure~(\ref{entropyhessianplot}) shows in gray the region of $(V_2,\, L)$ space for which the most positive eigenvalue of the Hessian is greater than zero, which is to say the region for which the entropy is no longer a convex function of $V_2$ and $L$\footnote{Note that since we are using formulae formally derived in the $V_2\rightarrow\infinity$ limit, we restrict our plot to large values of $V_2$.}.  The critical length that forms the phase transition boundary of this region is a function of the parameters $V_2$, $E$, and $T\times L$.  
In the right hand plot of figure~(\ref{entropyhessianplot}), the edge of the region for which the entropy is no longer concave is given by a function $L_c(E,\,V_2,\,T\times L)$ where $T\times L\approx 1$; i.e.\ the phase transition occurs when $L\approx 1/T$.
Therefore this geometrical phase transition can occur for a Bose system at arbitrarily large temperature.  We can conclude then that the phase transition is not one of Bose-Einstein condensation.  

\begin{figure}[!tbp]
	\centering
	\includegraphics[width=2.75in]{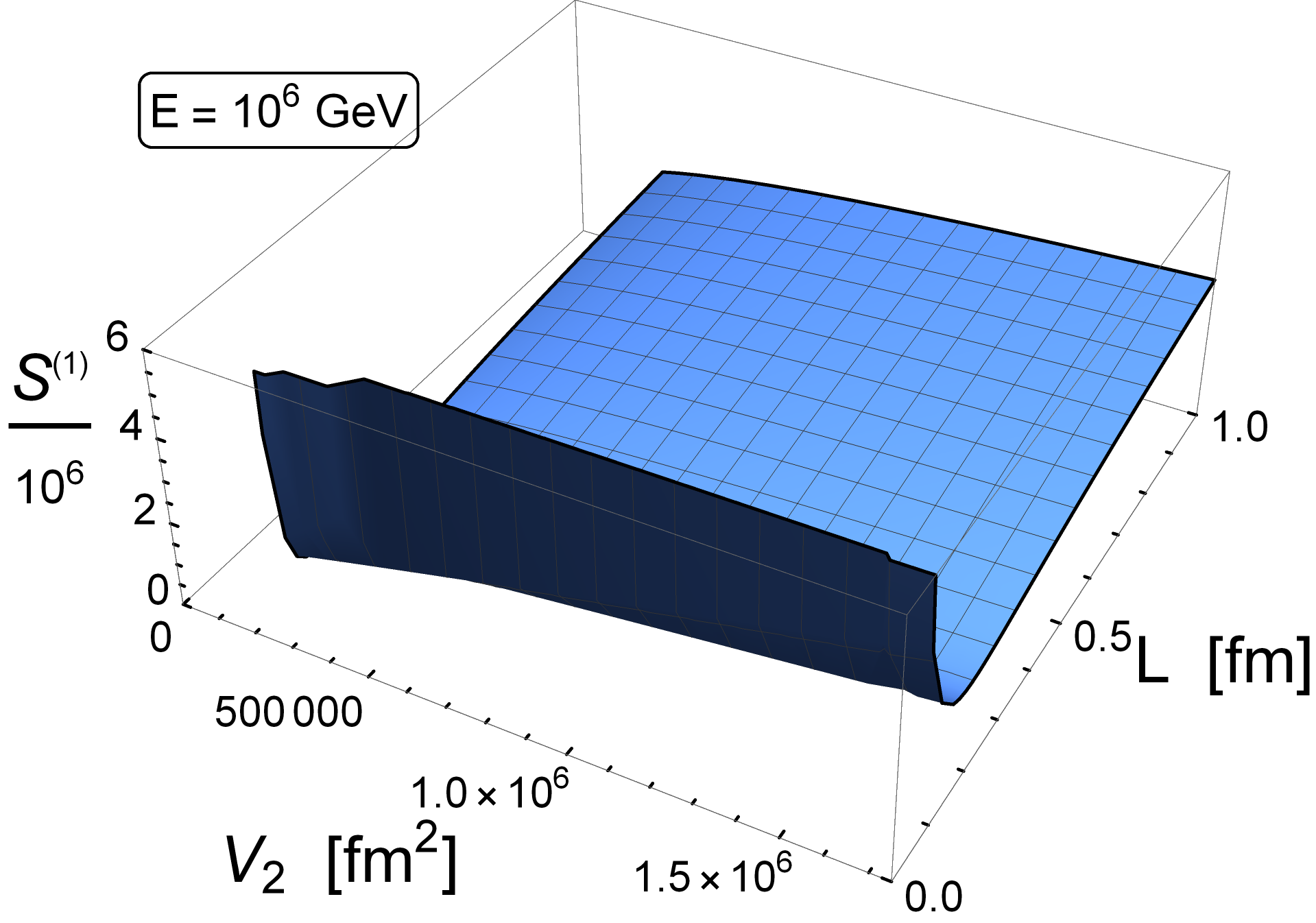} \hspace{.2in}
	\includegraphics[width=2.75in]{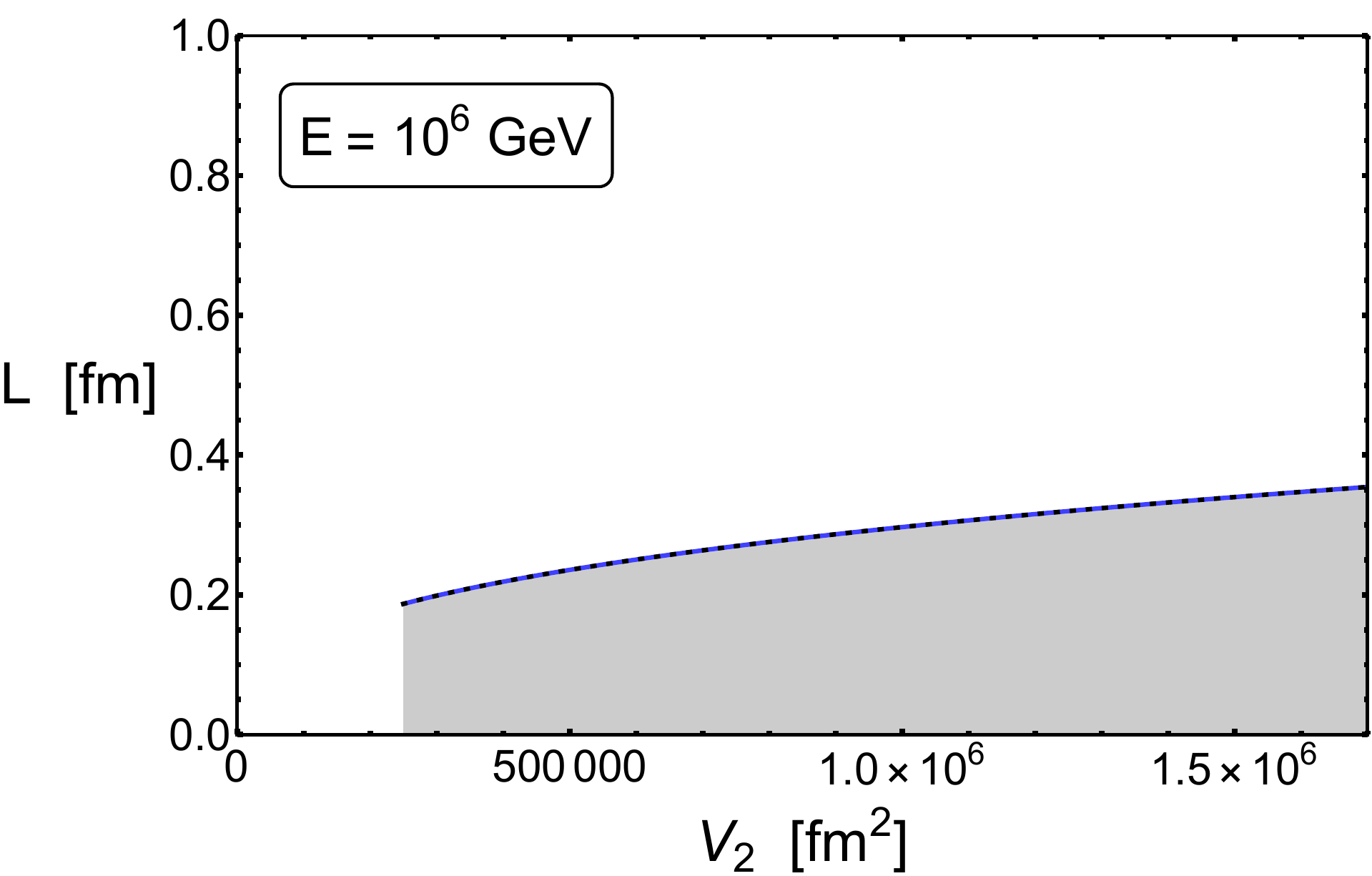}
	\caption{
		\label{entropyhessianplot}
		(Left) The entropy $S$ of a massless, noninteracting scalar field between two parallel plates of area $V_2$ in fm$^2$ separated by a distance $L$ in fm for energy $E=10^6$ GeV.  (Right) The region of $(V_2,\,L)$ space for which one of the eigenvalues of the Hessian of the entropy $S$ is positive is shaded in gray. The edge of the region is given by the equation $L_c(V_2,\ E = 10^6\,\,\mathrm{GeV},\,T\times L=1)$.
	}
\end{figure}

One may then naturally next ask if the phase transition can be found in a massless, noninteracting Fermi field.  We will show below that the phase transition does, indeed, persist for a massless, noninteracting Fermi field.  In order to repeat the above analysis for a Fermi field, we must slightly alter our quantization condition \cite{DeFrancia:1994us,DePaola:1999im}.  In order to prevent any Dirac current from leaking through the plates confining our system, we require that
\begin{equation}
	\eta^\mu\bar{\psi}\gamma_\mu\psi = 0,\qquad \eta^\mu = \left(0,\,\vec{\eta}\right)^\mu.
	\label{eq:diraccurrent}
\end{equation}
In order to satisfy eq.~(\ref{eq:diraccurrent}), one must take the momentum modes of our Dirac field to satisfy
\begin{equation}
	k_\ell = \left(\ell+\frac{1}{2}\right)\frac{\pi}{L},\qquad\ell=0,\,1,\,2,\ldots
\end{equation}
Following the methods of section~\secn{AppCanonicalComputation} the Fermion partition function for a massless, noninteracting Dirac field constrained between two parallel plates of area $V_2$ and kept a distance $L$ apart evaluates to
\begin{align}
	\label{eq:fermistart}
	\ln Z_F 
		& = 2V_2\sum_{\ell=0}^\infinity\int\frac{d^2k}{(2\pi)^2}\left[ \beta\sqrt{k^2+k_\ell^2} + 2\ln(1+e^{-\beta\sqrt{k^2+k_\ell^2}}) \right] \\
		& = \frac{\pi V_2}{L^2}\left[ \frac{7}{2}\frac{\tilde{\beta}}{1440} - \frac{2}{\tilde{\beta}^2}
			\sum_{\ell=0}^\infinity\left(\ell+\frac{1}{2}\right)\tilde{\beta}\Li_2\left(-e^{-\tilde{\beta}(\ell+\frac{1}{2})}\right) + \Li_3\big(-e^{-\tilde{\beta}\left(\ell+\frac{1}{2}\right)}\big) \right],
	\label{eq:fermipartition}
\end{align}
where $\tilde{\beta}\equiv\pi/(TL)$.  Note that the overall factor of 2 in eq.~(\ref{eq:fermistart}) is due to the spin-1/2 nature of the Dirac field.  The results of repeating the scalar field theory analysis are shown in figures~(\ref{pofLplotF}), (\ref{kappaofLplotF}), and (\ref{entropyhessianplotF}).  Qualitatively the picture is the same for the Dirac theory as for the scalar theory; quantitatively, one finds that the critical length separating the concave from convex region of the entropy $S$ as a function of plate area $V_2$ and separation length $L$ is related to a slightly higher temperature of the gas, $L_c\approx0.8/T$.  

\begin{figure}[!htbp]
	\centering
	\includegraphics[width=2.75in]{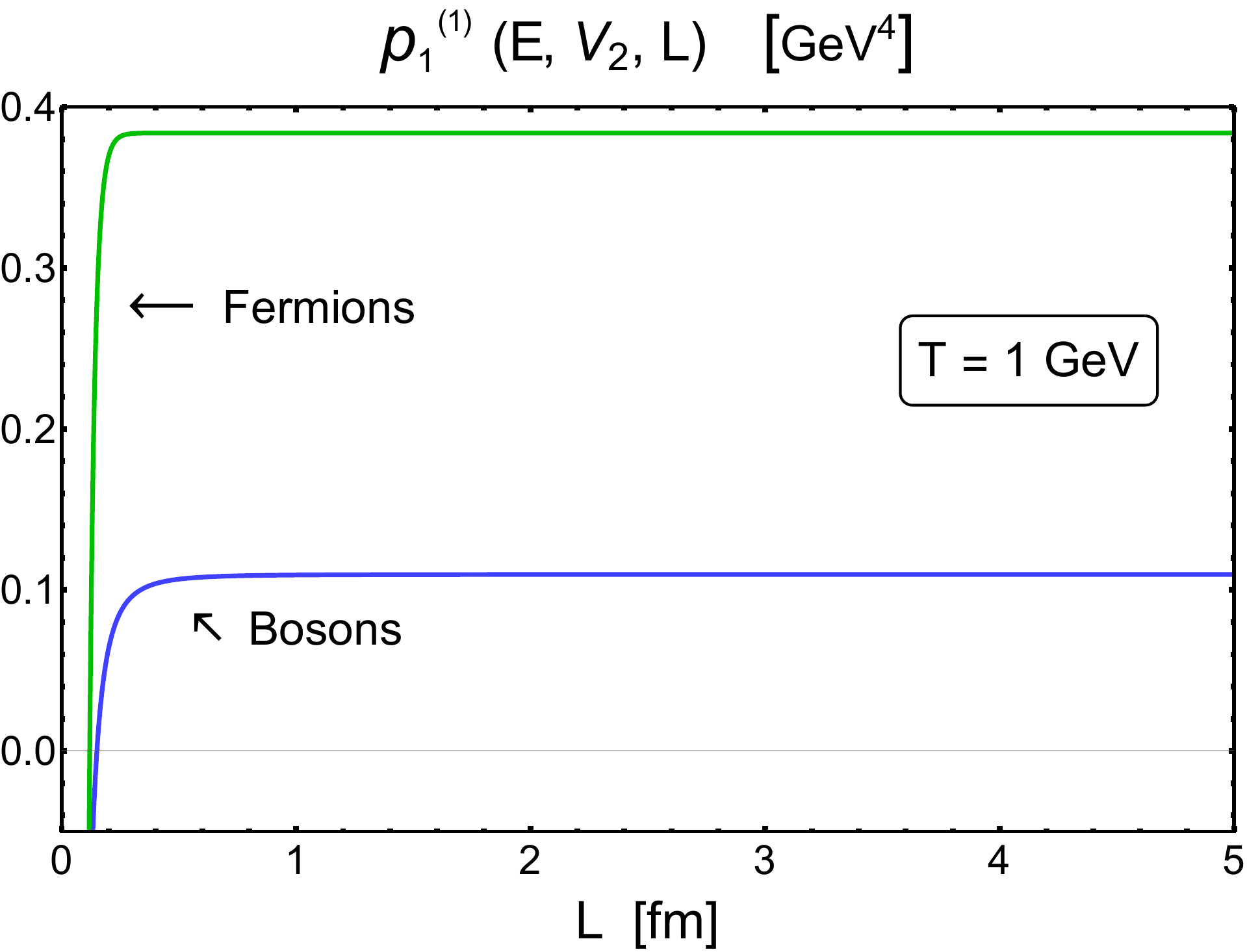} \hspace{.2in}
	\includegraphics[width=2.75in]{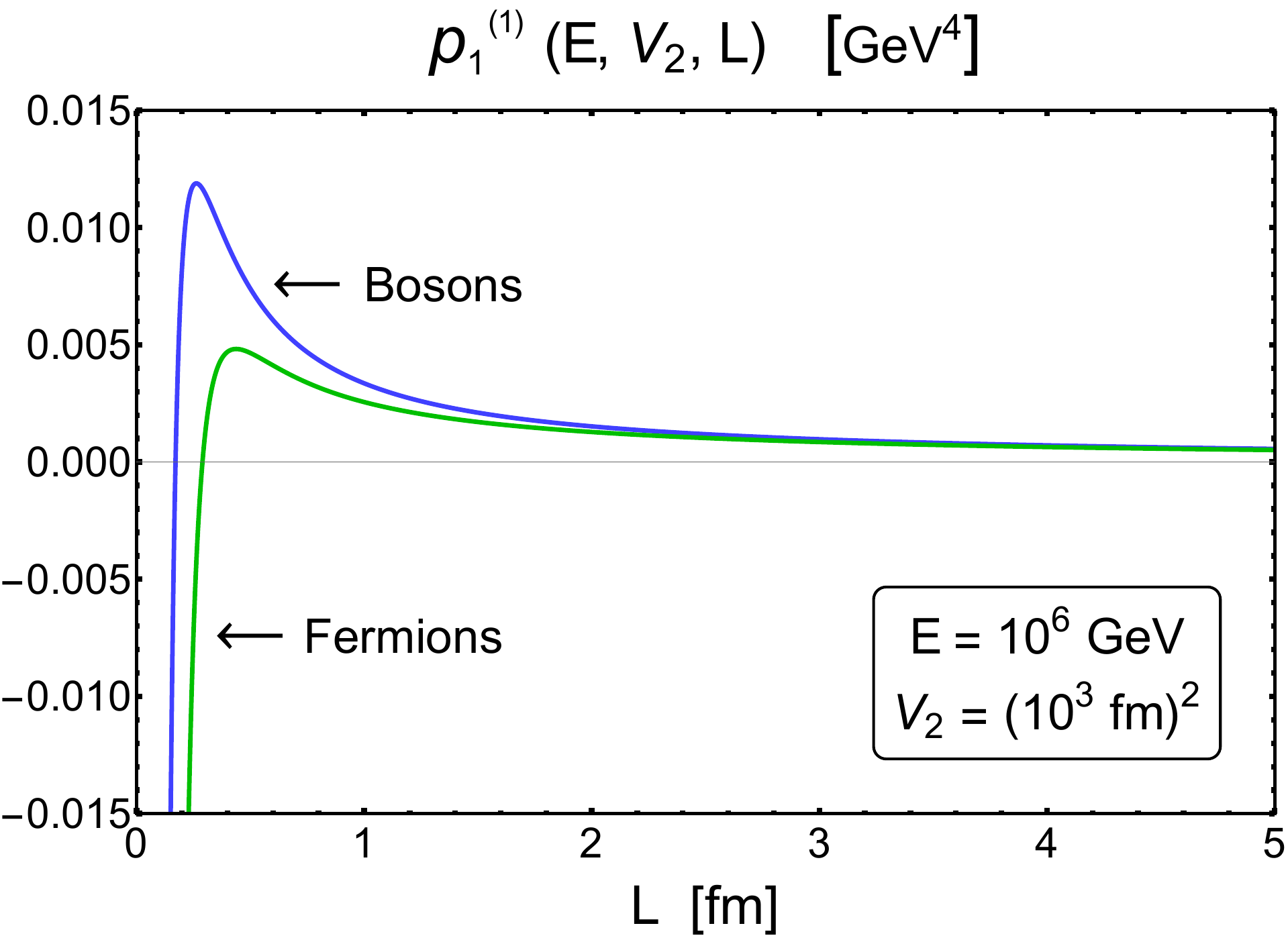}
	\caption{
		\label{pofLplotF}
		(Left) Pressure $p$ as a function of $L$ for a massless, noninteracting Dirac field between two parallel plates separated by a distance $L$ in contact with a thermal heat bath at a temperature $T = 1$ GeV.  (Right) Same but for an isolated system at constant energy $E = 10^6$ GeV and with parallel plates area $V_2=10^6$ fm$^2$.
	}
\end{figure}

\begin{figure}[!htbp]
	\centering
	\includegraphics[width=2.75in]{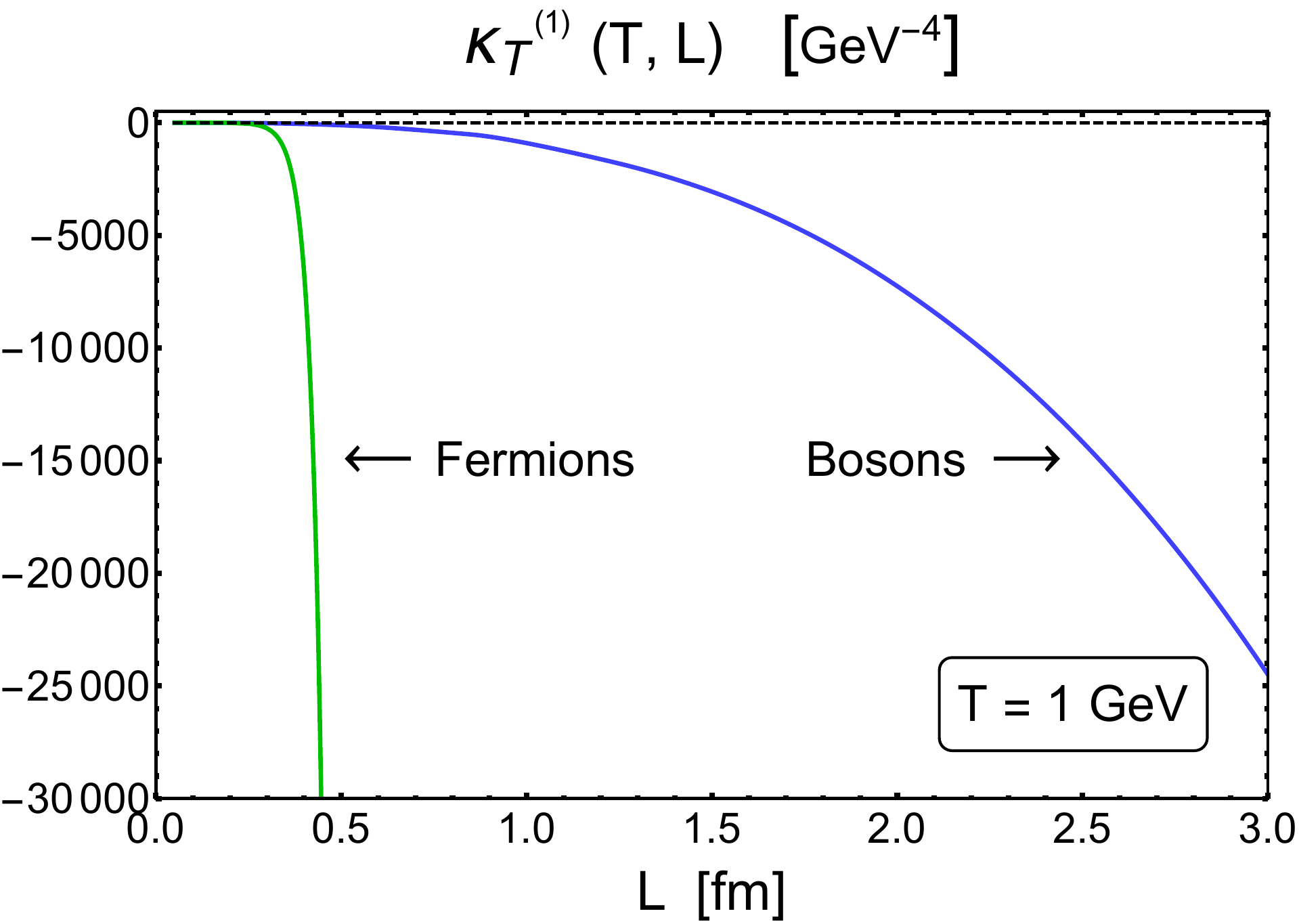} \hspace{.2in}
	\includegraphics[width=2.75in]{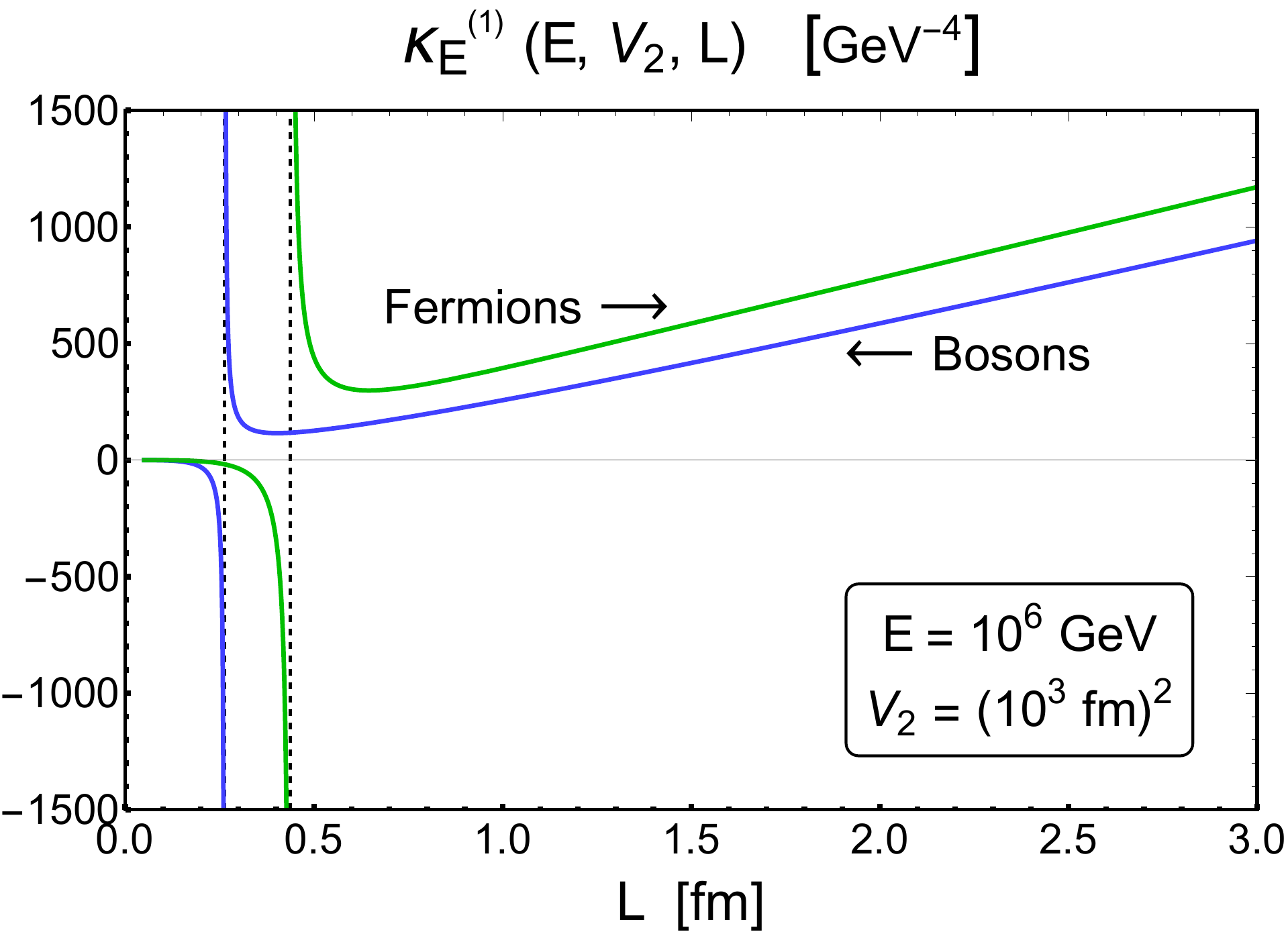}
	\caption{
		\label{kappaofLplotF}
		(Left) Isothermal compressibility $\kappa_T = (1/L)\partial L/\partial p|_T$ as a function of plate separation $L$ for a noninteracting, Dirac field between two parallel plates a distance $L$ apart and held at a temperature $T = 1$ GeV.  (Right) Isoenergetic compressibility $\kappa_E = (1/L)\partial L/\partial p|_E$ for the same scalar field system with constant energy $E = 10^6$ GeV and with parallel plates area $V_2=10^6$ fm$^2$.
	}
\end{figure}

\begin{figure}[!htbp]
	\centering
	\includegraphics[width=2.75in]{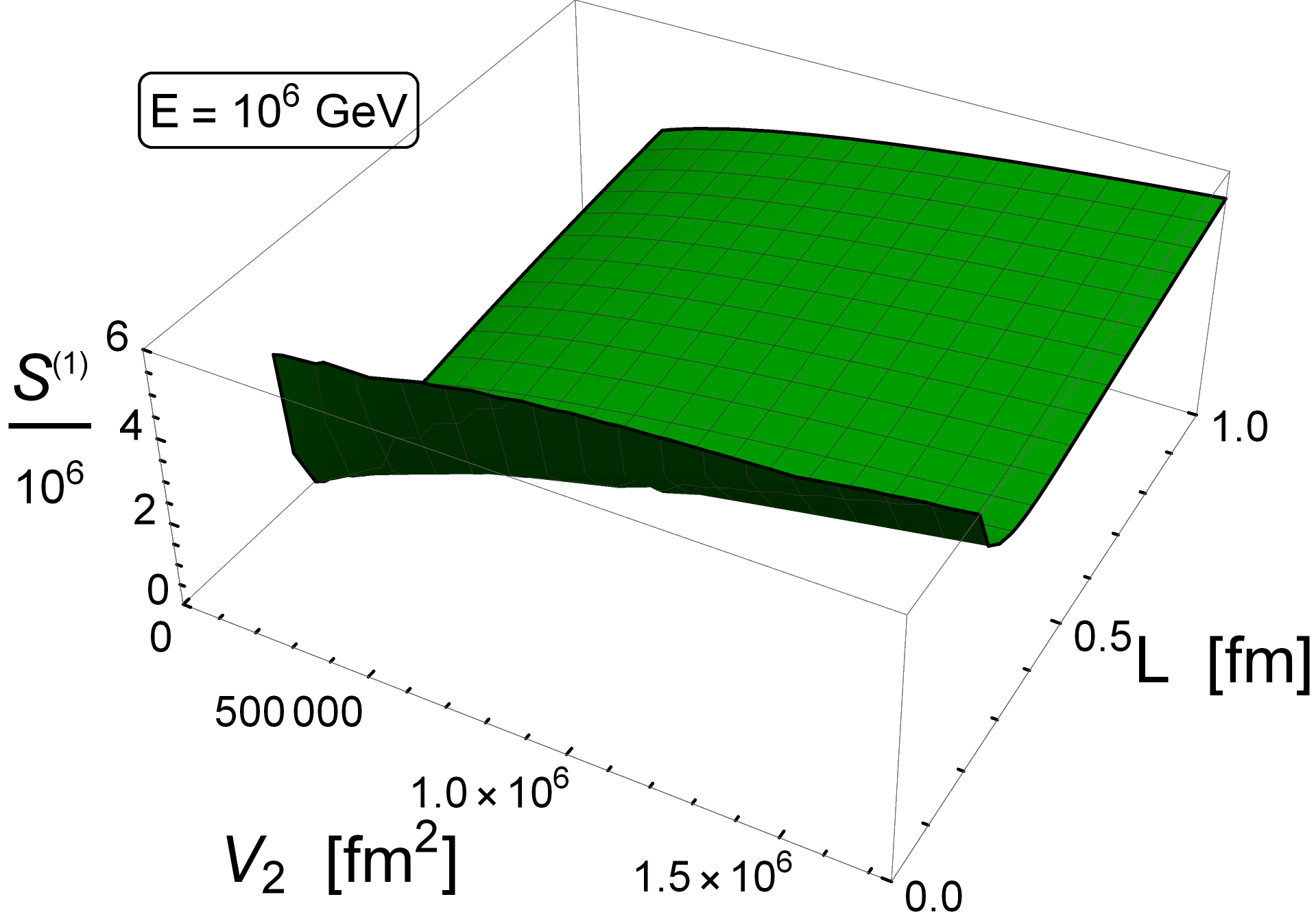} \hspace{.2in}
	\includegraphics[width=2.75in]{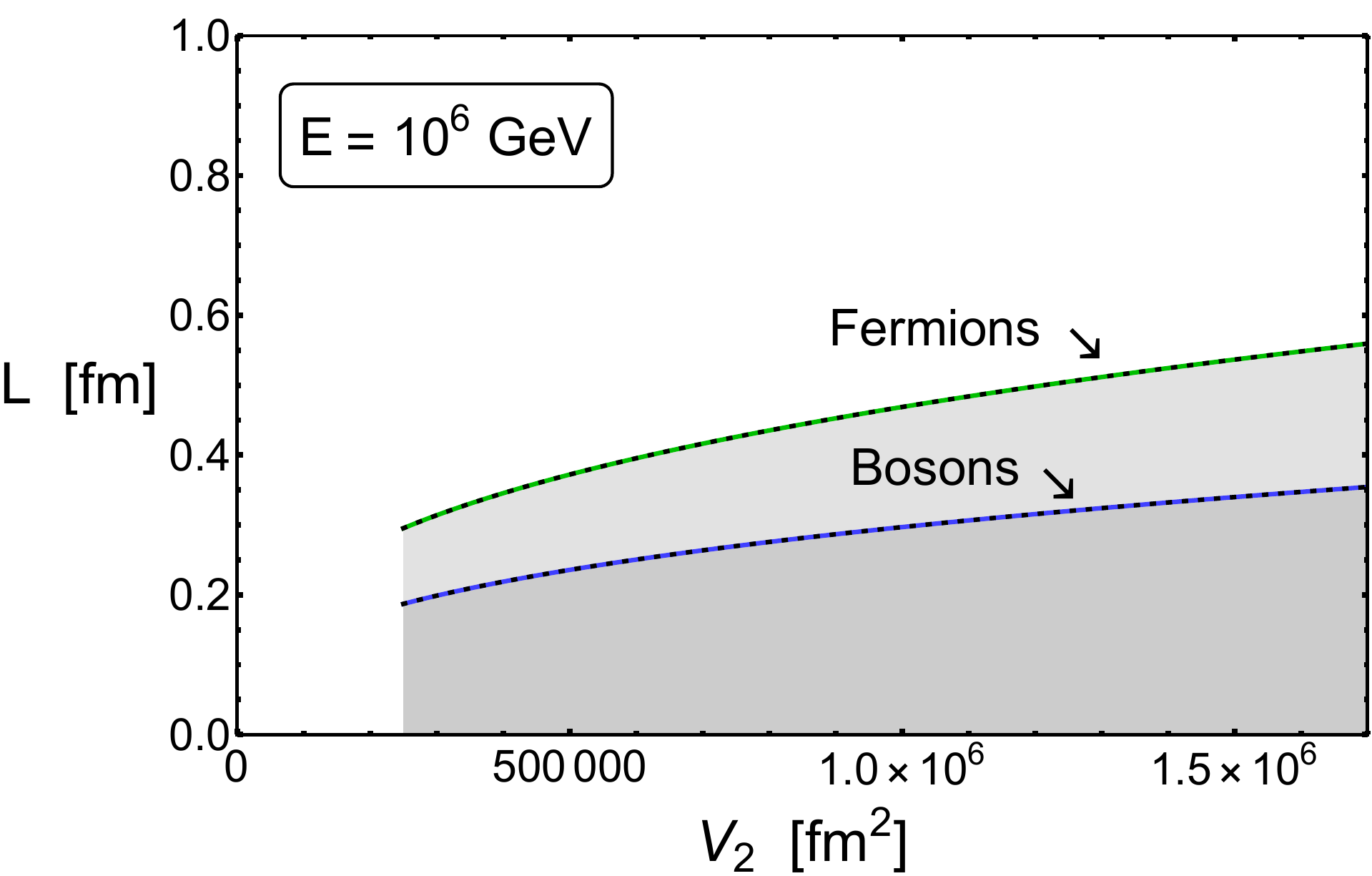}
	\caption{
		\label{entropyhessianplotF}
		(Left) The entropy $S$ of a massless, noninteracting Dirac field between two parallel plates of area $V_2$ in fm$^2$ separated by a distance $L$ in fm for energy $E=10^6$ GeV.  (Right) The region of $(V_2,\,L)$ space for which one of the eigenvalues of the Hessian of the entropy $S$ is positive is shaded in gray. The edge of the Fermion region is given by the equation $L_c(V_2,\ E = 10^6\,\,\mathrm{GeV},\,T\times L = 0.8)$.
	}
\end{figure}

One can even go so far as to show that the phase transition exists for massive, nonrelativistic Bosons and Fermions, too, although we leave the details for a future publication.
\section{Summary and prospects}\labsec{Conclu}

In this work we implemented the first step of a strategy designed to investigate the thermodynamics of small QGP systems, focusing on the finite size corrections to the usual Stefan-Boltzmann thermodynamic properties computed in thermal field theory and also demonstrating the emergence of a new geometric phase transition.  In particular, we set up a framework for probing relevant finite size effects by means of an actual spatial boundary for a geometric confinement, imposing Dirichlet boundary conditions. We argued in section~\secn{GeoConf} that Dirichlet BCs are the most appropriate for capturing the relevant finite size effects for heavy ion collisions and quark-gluon plasma phenomenology because these BCs prevent the weakly coupled quantum fields, supposed to account for the relevant degrees of freedom inside a QGP, from propagating outside of the geometric region defining this QGP system. For the sake of simplicity, we considered $D-1$ spatial dimensions with $c\le D-1$ dimensions of finite length $L_i$ arranged in standard rectilinear coordinates.  We then filled the space inside our geometrically confined region with a massless, noninteracting scalar field.  Such a model encodes the main aspects of a massless, noninteracting gas of gluons. In order to best make contact with lattice QCD results and also for simplicity we computed the thermodynamic quantities in the canonical ensemble, in the process extending standard thermal field theory techniques.  

The detailed analytical calculations found in the appendices are summarized in section~\secn{FieldTheoCalculations}.  The results were computed by two independent methods, resulting in two different, but numerically equivalent, infinite sums for each geometrically confined system.  The two different sums have very different numerical convergence properties: one converges exponentially fast for $T\times L_i<1$ while the other converges exponentially fast for $T\times L_i>1$.  

In section~\secn{FiniteSizeThermo} we discussed some important qualitative consequences of our derivation of the statistical mechanics of our system.  In particular, we saw that the First Law of Thermodynamics is generalized by the presence of different pressures: instead of a single scalar pressure as in the case of all spatial dimensions having infinite extent, the pressure in the $i^\mathrm{th}$ direction may be different from the pressure in the $j^\mathrm{th}$ direction depending on the various lengths $L_k$.  On the other hand, we found that the limited spatial extent of our system did not affect the Third Law of Thermodynamics.  Of critical importance, we computed the size of the fluctuations in energy away from the mean energy dictated by the contact of our system with a thermal heat bath.  Even though the systems of parallel plates or a tube have some but not all direction(s) of finite size, the volume of these systems is infinite.  Therefore the relative size of the energy fluctuations compared to the mean energy is 0.  As a result, the canonical ensemble can be used to compute the thermodynamic properties for isolated, constant energy parallel plates or infinite tube systems.  At the same time, the energy fluctuations in a finite volume box are \emph{not} zero, and there must be corrections in the application of the canonical ensemble to an isolated, finite volume box system.  

In section~\secn{ResultThermo} we presented a number of quantitative plots of the free energy, entropy, energy, pressure, and heat capacity for a massless, noninteracting scalar field contained inside infinite parallel plates, an infinite rectangular tube, and a finite-sized box. Of particular note was the surprisingly large finite size corrections for the pressure.  For a system of the approximate size and temperature of a high multiplicity proton-proton collision at LHC, which has been argued to exhibit hydrodynamic behavior based on thermodynamic quantities computed in the Stefan-Boltzmann limit \cite{Romatschke:2016hle}, we found $\sim20\%$ corrections for an infinite tube and $\sim40\%$ corrections for a symmetric, finite box.  Even for systems of the size of mid-central nucleus-nucleus collisions at LHC the corrections are of order $\sim10\%$.  Since the size of the azimuthal anisotropy measured in such collisions at LHC are of order $\sim10\%$ \cite{Bernhard:2016tnd}, these corrections to the equation of state due to the finite size of the system may have important implications for heavy ion phenomenology.  

In section~\secn{PhaseTransition} we discovered that an isolated system of noninteracting particles confined within parallel plates at temperature $T$ undergoes a second order phase transition at a critical length $L_c\sim1/T$: for $L>L_c$ the system is stable and resists compression; for $L<L_c$ the system collapses. The phase transition is not one of Bose-Einstein condensation as we showed that a noninteracting Dirac field also experiences such a phase transition.  Nor is the transition a relativistic effect as it is also experienced by massive, nonrelativistic Bosons and Fermions.  It is tempting to interpret our results as follows: a system of fixed energy can only resist so much external pressure before collapsing.  One can see that figures~(\ref{pofLplot}) and (\ref{pofLplotF}) support this interpretation: the system energy density is on the same order as the pressure, with $p\lesssim E/V$, when the system undergoes the phase transition at the critical length. Further clarity on this interpretation requires the use of a higher order ensemble in which the system is put in contact with a thermal pressure bath.  Such an investigation would also provide insight into the importance of the fluctuations in the separation distance between the parallel plates about the average, equilibrium length. Should this higher ensemble show that a pressure driven phase transition exists, one could use the higher ensemble to calculate the critical exponent for this second order transition.  Other work that has claimed the observation of a first order phase transition in small systems~\cite{Park:2008sk} performed the calculation in a finite volume box in the canonical ensemble; it is not clear that the energy fluctuations inherent in using the canonical ensemble invalidates the conclusions reached in that work. In the only other work that we are aware of that examines finite size driven phase transitions~\cite{Bausch2018}, the phase transition is due to self interactions of the system; since our system is noninteracting, the phase transition we see is due purely to geometric confinement effects.  Future work includes determining whether or not a similar phase transition exists for an isolated system with periodic boundary conditions in one direction or for a tube system, and numerical investigations of our conclusions in QCD \cite{Kitazawa:2019otp}.  

It is important for us to point out that geometrically confining a thermal quantum field into a certain region of space with Dirichlet boundary conditions (even with an infinite volume) appears to provide a solution to the infrared Linde problem~\cite{Linde:1980ts,Fraga:2016oul,Simonov:2016xaf}. The origin of the Linde problem is the existence of the zero mode in systems with, e.g., periodic boundary conditions: a momentum mode with zero energy.  Imposing Dirichlet boundary conditions naturally provides an infrared cut-off for the momentum modes $p>p_{\mathrm{min}}\sim1/L$ dictated by the system size $L$. We explicitly showed in this work that the Matsubara zero mode disappears from the leading order thermal field theory calculations.  It would be interesting to quantitatively check that in fact Dirichlet boundary conditions cure the zero mode Linde problem at higher orders in perturbation theory, too, and see that the perturbative series becomes analytic in the coupling.


\acknowledgments
The authors would like to thank Alexei Bazavov, Trambak Bhattacharyya, David Blaschke, J\'{e}r\^{o}me Charles, Jean Cleymans, Gerg\H{o} Endr\H{o}di, Zoltan Fodor, Charles Gale, Ioan Ghi\c{s}oiu, Hendrik van Hees, Joseph Kapusta, Dima Kharzeev, Klaus Kirsten, Masakiyo Kitazawa, Chris Korthals Altes, Aleksi Kurkela, Mikko Laine, Laurent Lellouch, Larry McLerran, Alexander Parvan, Andr\'{e} Peshier, Rob Pisarski, Dirk Rischke, Frederik Scholtz, York Schr\"{o}der, Mike Strickland, Nan Su, Derek Teaney, Giorgio Torrieri, Hugo Touchette, Raju Venugopalan, Aleksi Vuorinen, and many more not mentioned here for stimulating comments, discussions, and useful references.

S.~M.~would like to acknowledge the financial support from the Claude Leon Foundation.  Although S.~M.~was affiliated with UCT when the bulk of this research was completed, he is no longer affiliated with UCT at the time of publication.  I.~K.~and W.~A.~H.~wish to thank the SA-CERN Collaboration, the South African National Research Foundation, and the National Institute for Theoretical Physics for their support. I.~K.~wishes to further acknowledge Deutscher Akademischer Austauschdienst for supporting her work.  I.~K.~was further supported by the United States Department of Energy under grant number DE-FG02-00ER41132 and the Multifarious Minds grant provided by the Simons foundation during the final stages of this work.

\appendix
\section{Details of calculations}\labsec{AppDetailsCalculations}

In this section, we detail two different methods of computing the free energy density for a single neutral noninteracting massless scalar field in $c$ geometrically confined dimensions by deriving the logarithm of the partition function. It is important to note that the usual method for computing such a logarithm cannot be applied in a way that is necessarily straightforward since the compactification of any number of spatial dimensions leads to summations that are formally divergent.

For both methods we employ dimensional regularization~\cite{Leibbrandt:1975dj} within the modified minimal subtraction ($\MSbar$) scheme~\cite{Bardeen:1978yd} in order to regulate the divergences, but we also complement this regularization procedure through the use of zeta type of regularization whenever necessary. In the present massless case of interest, the usual (power like) divergences, if any, are set to zero by Dim.~Reg.\ so that there are no divergences that would need to be cured by means of renormalizing the vacuum~\cite{Kastening:1996nj}.

In subsection~\secn{AppCanonicalComputation} we follow the usual method as described by~\cite{Kapusta:2006pm,Laine:2016hma}.  In subsection~\secn{AppAlternativeComputation} we present an alternative calculation. Naturally, the two methods yield analytically equivalent results which then of course agree numerically---even though their different convergence properties, when the summations are truncated at some finite orders, make them conveniently suitable in different regimes.

It is important to note that we will make significant use of Epstein zeta functions and their various analytic representations.  The derivation of these representations often exploits the Poisson summation formula,
{\allowdisplaybreaks
\be
\sum_{s=-\infty}^\infty\left(e^{-x \alpha s^2}\right)=\sqrt{\frac{\pi}{x \alpha}}\times%
\sum_{s=-\infty}^\infty\left(e^{-\frac{\pi^2}{x \alpha}\left.s\right.^2}\right)%
.\labeq{PoissonResummation}%
\ee}%
We will not go into the detail of these derivations, but rather point the interested reader to the extensive literature~\cite{Ambjorn:1981xw,Ambjorn:1981xv,Elizalde:1989mr,Kirsten:1994yp,Elizalde:1994yz,Elizalde:1997jv,Elizalde:2007du,Kirsten:2010zp,Elizalde:2012zza} that is the result of three decades of intensive work on these special functions.

\subsection{Usual computation}\labsec{AppCanonicalComputation}

In this section, we compute the canonical representations for the free energy density in different scenarios, following and adapting the general methods of~\cite{Kapusta:2006pm,Laine:2016hma}. For pedagogical reasons, we present the case of $c=1$ in detail, but the result will then be easy to extend to arbitrary $c\leq D-1$. Our starting point is then either eq.~\eqn{ZM1final} or eq.~\eqn{LnZc}, leading to
{\allowdisplaybreaks
\be
\ln\mathcal{Z}^{(1)}=-\half\sum_{n\in\Z}\sum_{\ell_1\in\N}\sum_{\vect{n}_k\in\Z^{D-2}}%
\ln\bigg[\beta^2\left(\omega_{n}^2+\omega_{\ell_1}^2+\vect{\omega}_k^2+m^2\right)\bigg]%
.\labeq{ZM1Start}
\ee}

We then consider the following two identities
{\allowdisplaybreaks
\bea
\ln\bigg\{(2\pi n)^2+\beta^2\chi^2\bigg\}&=&\int_{1}^{(\beta\chi)^2}\!\!\frac{d\theta^2}{\theta^2+(2\pi n)^2}+\ln\bigg\{1+(2\pi n)^2\bigg\},\labeq{ID1}\\%
\sum_{n=-\infty}^{+\infty}\frac{1}{(\sfrac{\theta}{2\pi})^2+n^2}&=&\frac{2\pi^2}{\theta}\left(1+\frac{2}{\e^{\theta}-1}\right),\labeq{ID2}%
\eea}
\HS{0.12}and after substituting eq.~\eqn{ID1} into eq.~\eqn{ZM1Start}, defining for brevity $\omega^2\equiv\vect{\omega}_k^2+m^2$, dropping an infinite $T$- and $L_i$-independent term, and employing eq.~\eqn{ID2}, we obtain
{\allowdisplaybreaks
\bea
\ln\mathcal{Z}^{(1)}&=&-\sum_{\ell_1\in\N}\sum_{\vect{n}_k\in\Z^{D-2}}\left[%
\int_{1}^{\beta\sqrt{\omega_{\ell_1}^2+\omega^2}}\!\!d\theta~\bigg(\frac{1}{2}+\frac{1}{\e^{\theta}-1}\bigg)\right]\\%
&=&-\sum_{\ell_1\in\N}\sum_{\vect{n}_k\in\Z^{D-2}}\Bigg[%
\frac{\beta}{2}\sqrt{\omega_{\ell_1}^2+\omega^2}+\ln\left\{1-\e^{-\beta\sqrt{\omega_{\ell_1}^2+\omega^2}}\right\}\Bigg]%
,\labeq{LnZM12}%
\eea}
\HS{0.12}where we again dropped infinite $T$- and $L_i$-independent terms.

Assessing the case $c=1$, it is time to release the momentary (periodic) compactifications of the $D-2$ other spatial dimensions, sending each of the $R$ compactification lengths to its appropriate value (presently $L_2$ or $L_3$) each being asymptotically large, and get
{\allowdisplaybreaks
\be
\ln\mathcal{Z}^{(1)}=-L_2L_3\prefacmsb\sum_{\ell_1\in\N}\int\frac{d^{D-2}\vect{k}}{(2\pi)^{D-2}}~\Bigg[%
\frac{\beta}{2}\sqrt{\omega_{\ell_1}^2+\omega^2}+\ln\left\{1-\e^{-\beta\sqrt{\omega_{\ell_1}^2+\omega^2}}\right\}\Bigg]%
,\labeq{LnZM13}%
\ee}
\HS{0.12}where $\omega^2\rightarrow\vect{k}^2+m^2$, $D\equiv 4-2\epsilon$ being now a parameter ($\epsilon$ will be set to zero at the end), and we introduced an arbitrary regularization scale ($\Lambda^2\equiv \lmsb^{2} e^{\gamma_\rmi{\tiny E}}/4\pi$) in order to keep the proper dimensionality when $D\neq 4$, since we employ Dim.~Reg. within the $\MSbar$ scheme.

We now focus on the computation of the second term of eq.~\eqn{LnZM13}. To do so, recall that the volume of an $N$-dimensional unit sphere is given by $\Omega_N=2\pi^{N/2}/\Gamma\left(N/2\right)$, and consider the integral
{\allowdisplaybreaks
\be
J\left(a,N\right)\equiv\int\frac{d^Np}{(2\pi)^N}~\ln\left\{1-\e^{-\beta\sqrt{p^2+a^2}}\right\}.%
\ee}
\HS{0.12}Making the substitutions $p\rightarrow Tx$ and $a=Tb$ in the above integral, Taylor expanding the logarithm, and rewriting the integration measure, we obtain
{\allowdisplaybreaks
\be
J\left(a,N\right)=-\frac{\Omega_{N}}{(2\pi/T)^N}\sum_{s=1}^{\infty}\Bigg[\frac{1}{s}\int_0^{\infty}dx~x^{N-1}~\e^{-s\sqrt{x^2+b^2}}\Bigg].%
\ee}
\HS{0.12}Then, the substitution $z=\sqrt{x^2+b^2}$ leads to an integral representation of the modified Bessel function of the second kind, namely $K$, and we get
{\allowdisplaybreaks
\bea
J\left(a,N\right)&=&-\frac{2\pi^{\sfrac{N}{2}}}{(2\pi/T)^N\Gamma\left(\sfrac{N}{2}\right)}~\sum_{s=1}^{\infty}%
\frac{1}{s}\int_b^{\infty}\!\! dz~z\left(z^2-b^2\right)^{\frac{N}{2}-1}~\e^{-sz}\\%
&=&-\frac{2~T^N}{(2\pi)^{(N+1)/2}}~\sum_{s=1}^{\infty}\Bigg[\left(\frac{b}{s}\right)^{\frac{N+1}{2}}K_{\frac{N+1}{2}}(bs)\Bigg].%
\eea}
\HS{0.12}Therefore, in order to compute the second term of eq.~\eqn{LnZM13}, we need the following result
{\allowdisplaybreaks
\be
J\left(\sqrt{\omega_{\ell_1}^2+m^2},D-2\right)=-\frac{2~T^{D-2}}{(2\pi)^{\frac{D-1}{2}}}\sum_{s=1}^{\infty}\left[%
\left(\frac{\sqrt{\omega_{\ell_1}^2+m^2}}{sT}\right)^{\frac{D-1}{2}}~K_{\frac{D-1}{2}}\bigg(\frac{s}{T}\sqrt{\omega_{\ell_1}^2+m^2}\bigg)\right]%
.\labeq{IM1}%
\ee}
\HS{0.12}We further notice that the above result is indeed finite, as the second term of eq.~\eqn{LnZM13} contains neither Ultra-Violet (UV) nor Infra-Red (IR) divergences.

We focus now on the computation of the first term of eq.~\eqn{LnZM13}, which obviously contains a UV divergence. There are a number of different ways of regularizing this divergence, and we follow more specifically~\cite{Laine:2016hma}, employing again Dim.~Reg. within the $\MSbar$ scheme. We then have
{\allowdisplaybreaks
\bea
\int\frac{d^{D-2}\vect{k}}{(2\pi)^{D-2}}~\sqrt{\omega_{\ell_1}^2+\vect{k}^2+m^2}=\frac{\Gamma\left(-\frac{D-1}{2}\right)}%
{(4\pi)^{\frac{D-2}{2}}\Gamma\left(-\half\right)}\left[\left(\frac{\pi \ell_1}{L_1}\right)^2+m^2\right]^{\frac{D-1}{2}}%
.\labeq{PhiM1}%
\eea}
\HS{0.12}We notice an important point related to the above calculation: the integral does not feature any IR divergences in the massless limit as long as $L_1$ is finite, because $\ell_1$ never vanishes.

Combining eq.~\eqn{LnZM13} with eqs.~\eqn{IM1} and~\eqn{PhiM1}, we may write the free energy density, $f^{(c=1)}(T,L_1)$, of a system with one geometrically confined dimension as
{\allowdisplaybreaks
\bea
f^{(1)}&=&\frac{\Gamma\left(-\frac{D-1}{2}\right)}{2(4\pi)^{\frac{D-2}{2}}L_1\Gamma\left(-\half\right)}%
\prefacmsb\sum_{\ell_1\in\N}~\Bigg[\frac{\pi \ell_1}{L_1}\Bigg]^{D-1}\nn\\%
&-&\frac{2~T^{D-1}}{(2\pi)^{\frac{D-1}{2}}L_1}\prefacmsb\sum_{s=1}^{\infty}\sum_{\ell_1\in\N}~\Bigg[\left(\frac{\pi\ell_1}{sTL_1}\right)^{\frac{D-1}{2}}%
K_{\frac{D-1}{2}}\bigg(\frac{\pi}{TL_1}~s\ell_1\bigg)\Bigg]%
,\labeq{LnZM1final}%
\eea}
\HS{0.12}where, of course, we took the massless limit. It is then straightforward to extend the above procedure to an arbitrary number $c$ of compactified dimensions, and obtain the result
{\allowdisplaybreaks
\bea
\label{app:eqgam}
f^{(c)}&=&\frac{\Gamma\left(-\frac{D-c}{2}\right)}{2(4\pi)^{\frac{D-1}{2}}\prod_{i=1}^c\!(L_i)\Gamma\left(-\half\right)}%
\prefacmsb\!\!\sum_{\vect{\ell}\in\N^c}~\Bigg[\sum_{i=1}^{c}\left(\frac{\pi \ell_i}{L_i}\right)^2\Bigg]^{\frac{D-c}{2}}\labeq{LnZcfinal1}\\%
&-&\frac{2~T^{D-c}\prefacmsb}{(2\pi)^{\frac{D-1}{2}}\prod_{i=1}^c\!(L_i)}%
\sum_{s=1}^{\infty}\sum_{\vect{\ell}\in\N^c}\left[\left(\frac{\sum_{i=1}^{c}\left(\frac{\pi \ell_i}{L_i}\right)^2}{(sT)^2}\right)^{\frac{D-1}{4}}%
\!\!\!\!K_{\frac{D-c}{2}}\left(\frac{s}{T}\sqrt{\sum_{i=1}^{c}\left(\frac{\pi \ell_i}{L_i}\right)^2}\right)\right],\nn%
\eea}
\HS{0.12}for which we notice that Dim.~Reg. took care of the non logarithmic UV divergence, and that the above result will not need to be renormalized\footnote{Note that for certain values of $D-c$ the gamma function in the numerator of eq.~\protect\ref{app:eqgam} will have a pole.  For precisely those values of $D-c$ the Epstein zeta function will multiply these infinities by trivial zeros thus rendering the divergences harmless.}. 

Computing the summand in the first term of eq.~\eqn{LnZcfinal1} will then require some Epstein-Zeta regularization. Note the present Epstein-Zeta function of interest is defined~\cite{Kirsten:1994yp} as
{\allowdisplaybreaks
\be
E_N^{m^2}(s;a_1,\dots,a_N)\equiv\sum_{(n_1,\dots,n_N)\in\N^{N}}\Big(a_1n_1^2+\dots+a_Nn_N^2+m^2\Big)^{-s}%
\quad\forall a_1,\dots,a_N,m^2>0.\quad%
\ee}
\HS{0.12}We reproduce now the result for the analytic continuation of the above function (see~\cite{Kirsten:1994yp} for more details), in which the following recursion relation is useful
{\allowdisplaybreaks
\bea
E_N^{m^2}(s;a_1,\dots,a_N)&=&-\half E_{N-1}^{m^2}(s;a_1,\dots,a_{N-1})\nn\\%
&+&\half\sqrt{\frac{\pi}{a_N}}\frac{\Gamma(s-\half)}{\Gamma(s)}~E_{N-1}^{m^2}(s-\shalf;a_1,\dots,a_{N-1})\nn\\%
&+&\frac{2\pi^s}{\Gamma(s)}a_N^{-\half\left(s+\half\right)}\!\!\!\sum_{(n_1,\dots,n_N)\in\N^{N}}%
n_N^{s-\half}\Big(a_1n_1^2+\dots+a_{N-1}n_{N-1}^2+m^2\Big)^{\half\left(\half-s\right)}\times\nn\\%
&&\hspa{2.5}\times K_{s-\half}\left(\frac{2\pi n_N}{\sqrt{a_N}}\sqrt{a_1n_1^2+\dots+a_{N-1}n_{N-1}^2+m^2}\right)%
,\labeq{EN}%
\eea}
\HS{0.12}with
{\allowdisplaybreaks
\be
E_1^{m^2}(s;a)=-\frac{1}{2m^{2s}}+\left(\frac{\pi}{a}\right)^{\half}\frac{1}{2m^{2s-1}\Gamma(s)}\left(\Gamma\left(s-\half\right)+4\sum_{\ell\in\N}%
\left[\frac{a^{\frac{1}{4}-\frac{s}{2}}}{(\pi\ell m)^{\half-s}}K_{s-\half}\left(\frac{2\pi\ell m}{\sqrt{a}}\right)\right]\right)%
,\labeq{E1}%
\ee}
\HS{0.12}which reduces to $E_1^{0}(s;a)=a^{-s}\zeta(2s)$ in the limit $m\rightarrow 0$, providing $\Rea{s}<1/2$. We may now use the expressions~\eqn{EN} and~\eqn{E1} in order to rewrite eq.~\eqn{LnZcfinal1} as
{\allowdisplaybreaks
\bea
f^{(c)}&=&\frac{\Gamma\left(-\frac{D-c}{2}\right)\prefacmsb}{2(4\pi)^{\frac{D-1}{2}}\prod_{i=1}^c\!(L_i)\Gamma\left(-\half\right)}%
~E_c^{0}\left(-\frac{D-c}{2};\left(\frac{\pi}{L_1}\right)^2,\dots,\left(\frac{\pi}{L_c}\right)^2\right)\labeq{LnZcfinal2}\\%
&-&\frac{2~T^{D-c}\prefacmsb}{(2\pi)^{\frac{D-1}{2}}\prod_{i=1}^c\!(L_i)}%
\!\!\sum_{s=1}^{\infty}\sum_{\vect{\ell}\in\N^c}\left[\left(\frac{\sum_{i=1}^{c}\left(\frac{\pi \ell_i}{L_i}\right)^2}{(sT)^2}\right)^{\frac{D-c}{4}}%
\!\!\!\!K_{\frac{D-c}{2}}\bigg(\frac{s}{T}\sqrt{\sum_{i=1}^{c}\left(\frac{\pi \ell_i}{L_i}\right)^2}\bigg)\right],\nn%
\eea}
\HS{0.12}which is our canonical result for the free energy density with $c$ geometrically compactified dimensions. Notice that the first line in eq.~\eqn{LnZcfinal2} may now be explicitly computed using eqs.~\eqn{EN} and~\eqn{E1}, for any number of compactified dimensions.  

As a pedagogical example, we will show here how one arrives at the result for the case of one compactified direction.  We may informally drop the Dim. Reg factor $ \prefacmsb $ since we will take $ D=4-2\epsilon $ and then expand around $ \epsilon\rightarrow 0 $ so that this prefactor will fall out, as it was constructed to do.  So, taking $ c=1 $, eq.\eqn{LnZcfinal2} gives
	\begin{align}
		f^{(1)}
			=&\frac{1}{L_1}\frac{\Gamma\left(-\frac{D-1}{2}\right)}{2(4\pi)^{\frac{D-1}{2}}}\frac{1}{\Gamma\left(-\half\right)}
					E_1^0\left(-\frac{D-1}{2};\left(\frac{\pi}{L_1}\right)^2\right)\nonumber\\
				&-\frac{2T^{D-1}}{L_1}\frac{1}{(2\pi)^{\frac{D-1}{2}}}
					\sum_{\ell=1}^{\infty}\sum_{\ell_1=1}^{\infty}\left[
					\left(\left(\frac{\pi\ell}{L_1}\right)^2\left(\frac{1}{\ell T}\right)^2\right)^{\frac{D-1}{4}}
					K_{\frac{D-1}{2}}\left(\frac{\ell}{T}\frac{\pi\ell_1}{L_1}\right)	\right]\nonumber\\
			=&\frac{1}{L_1}\frac{\Gamma\left(-\frac{D-1}{2}\right)}{2(4\pi)^{\frac{D-1}{2}}}\frac{1}{\Gamma\left(-\half\right)}
					\left(\frac{\pi}{L_1}\right)^{D-1}\zeta\left(1-D\right)\nonumber\\
				&-\frac{2\,T^{D-1}}{L_1}\frac{1}{(2\pi)^{\frac{D-1}{2}}}	
					\sum_{\ell=1}^{\infty}\sum_{\ell_1=1}^{\infty}\left[
					\left(\frac{1}{\ell}\frac{\pi\ell_1}{T L_1}\right)^{\frac{D-1}{2}}K_{\frac{D-1}{2}}\left(\ell\frac{\pi\ell_1}{T L_1}\right)
					\right].			
	\end{align}
Now take $ D-1=3-2\epsilon $ and perform a systematic expansion around $ \epsilon\rightarrow 0 $, discarding all $ \mathcal{O}(\epsilon) $ terms:
	\begin{align}
		f^{(1)}	
			=&\frac{1}{L_1}\frac{\Gamma\left(-\frac{3-2\epsilon}{2}\right)}{2(4\pi)^{\frac{2-2\epsilon}{2}}}\frac{1}{\Gamma\left(-\half\right)}
					\left(\frac{\pi}{L_1}\right)^{\frac{3-2\epsilon}{2}}\zeta(-3+2\epsilon)\nonumber\\
				&-\frac{2T^{3-2\epsilon}}{L_1}\frac{1}{(2\pi)^{\frac{3-2\epsilon}{2}}}
					\sum_{\ell=1}^{\infty}\sum_{\ell_1=1}^{\infty}\left[
					\left(\frac{1}{\ell}\frac{\pi\ell_1}{T L_1}\right)^{\frac{3-2\epsilon}{2}}
					K_\frac{3-2\epsilon}{2}\left(\ell\frac{\pi\ell_1}{T L_1}\right)
					\right]\nonumber\\
			=&-\frac{\pi^2}{1440}\frac{1}{L_1^4}
					-\frac{T^3}{L_1}\frac{1}{\sqrt{2}}\frac{1}{\pi^{\frac{3}{2}}}
					\sum_{\ell=1}^{\infty}\sum_{\ell_1=1}^{\infty}\left[
					\left(\frac{\pi\ell}{T L_1}\frac{1}{\ell}\right)^{\frac{3}{2}}
					K_{\frac{3}{2}}\left(\ell\frac{\pi\ell_1}{T L_1}\right)
					\right].
	\end{align}
We may go one step further by noticing that the sum over $ \ell $ has a closed form in terms of the polylogarithm functions so that
	\begin{align}
		f^{(1)}	
			=&-\frac{\pi^2}{1440}\frac{1}{L_1^4}
				-\frac{T^3}{L_1}\frac{1}{2\pi}\sum_{\ell_1=1}^{\infty}\left[
				\frac{\pi\ell_1}{T L_1}\Li_2\,\e^{-\frac{\pi\ell_1}{T L_1}}
				+\Li_3\,\e^{-\frac{\pi\ell_1}{T L_1}}
				\right].\label{f1app}
	\end{align}	

Careful application of eq.~\eqn{EN} and similar manipulations will yield the final results for the plates and tube case as well.  

It is worth showing explicitly how eq.~\eqref{f1app} reduces to the appropriate Stefan-Boltzmann limit.  The ``infinite volume'' limit is really the limit in which the de Broglie thermal wavelength becomes infinitesimally small compared to the relevant length scale of the problem; i.e.\ in the limit that the dimensionless quantity $T L_1\rightarrow\infty$.  To see how the Stefan-Boltzmann limit is recovered, recall that the Riemann integral is defined by
	\begin{align}
		\int_{a}^{b}f(x)dx
			=\lim_{\Delta x\rightarrow 0}\sum_{i=1}^{N}\Delta xf(x_i),
	\end{align}
and that the polylogarithms have the following properties in terms of zeta functions
	\begin{align}
		\int_{0}^{\infty}dx\,\Li_3 \,\e^{-x}
			=&\int_{0}^{\infty}dx\,\sum_{j=1}^{\infty}\frac{1}{j^3}\,\e^{-jx}
				=\sum_{j=1}^{\infty}\frac{1}{j^4}=\frac{\pi^4}{90},\nonumber\\
		\int_{0}^{\infty}dx\,x\,\Li_2 \,\e^{-x}
			=&\int_{0}^{\infty}dx\,x\,\sum_{j=1}^{\infty}\frac{1}{j^2}\,\e^{-jx}
				=\sum_{j=1}^{\infty}\frac{1}{j^4}=\frac{\pi^4}{90}.	
	\end{align}

We therefore have that
	\begin{align}
		\lim_{TL_1\rightarrow \infty}
			&\left[\sum_{j=1}^{\infty}\frac{\pi}{TL_1}\Li_3\left(\e^{-\frac{\pi\ell}{TL_1}}\right)
			\right]
			=\int_{0}^{\infty}dx\,\Li_3\,\e^{-\ell x}=\frac{\pi^4}{90}\\
		\lim_{TL_1\rightarrow \infty}
		&\left[\sum_{j=1}^{\infty}\frac{\pi}{TL_1}\frac{\pi}{TL_1}\Li_2\left(\e^{-\frac{\pi\ell}{TL_1}}\right)
		\right]
			=\int_{0}^{\infty}dx\,x\,\Li_2\,\e^{-\ell x}=\frac{\pi^4}{90}.
	\end{align}

It is then clear that, in the limit $ TL_1\rightarrow\infty $, the first term in eq.~\eqref{f1app} vanishes and the second term gives the expected value of $ -\frac{\pi^2}{90}T^4 $.  One may perform similar checks of the tube and box cases if one keeps in mind that only the terms that carry a $ T $-dependence will contribute in the limit that $ TL_i\rightarrow\infty $ and that integrals over the Bessel functions also have closed forms in terms of zeta functions.
\subsection{Alternative computation}\labsec{AppAlternativeComputation}

In this section, we compute the alternative representations for the free energy density in different scenarios. We generally follow the methods of~\cite{Kapusta:2006pm,Laine:2016hma} but take great care with the regularization procedure; given the spatial compactification(s), we avoid the formal manipulation of divergences. Moreover, we notice that certain massive results will be given as byproducts, even though these are of no interest for the present work.

In fact, straightforwardly taking the asymptotically large $R$ limits in eq.~\eqn{LnZc} could appear problematic for some values of $c$. Indeed, the subsequent integral and the remaining infinite sums may not have a common strip of convergence. To show that this would not be a problem, we could make use of the monotone convergence (or Beppo–Levi) theorem. Given the asymptotically large $R$ limits and since the infinite summation kernel of eq.~\eqn{LnZc} would then still be positive definite and monotonically increasing\footnote{The $\beta^2$ normalization of the logarithm argument could be changed without modifying the final result, so to keep this statement true for all values of $T$ and $L_i$.}, one could take the infinite series as a set of partial summations under a limit and exchange the order between the limit and the integral. This could allow for an analytic continuation without spoiling the convergence of the expression, thereby asserting the existence of the dimensionally regularized free energy density for all $c$. However, we could, instead, proceed with a convenient trick, making use of the following identity
{\allowdisplaybreaks
\be
\ln\left(A\right)\equiv\left.-\pard{}{\alpha}\left(\frac{1}{A^{\alpha}}\right)\right|_{\alpha\rightarrow 0},\labeq{lnA}%
\ee}
\HS{0.12}where $A$ is positive definite and $\alpha$ is real. $1/A^\alpha$ can serve the purpose of defining a master expression that can be analytically continued as a function of the dimension $D$, before applying \eq{lnA} to find the log. Doing so and momentarily assuming a convenient range for $\alpha$, allows us to work with convergent expressions. Using this identity, and sending all $R$ to their asymptotically large values, we define the following master sum-integral\footnote{We recall that we use Dim.~Reg.~within the $\MSbar$ scheme, and refer to the beginning of subsection~\secn{AppCanonicalComputation} for more details on this scheme and the induced modifications on the dimensionality of the expressions.}
{\allowdisplaybreaks
\be
I^{(c)}_{\alpha}\equiv -\frac{T^{1+2\alpha}}{2\prod_{i=1}^{c}\big(L_i\big)}\prefacmsb\!\!%
\sum_{n\in\Z}\sum_{\vect{\ell}\in\N^{c}}\minte{k}{D-1-c}\left[\frac{1}{\left(\omega_n^{2}+\sum_{i=1}^{c}%
\omega_{\ell_i}^2+\vect{k}^2+m^2\right)^\alpha}\right],\labeq{MasterIAlpha}%
\ee}
\HS{0.12}related to the free energy density in $D$ dimensions, after analytic continuation in terms of the complex $D$ variable, via the following identity
{\allowdisplaybreaks
\be
f^{(c)}=\pard{}{\alpha}\bigg[I^{(c)}_{\alpha}\bigg]\bigg|_{\alpha\rightarrow 0}%
.\labeq{FreeEnergyLogIdentity}%
\ee}

We will then first focus on the evaluation of eq.~\eqn{MasterIAlpha}. Prior to setting the number of compactified spatial dimensions $c$ to some value in order to do so, and thus to compute the free energy density, let us perform the continuous momentum integration. By doing so, we will be able to analytically continue the expression as a function of the dimensional parameter $D$, and evaluate the infinite summations outside of their original radii of convergence when needed. For now, we shall assume that there exists some strip (to be specified below) for the complex variable $D$, in which eq.~\eqn{MasterIAlpha} is convergent. We then proceed to rescale the continuous momentum $\vect{k}$, in order to factorize all the frequencies and the mass. We then obtain the following expression
{\allowdisplaybreaks
\bea
I^{(c)}_{\alpha}&=&-\frac{T^{1+2\alpha}~\prefacmsb}{2~(4\pi)^{2-\frac{D}{2}}\prod_{i=1}^{c}\big(L_i\big)}~%
\minte{k}{D-1-c}\Bigg[\left(1+\vect{k}^2\right)^{-\alpha}\Bigg]\times\nn\\%
&&\hspa{3.5}\times\sum_{n\in\Z}~\sum_{\vect{\ell}\in\N^{c}}\left(\omega_n^{2}+\sum_{i=1}^{c}\omega_{\ell_i}^2+m^2\right)^{\frac{D-1-c}{2}-\alpha}%
.\labeq{MasterIAlpha2}%
\eea}
\HS{0.12}In order to assure the existence of the strip we take $\alpha>0$.  Then the integral above is clearly convergent in the $D$-strip such that $\Rea{D}-1-c\in\left(0,2\alpha\right)$. The set of infinite summations, being some Epstein zeta type of function, naturally converges in a $D$-strip such that $2\alpha+1+c-\Rea{D}>c$. Consequently, all we need to assume for now is that $\alpha$ and $D$ satisfy the following conditions: $2\alpha>\Rea{D}-1\geq \Rea{D}-1-c>0$, the middle one being automatically satisfied. We shall then work under this assumption, until we are able to perform an analytic continuation in the dimensional parameter $D$. Then, the constraint on $\alpha$ will be released, and we will be able to compute the free energy density according to eq.~\eqn{FreeEnergyLogIdentity}, in all the possible spatial compactification cases.

\HS{0.12}Next, we use standard techniques~\cite{Kapusta:2006pm,Laine:2016hma} to perform the continuous momentum integration in eq.~\eqn{MasterIAlpha2}, and we get
{\allowdisplaybreaks
\bea
I^{(c)}_{\alpha}&=&-\frac{T^{1+2\alpha}~\prefacmsb\Gamma\left(\alpha+\frac{1+c}{2}-\frac{D}{2}\right)}%
{2^{4-c}~\pi^{\frac{3-c}{2}}\prod_{i=1}^{c}\big(L_i\big)\Gamma\left(\alpha\right)}~\times\nn\\%
&&\hspa{3.5}\times\sum_{n\in\Z}\sum_{\vect{\ell}\in\N^{c}}\left(\omega_n^{2}+\sum_{i=1}^{c}\omega_{\ell_i}^2+m^2\right)%
^{\frac{D-1-c}{2}-\alpha}%
,\labeq{MasterIAlpha3}%
\eea}
\HS{0.12}being left with the evaluation of the following $(1+c)$-dimensional Epstein zeta function
{\allowdisplaybreaks
\be
Z_\rmi{1+c}\left(\alpha+\frac{1+c-D}{2}\right)\equiv\sum_{n\in\Z}\sum_{\vect{\ell}\in\N^{c}}%
\left(\omega_n^{2}+\sum_{i=1}^{c}\omega_{\ell_i}^2+m^2\right)^{\frac{D-1-c}{2}-\alpha}%
,\labeq{EpsteinZeta1}%
\ee}
\HS{0.12}for which we need to set $c$ to some particular value prior to any evaluation.

\paragraph{Two infinite parallel plates\\*}

In this subsection, we shall analytically continue eq.~\eqn{EpsteinZeta1} in order to analytically continue eq.~\eqn{MasterIAlpha3}.  Taking $c=1$
{\allowdisplaybreaks
\be
Z_\rmi{2}\Big(1+\alpha-D/2\Big)=\sum_{n\in\Z}\sum_{\ell_1\in\N}\left(\omega_n^{2}+\omega_{\ell_1}^2+m^2\right)^{\frac{D}{2}-\alpha-1}%
.\labeq{EpsteinZeta1D1}%
\ee}%

We then split the Matsubara sum into a zero and a nonzero mode contribution, following
{\allowdisplaybreaks
\be
G_{1+1}\equiv\sum_{n\in\Z}~\sum_{\ell_1\in\N}g\left(n^2,\ell_1^2\right)=\sum_{\ell_1=1}^{+\infty}%
g\left(0,\ell_1^2\right)+2\sum_{n,\ell_1=1}^{+\infty}g\left(n^2,\ell_1^2\right)%
,\labeq{SplitMatsuContrib1}%
\ee}
\HS{0.12}and make use of the two-dimensional Epstein zeta function representation~\cite{Elizalde:1989mr}
{\allowdisplaybreaks
\bea
\sum_{n,\ell=1}^{+\infty}\bigg[\left(a \ell^{2}+b n^{2}\right)^{-s}\bigg]&=&-\frac{b^{-s}}{2}~\zeta\left(2s\right)%
+\frac{\sqrt{\pi}}{2}~\frac{b^{1/2-s}}{\sqrt{a}}~\frac{\Gamma\left(s-1/2\right)}{\Gamma\left(s\right)}~\zeta\left(2s-1\right)\nn\\%
&+&\frac{2\pi^{s}}{\Gamma\left(s\right)}~\frac{(b/a)^{1/4}}{(a b)^{s/2}}\sum_{n,\ell=1}^{+\infty}%
\left[\left(\frac{\ell}{n}\right)^{s-1/2}K_{s-1/2}\bigg(2\pi\sqrt{\frac{b}{a}}~n \ell\bigg)\right]%
,\labeq{EpsteinZetaTwoDimMassZeroAnaConti}%
\eea}
\HS{0.12}for the second term, the first one being the usual Riemann zeta function. Analytic continuation of the above formula readily follows from the straightforward continuation of the Euler gamma function, the Riemann zeta and modified Bessel functions, or simply using the functional equation for the corresponding Epstein zeta function~\cite{Elizalde:1989mr}.

As it turns out, given a certain choice for $a$ and $b$, the limit of vanishing $T$ or asymptotically high $L_1$ appears explicitly. Note also the change in the argument of the Bessel function, upon an exchange between $a$ and $b$. This gives the possibility to slightly enhance the convergence of the remaining two-fold sum, as depending on the choice for $a$ and $b$, a factor of $\sqrt{b/a}$ or $\sqrt{a/b}$ (giving either a $2TL_1$ or its inverse) appears in the argument of the Bessel function. And given that the larger the argument the better the convergence, we choose the former possibility in order to easily reach the thermodynamic, Stefan-Boltzmann limit. 
Therefore, the asymptotically high $L_1$ limit (that is the usual Stefan-Boltzmann finite temperature result) will appear explicitly in the analytically continued result. Using then eq.~\eqn{EpsteinZetaTwoDimMassZeroAnaConti} with $a\equiv (\pi/L_1)^{2}$ and $b\equiv (2\pi T)^{2}$, we further analytically continue eq.~\eqn{EpsteinZeta1D1} and can finally extend it together with eq.~\eqn{MasterIAlpha3} for $c=1$, as functions of $D$ from the original $D$-strip to the whole complex plane. Gathering all together, we get the following result
{\allowdisplaybreaks
\bea
I^{(1)}_{\alpha}&=&-\frac{T^{1+2\alpha}\left(\lmsb^{2} e^{\gamma_\rmi{\tiny E}}\right)^{\,2-\frac{D}{2}}}{8\pi L_1~\Gamma\left(\alpha\right)}%
\times\Bigg\{\left(\frac{\pi}{L_1}\right)^{D-2\alpha-2}\!\!\!\!\!\!\zeta\Big(2+2\alpha-D\Big)~\Gamma\left(1+\alpha-\frac{D}{2}\right)\nn\\%
&+&\frac{2\sqrt{\pi}~T L_1}{\left(2\pi T\right)^{2+2\alpha-D}}~~\zeta\Big(1+2\alpha-D\Big)~~\Gamma\left(\frac{1}{2}+\alpha-\frac{D}{2}\right)\nn\\%
&-&\left(2\pi T\right)^{D-2\alpha-2}~\zeta\Big(2+2\alpha-D\Big)~\Gamma\left(1+\alpha-\frac{D}{2}\right)\nn\\%
&+&\frac{4\sqrt{2}~L_1\left(T/L_1\right)^{\frac{D-1}{2}-\alpha}}{\left(2\pi\right)^{1+\alpha-\frac{D}{2}}}\!\!\sum_{n,\ell_1=1}^{+\infty}%
\left[\left(\frac{n}{\ell_1}\right)^{\frac{D-1}{2}-\alpha}\!\!\!\!K_{\frac{D-1}{2}-\alpha}\left(4\pi T L_1 n\ell_1\right)\right]~\!\Bigg\}%
,\labeq{MasslessI1Result}%
\eea}
\HS{0.12}for the corresponding eq.~\eqn{MasterIAlpha3} with $c=1$ and $m=0$, valid on the whole $D$-complex plane. This means that we can actually probe the above result for any value of $\alpha$. We then apply the relation~\eqn{FreeEnergyLogIdentity}, and since there is no pole around $D=4$ (i.e., no logarithmic UV divergence), we readily set the dimension to four and get the renormalized\footnote{No counter term was needed, yet Dim.\ Reg.\ set the non logarithmic type of divergence to zero which makes the result formally renormalized.} result
{\allowdisplaybreaks
\be
f^{(1)}=-\frac{\pi^2 T^4}{90}+\frac{T^3\zeta(3)}{4\pi L_1}-\frac{T\zeta(3)}{16\pi L_1^3}-\frac{\sqrt{2}~T^{5/2}}{L_1^{3/2}}%
\sum_{n,\ell_1=1}^{+\infty}\left[\left(\frac{n}{\ell_1}\right)^{\frac{3}{2}}\!\!\!K_{\frac{3}{2}}\left(4\pi TL_1 n\ell_1\right)\right],%
\ee}
\HS{0.12}for the corresponding free energy density. We notice indeed that the first term is the usual Stefan-Boltzmann finite temperature result, relevant to a massless scalar field in a noncompactified space. In addition, we see that the above result reduces to the Stefan-Boltzmann one when sending $L_1$ to infinity. Note however, that the zero temperature result relevant to the usual Casimir effect, and equal to $-\pi^2/1440L_1^{4}$~\cite{Elizalde:1989mr}, appears only explicitly when using the same analytical representation for the two-dimensional Epstein zeta function, but done with $a=(2\pi T)^{2}$ and $b=(\pi/L_1)^{2}$ instead of the present choice. Our result contains a two-fold infinite summation which is of course convergent. And in fact, one can also get a closed form when performing one (or the other) summation, with once again two different looking yet equivalent representations, depending on which summation is explicitly done. We choose to explicitly perform the summation over the $\ell_1$ variable, for convenience when further enhancing the convergence as we are going to explain below.\footnote{Notice that both the Matsubara and the spatial summations have been done. Even though we keep the same dummy variables $n$ and $\ell_1$, they do not correspond anymore to the original modes which have been summed over.} Doing so, we end up with the following one-fold representation
{\allowdisplaybreaks
\be
f^{(1)}=-\frac{\pi^2 T^4}{90}-\frac{T^2}{2 L_1^2}\sum_{n=1}^{+\infty}\bigg[n~\text{Li}_{2}\left(e^{-4\pi TL_1 n}\right)\bigg]%
+\frac{T^3\zeta(3)}{4\pi L_1}-\frac{T\zeta(3)}{16\pi L_1^3}-\frac{T}{8\pi L_1^3}\sum_{n=1}^{+\infty}\text{Li}_{3}\left(e^{-4\pi TL_1 n}\right),%
\ee}
\HS{0.12}involving, again, two polylogarithm functions. With the above representation, one would need to change the remaining infinite summations by the corresponding integrals, in order to probe the asymptotic $T=0$ limit where the $n$ variable becomes then continuous. Note that this representation, when truncated to some finite order, is relatively well behaved in terms of convergence. However, the situation can easily be further improved. Indeed, making use of a contour integral representation for the polylogarithm functions, one can replace the summations above with another set of summations with much better convergence properties. This new one-fold representation happens to be exponentially enhanced, as can be easily checked. Relabeling the dummy variables, our final result for the free energy density of one neutral massless noninteracting scalar field is
{\allowdisplaybreaks
\bea
f^{(1)}&=&-\frac{\pi^2T^4}{90}+\frac{\zeta(3) T^3}{4\pi L_1}-\frac{T^2}{8 L_1^2}\sum_{\ell=1}^{+\infty}\bigg[\frac{\text{csch}^{2}%
\left(2\pi TL_1 \ell\right)}{\ell^{2}}\bigg]\nn\\%
&-&\frac{\zeta(3) T}{16\pi L_1^3}-\frac{T}{16\pi L_1^3}\sum_{\ell=1}^{+\infty}%
\bigg[\frac{\coth\left(2\pi TL_1 \ell\right)-1}{\ell^{3}}\bigg]%
.\labeq{FinalResultfMasslessL1}%
\eea}
\HS{0.12}As a matter of fact, this new representation allows for both the $T=0$ and $L_1=\infty$ limits to be carried out, which is relevant to the corresponding quantum field when coupled to a heat bath at temperature $T$, and geometrically confined in between two infinite parallel plates separated by a distance $L_1$. We notice that the term proportional to $T L_1^{-3}$ actually cancels the $s^{-3}$ term in the second sum, but it is preferable to keep the result as such. The above second sum indeed converges more rapidly. 

\paragraph{Infinite rectangular tube\\*}

We shall now analytically continue eq.~\eqn{EpsteinZeta1} with $c=2$, that is
{\allowdisplaybreaks
\be
Z_\rmi{3}\left(\frac{3}{2}+\alpha-\frac{D}{2}\right)=\sum_{n\in\Z}\sum_{\ell_1\in\N}\sum_{\ell_2\in\N}%
\left(\omega_n^{2}+\omega_{\ell_1}^2+\omega_{\ell_2}^2+m^2\right)^{\frac{D}{2}-\alpha-\frac{3}{2}}%
,\labeq{EpsteinZeta2D1}%
\ee}
\HS{0.12}which will allow us to investigate the free energy density of a neutral noninteracting scalar field coupled to a heat bath at temperature $T$, and geometrically confined inside an infinite tube of rectangular section with two sides of respective finite lengths $L_1$ and $L_2$. For computational convenience, let us keep the mass nonzero for a little longer.

We then notice that we can use the following identity
{\allowdisplaybreaks
\bea
G_{1+2}&\equiv&\sum_{n\in\Z}\sum_{\ell_1\in\N}\sum_{\ell_2\in\N}g\left(n^2,\ell_1^2,\ell_2^2;m\right)\nn\\%
&=&\frac{1}{4}\sum_{n\in\Z}\sum_{\ell_1\in\Z}\sum_{\ell_2\in\Z}g\left(n^2,\ell_1^2,\ell_2^2;m\right)%
-\frac{1}{4}\sum_{n\in\Z}\sum_{\ell_1\in\Z}g\left(n^2,\ell_1^2,0;m\right)\nn\\%
&-&\frac{1}{4}\sum_{n\in\Z}\sum_{\ell_2\in\Z}g\left(n^2,0,\ell_2^2;m\right)+\frac{1}{4}\sum_{n\in\Z}g\left(n^2,0,0;m\right)%
,\labeq{SplitMatsuContrib2}%
\eea}
\HS{0.12}in order to rewrite the multifold summation, in eq.~\eqn{EpsteinZeta2D1}, in a more convenient way. For the sake of analytically continuing the corresponding Epstein zeta function, let us make use of the following representation~\cite{Kirsten:2010zp}
{\allowdisplaybreaks
\bea
E_{1+c}^{m^2}&=&\sum_{(n,\vect{\ell})\in\Z^{1+c}}\Big(a_{0}^2n^2+a_{1}^2\ell_1^2+...+a_{c}^2\ell_c^2+m^2\Big)^{-s}\nn\\
&=&\frac{\pi^{\frac{1+c}{2}}\Gamma\left(s-\frac{1+c}{2}\right)(m^2)^{\frac{1+c}{2}-s}}{a_{0}a_{1}...a_{c}\Gamma\left(s\right)}\nn\\%
&+&\frac{2\pi^s(m^2)^{\frac{1+c}{4}-\frac{s}{2}}}{a_{0}a_{1}...a_{c}\Gamma\left(s\right)}\sum_{(n,\vect{\ell})\in\Z^{1+c}\setminus\{\vect{0}\}}%
\left[\frac{K_{\frac{1+c}{2}-s}\bigg(2\pi m\sqrt{\frac{n^2}{a_{0}^2}+\frac{\ell_1^2}{a_{1}^2}+...+\frac{\ell_c^2}{a_{c}^2}}\bigg)}%
{\left(\frac{n^2}{a_{0}^2}+\frac{\ell_1^2}{a_{1}^2}+...+\frac{\ell_c^2}{a_{c}^2}\right)^{\frac{1+c}{4}-\frac{s}{2}}}\right]%
,\labeq{EpsteinZetaNDimMassiveAnaConti}%
\eea}
\HS{0.12}noticing the difference of definition for the $a_i$ coefficients, with respect to eq.~\eqn{EN}. Thus, with the help of this above representation, together with the identity~\eqn{SplitMatsuContrib2}, we can obtain an analytic and symmetric representation for the corresponding master sum-integral $I^{(2)}_{\alpha}$, and further apply the relation~\eqn{FreeEnergyLogIdentity} for obtaining the free energy density. Finally, given that $D=4-2\epsilon$, we Laurent expand the resulting expression around $\epsilon=0$. After a few more steps, which we will avoid here for the sake of brevity, we finally obtain
{\allowdisplaybreaks
\bea
f^{(2)}&=&-\frac{m^4}{64\pi^{2}~\epsilon}+\frac{m^2}{32\pi L_1L_2~\epsilon}+\frac{m^4}{128\pi^2}%
\bigg[4~\ln\left(\frac{m}{\lmsb}\right)-3\bigg]-\frac{m^2}{32\pi L_1L_2}\bigg[2~\ln\left(\frac{m}{\lmsb}\right)-1\bigg]\nn\\%
&+&\frac{m^3}{24\pi}\left(\frac{L_1+L_2}{L_1L_2}\right)-\frac{mT}{4\pi L_1 L_2}%
\!\sum_{n=1}^{+\infty}\left[\frac{1}{n}~K_{1}\Big(\frac{m}{T} n\Big)\right]\nn\\%
&+&\frac{T^{3}}{8\pi L_2}\!\sum_{(n,\ell_1)\in\Z^{2}\setminus\{\vect{0}\}}\left[\frac{e^{-\frac{m}{T}\sqrt{n^2+(2TL_1)^2\ell_1^2}}%
\left(1+\frac{m}{T}\sqrt{n^2+(2TL_1)^2\ell_1^2}\right)}{\Big(n^2+(2TL_1)^2\ell_1^2\Big)^{3/2}}\right]\nn\\%
&+&\frac{T^{3}}{8\pi L_1}\!\sum_{(n,\ell_2)\in\Z^{2}\setminus\{\vect{0}\}}\left[\frac{e^{-\frac{m}{T}\sqrt{n^2+(2TL_2)^2\ell_2^2}}%
\left(1+\frac{m}{T}\sqrt{n^2+(2TL_2)^2\ell_2^2}\right)}{\Big(n^2+(2TL_2)^2\ell_2^2\Big)^{3/2}}\right]\nn\\%
&-&\frac{m^2T^2}{4\pi^2}\!\sum_{(n,\ell_1,\ell_2)\in\Z^{3}\setminus\{\vect{0}\}}\left[\frac{K_{2}\left(\frac{m}{T}%
\sqrt{n^2+(2TL_1)^2\ell_1^2+(2TL_2)^2\ell_2^2}\right)}{\Big(n^2+(2TL_1)^2\ell_1^2+(2TL_2)^2\ell_2^2\Big)}\right]+{\cal O}\left(\epsilon^1\right)%
.\labeq{FreeEnergyDensityMassiveL1L2}%
\eea}

Let us now simply comment on the massless limit, given the unusual UV structure of the above expression whose renormalization techniques, if necessary, can be found at~\cite{Bordag:2009zzd,Bordag:2001qi} by means of background field methods. However, we notice that under the limit $m\rightarrow 0$, the above result becomes convergent. Applying this limit, and performing the summations whenever it is possible to get a closed form, we obtain the following massless finite result
{\allowdisplaybreaks
\bea
\lim_{m\rightarrow 0}\left(f^{(2)}\right)&=&-\frac{\pi T^2}{24L_1L_2}-\frac{T^4}{2\pi^2}\!\sum_{(n,\ell_1,\ell_2)%
\in\Z^{3}\setminus\{\vect{0}\}}\Big(n^2+(2TL_1)^2\ell_1^2+(2TL_2)^2\ell_2^2\Big)^{-2}\nn\\%
&+&\frac{T^{3}}{8\pi L_2}\!\sum_{(n,\ell_1)\in\Z^{2}\setminus\{\vect{0}\}}\Big(n^2+(2TL_1)^2\ell_1^2\Big)^{-3/2}\nn\\%
&+&\frac{T^{3}}{8\pi L_1}\!\sum_{(n,\ell_2)\in\Z^{2}\setminus\{\vect{0}\}}\Big(n^2+(2TL_2)^2\ell_2^2\Big)^{-3/2}%
.\labeq{FreeEnergyDensityMasslessL1L2Pre1}%
\eea}
\HS{0.12}In addition, the first set of summations can be conveniently rewritten in a symmetric fashion with
{\allowdisplaybreaks
\bea
S_3&\equiv&-\frac{T^4}{2\pi^2}\!\sum_{(n,\ell_1,\ell_2)\in\Z^{3}\setminus\{\vect{0}\}}\Big(n^2+(2TL_1)^2\ell_1^2+(2TL_2)^2\ell_2^2\Big)^{-2}\nn\\%
&=&-\frac{T^4}{2\pi^2}\times\frac{2}{2}\times\!\sum_{\ell_1=1}^{+\infty}~\sum_{(n,\ell_2)\in\Z^{2}}\Big(n^2+(2TL_1)^2\ell_1^2+(2TL_2)^2\ell_2^2\Big)^{-2}\nn\\%
&-&\frac{T^4}{2\pi^2}\times\frac{2}{2}\times\!\sum_{\ell_2=1}^{+\infty}~\sum_{(n,\ell_1)\in\Z^{2}}\Big(n^2+(2TL_1)^2\ell_1^2+(2TL_2)^2\ell_2^2\Big)^{-2}\nn\\%
&-&\frac{T^4}{2\pi^2}\times\frac{1}{2}\times\!\sum_{(n,\ell_2)\in\Z^{2}\setminus\{\vect{0}\}}\Big(n^2+(2TL_2)^2\ell_2^2\Big)^{-2}\nn\\%
&-&\frac{T^4}{2\pi^2}\times\frac{1}{2}\times\!\sum_{(n,\ell_1)\in\Z^{2}\setminus\{\vect{0}\}}\Big(n^2+(2TL_1)^2\ell_1^2\Big)^{-2}%
.\labeq{S3Summation1}%
\eea}
\HS{0.12}While the above expressions are finite, as we just mentioned, their convergence is rather slow, but can be enhanced using some Poisson resummation formulas such as in eq.~\eqn{PoissonResummation}. Thus, let us first use eq.~\eqn{EpsteinZetaTwoDimMassZeroAnaConti} in order to improve the convergence properties of the two-fold summations, once properly re-expressed using eq.~\eqn{SplitMatsuContrib2}. Then, we can proceed in implementing a specific resummation to the remaining three-fold summations, following
{\allowdisplaybreaks
\bea
S_{3;(i,j)}&\equiv&-\frac{T^4}{2\pi^2}\!\sum_{\ell_i=1}^{+\infty}\sum_{(n,\ell_j)\in\Z^{2}}\Big(n^2+(2TL_i)^2\ell_i^2+(2TL_j)^2\ell_j^2\Big)^{-2}\nn\\%
&=&-\frac{T^4}{2\pi^2}\!\inte{t}{0}{+\infty}\left[t\sum_{\ell_i=1}^{+\infty}\left(e^{-t (2TL_i)^2\ell_i^2}\right)\sum_{n=-\infty}^{+\infty}%
\left(e^{-t n^2}\right)\sum_{\ell_j=-\infty}^{+\infty}\left(e^{-t (2TL_j)^2\ell_j^2}\right)\right]\nn\\%
&=&-\frac{T^3}{4\pi L_j}\!\sum_{\ell_i=1}^{+\infty}\sum_{(n,\ell_j)\in\Z^{2}\setminus\{\vect{0}\}}\inte{t}{0}{+\infty}%
e^{-t (2TL_i)^2\ell_i^2-\frac{\pi^2}{t}\left(n^2+\frac{\ell_j^2}{(2TL_j)^2}\right)}\nn\\%
&-&\frac{T^3}{4\pi L_j}\!\sum_{\ell_i=1}^{+\infty}\inte{t}{0}{+\infty}e^{-t (2TL_i)^2\ell_i^2}\nn\\%
&=&-\frac{T}{8L_i L_j^2}\!\sum_{\ell_i=1}^{+\infty}\sum_{(n,\ell_j)\in\Z^{2}\setminus\{\vect{0}\}}%
\Bigg[\frac{\sqrt{\ell_j^2+(2TL_j)^2n^2}}{\ell_i} K_{1}\left(\frac{2\pi L_i}{L_j}~\ell_i~\sqrt{\ell_j^2+(2TL_j)^2n^2}\right)\Bigg]\nn\\%
&-&\frac{\pi T}{96L_j L_i^2}~,%
\eea}
\HS{0.12}where from the second to the third line, we used eq.~\eqn{PoissonResummation} for the $n$- and $\ell_j$-summations, and singled out the $\ell_j=n=0$ term. From the third to the last line, we used the integral representation of the modified Bessel function of the second kind.

Bringing all the results together, performing some of the summations to get closed forms when it is possible, and relabeling some of the variables, we finally arrive to the following
{\allowdisplaybreaks
\bea
f^{(2)}&=&-\frac{\pi^2 T^4}{90}+\frac{\zeta\left(3\right)T^3(L_1+L_2)}{4\pi L_1L_2}%
-\frac{\pi T^2}{24L_1 L_2}+\frac{\pi T(L_1+L_2)}{96L_1^2 L_2^2}-\frac{\zeta\left(3\right)T(L_1^3+L_2^3)}{32\pi L_1^3 L_2^3}\nn\\%
&-&\frac{T^2}{4L_1^2}\!\sum_{\ell=1}^{+\infty}\Bigg[\ell~\text{Li}_{2}\left(e^{-4\pi TL_1 \ell}\right)\Bigg]%
-\frac{T^2}{4L_2^2}\!\sum_{\ell=1}^{+\infty}\Bigg[\ell~\text{Li}_{2}\left(e^{-4\pi TL_2 \ell}\right)\Bigg]\nn\\%
&-&\frac{T}{16\pi L_1^3}\!\sum_{\ell=1}^{+\infty}\text{Li}_{3}\left(e^{-4\pi TL_1 \ell}\right)%
-\frac{T}{16\pi L_2^3}\!\sum_{\ell=1}^{+\infty}\text{Li}_{3}\left(e^{-4\pi TL_2 \ell}\right)\nn\\%
&+&\frac{T^2}{L_1L_2}\!\sum_{n,\ell=1}^{+\infty}\Bigg[\frac{n}{\ell}~K_{1}\left(4\pi TL_1 n\ell\right)%
+\frac{n}{\ell}~K_{1}\left(4\pi TL_2 n\ell\right)\Bigg]\\%
&-&\frac{T}{8L_1 L_2^2}\!\sum_{\ell_1=1}^{+\infty}~\sum_{(n,\ell_2)\in\Z^{2}\setminus\{\vect{0}\}}%
\left[\frac{\sqrt{\ell_2^2+(2TL_2)^2n^2}}{\ell_1}~K_{1}\left(\frac{2\pi L_1}{L_2}~\ell_1~\sqrt{\ell_2^2+(2TL_2)^2n^2}\right)\right]\nn\\%
&-&\frac{T}{8L_2 L_1^2}\!\sum_{\ell_2=1}^{+\infty}~\sum_{(n,\ell_1)\in\Z^{2}\setminus\{\vect{0}\}}%
\left[\frac{\sqrt{\ell_1^2+(2TL_1)^2n^2}}{\ell_2}~K_{1}\left(\frac{2\pi L_2}{L_1}~\ell_2~\sqrt{\ell_1^2+(2TL_1)^2n^2}\right)\right]\nn%
,\labeq{FinalResultfMasslessL1L2}%
\eea}
\HS{0.12}which is the symmetrized and renormalized result for the free energy density of a noninteracting massless neutral scalar field coupled to a heat bath at temperature $T$, and geometrically confined inside an infinite tube of rectangular section with two sides of respective lengths $L_1$ and $L_2$.

\paragraph{Finite volume box\\*}

Let us finally take care of the last case, analytically continuing eq.~\eqn{EpsteinZeta1} with $c=3$, that is
{\allowdisplaybreaks
\be
Z_\rmi{4}\left(2+\alpha-\frac{D}{2}\right)=\sum_{n\in\Z}~\sum_{\ell_1\in\N}~\sum_{\ell_2\in\N}~\sum_{\ell_3\in\N}%
\left(\omega_n^{2}+\omega_{\ell_1}^2+\omega_{\ell_2}^2+\omega_{\ell_3}^2+m^2\right)^{\frac{D}{2}-\alpha-2}%
,\labeq{EpsteinZeta3D1}
\ee}
\HS{0.12}in order to investigate the free energy density of a neutral noninteracting scalar field coupled to a heat bath at temperature $T$, and geometrically confined inside a finite volume box with sides of respective finite lengths $L_1$, $L_2$, and $L_3$.

As previously, notice again that we can use a convenient identity, which in this case reads
{\allowdisplaybreaks
\bea
G_{1+3}&\equiv&\sum_{n\in\Z}\sum_{\ell_1\in\N}\sum_{\ell_2\in\N}\sum_{\ell_3\in\N}g\left(n^2,\ell_1^2,\ell_2^2,\ell_3^2;m\right)=%
\frac{1}{8}\sum_{n\in\Z}\sum_{\ell_1\in\Z}\sum_{\ell_2\in\Z}\sum_{\ell_3\in\Z}g\left(n^2,\ell_1^2,\ell_2^2,\ell_3^2;m\right)\nn\\
&-&\frac{1}{8}\sum_{n\in\Z}\sum_{\ell_1\in\Z}\sum_{\ell_2\in\Z}g\left(n^2,\ell_1^2,\ell_2^2,0;m\right)%
-\frac{1}{8}\sum_{n\in\Z}\sum_{\ell_1\in\Z}\sum_{\ell_3\in\Z}g\left(n^2,\ell_1^2,0,\ell_3^2;m\right)\nn\\
&-&\frac{1}{8}\sum_{n\in\Z}\sum_{\ell_2\in\Z}\sum_{\ell_3\in\Z}g\left(n^2,0,\ell_2^2,\ell_3^2;m\right)%
+\frac{1}{8}\sum_{n\in\Z}\sum_{\ell_1\in\Z}g\left(n^2,\ell_1^2,0,0;m\right)\nn\\
&+&\frac{1}{8}\sum_{n\in\Z}\sum_{\ell_2\in\Z}g\left(n^2,0,\ell_2^2,0;m\right)%
+\frac{1}{8}\sum_{n\in\Z}\sum_{\ell_3\in\Z}g\left(n^2,0,0,\ell_3^2;m\right)\nn\\%
&-&\frac{1}{8}\sum_{n\in\Z}g\left(n^2,0,0,0;m\right)%
,\labeq{SplitMatsuContrib3}%
\eea}
\HS{0.12}in order to rewrite the multifold summation in eq.~\eqn{EpsteinZeta3D1}, and make use of an appropriate representation for computing the free energy density with the help of eq.~\eqn{EpsteinZetaNDimMassiveAnaConti}. Doing so allows us to obtain an analytic and symmetric representation for corresponding master sum-integral $I^{(3)}_{\alpha}$, which we do not display here for the sake of brevity. Further applying the relation~\eqn{FreeEnergyLogIdentity}, in order to get the free energy density, we Laurent expand the resulting expression around $\epsilon=0$. Again, skipping some minor technical details, we finally obtain
{\allowdisplaybreaks
\bea
f^{(3)}&=&-\frac{m^4}{64\pi^{2}~\epsilon}+\frac{m^2}{32\pi~\epsilon}\left(\frac{L_1+L_2+L_3}{L_1L_2L_3}\right)%
+\frac{m^4}{128\pi^2}\bigg[4~\ln\left(\frac{m}{\lmsb}\right)-3\bigg]\nn\\%
&-&\frac{m}{16L_1L_2L_3}-\frac{m^2}{32\pi}\left(\frac{L_1+L_2+L_3}{L_1L_2L_3}\right)\bigg[%
2~\ln\left(\frac{m}{\lmsb}\right)-1\bigg]+\frac{m^3}{24\pi}\left(\frac{L_1L_2+L_1L_3+L_2L_3}{L_1L_2L_3}\right)\nn\\%
&-&\frac{T}{8L_1L_2L_3}\log\Big(1-e^{-\frac{m}{T}}\Big)%
-\frac{mT}{8\pi L_1L_2}\!\sum_{(n,\ell_3)\in\Z^{2}\setminus\{\vect{0}\}}\!\!\left[%
\frac{K_{1}\left(\frac{m}{T}\sqrt{n^2+(2TL_3)^2\ell_3^2}\right)}{\sqrt{n^2+(2TL_3)^2\ell_3^2}}\right]\nn\\%
&-&\frac{mT}{8\pi L_1L_3}\!\sum_{(n,\ell_2)\in\Z^{2}\setminus\{\vect{0}\}}\!\!\left[%
\frac{K_{1}\left(\frac{m}{T}\sqrt{n^2+(2TL_2)^2\ell_2^2}\right)}{\sqrt{n^2+(2TL_2)^2\ell_2^2}}\right]\nn\\%
&-&\frac{mT}{8\pi L_2L_3}\!\sum_{(n,\ell_1)\in\Z^{2}\setminus\{\vect{0}\}}\!\!\left[%
\frac{K_{1}\left(\frac{m}{T}\sqrt{n^2+(2TL_1)^2\ell_1^2}\right)}{\sqrt{n^2+(2TL_1)^2\ell_1^2}}\right]\labeq{FreeEnergyDensityMassiveL1L2L3}\\
&+&\frac{T^{3}}{8\pi L_1}\!\sum_{(n,\ell_2,\ell_3)\in\Z^{3}\setminus\{\vect{0}\}}\!\!\left[%
\frac{e^{-\frac{m}{T}\sqrt{n^2+(2TL_2)^2\ell_2^2+(2TL_3)^2\ell_3^2}}%
\left(1+\frac{m}{T}\sqrt{n^2+(2TL_2)^2\ell_2^2+(2TL_3)^2\ell_3^2}\right)}{\Big(n^2+(2TL_2)^2\ell_2^2+(2TL_3)^2\ell_3^2\Big)^{3/2}}\right]\nn\\%
&+&\frac{T^{3}}{8\pi L_2}\!\sum_{(n,\ell_1,\ell_3)\in\Z^{3}\setminus\{\vect{0}\}}\!\!\left[%
\frac{e^{-\frac{m}{T}\sqrt{n^2+(2TL_1)^2\ell_1^2+(2TL_3)^2\ell_3^2}}%
\left(1+\frac{m}{T}\sqrt{n^2+(2TL_1)^2\ell_1^2+(2TL_3)^2\ell_3^2}\right)}{\Big(n^2+(2TL_1)^2\ell_1^2+(2TL_3)^2\ell_3^2\Big)^{3/2}}\right]\nn\\%
&+&\frac{T^{3}}{8\pi L_3}\!\sum_{(n,\ell_1,\ell_2)\in\Z^{3}\setminus\{\vect{0}\}}\!\!\left[%
\frac{e^{-\frac{m}{T}\sqrt{n^2+(2TL_1)^2\ell_1^2+(2TL_2)^2\ell_2^2}}%
\left(1+\frac{m}{T}\sqrt{n^2+(2TL_1)^2\ell_1^2+(2TL_2)^2\ell_2^2}\right)}{\Big(n^2+(2TL_1)^2\ell_1^2+(2TL_2)^2\ell_2^2\Big)^{3/2}}\right]\nn\\%
&-&\frac{m^2T^2}{4\pi^2}\!\sum_{(n,\ell_1,\ell_2,\ell_3)\in\Z^{4}\setminus\{\vect{0}\}}\left[%
\frac{K_{2}\left(\frac{m}{T}\sqrt{n^2+(2TL_1)^2\ell_1^2+(2TL_2)^2\ell_2^2+(2TL_3)^2\ell_3^2}\right)}%
{\Big(n^2+(2TL_1)^2\ell_1^2+(2TL_2)^2\ell_2^2+(2TL_3)^2\ell_3^2\Big)}\right]%
+{\cal O}\left(\epsilon^1\right).\nn%
\eea}

Unlike in the case of the tube geometry, we cannot here simply come back to an earlier version of the massive result such as eq.~\eqn{FreeEnergyDensityMassiveL1L2L3}. Indeed, even though there is no infrared (or even UV) divergence, and the application of the massless limit upon eq.~\eqn{FreeEnergyDensityMassiveL1L2L3} would leave an expression which is overall finite, each of the summations therein produces, in fact, a divergence. And only the sum of all of these divergences actually vanishes. The reason is that when $m=0$, each of the sums in eq.~\eqn{FreeEnergyDensityMassiveL1L2L3} can be represented by a certain zeta function, and we hit the corresponding pole in the $\epsilon\rightarrow 0$ limit. Therefore, we will implement the limit $m\rightarrow 0$, keeping $\epsilon$ nonzero for the sake of regularizing intermediate stage divergences. Doing so, we find
{\allowdisplaybreaks
\begin{eqnarray*}
\lim_{m\rightarrow 0}\left(f^{(3)}\right)&=&%
\SPEprefacmsb{3}{1/2}\left(\frac{T}{L_1L_2L_3}\right)\Gamma\left(\frac{1}{2}-\epsilon\right)\zeta\big(1-2\epsilon\big)\nn\\%
&-&\SPEprefacmsb{3}{}\left(\frac{T^2}{L_1L_2}\right)\Gamma\big(1-\epsilon\big)%
\!\!\!\!\!\!\sum_{(n,\ell_3)\in\Z^{2}\setminus\{\vect{0}\}}\!\!%
\Big(n^2+(2TL_3)^2\ell_3^2\Big)^{\epsilon-1}\nn\\%
&-&\SPEprefacmsb{3}{}\left(\frac{T^2}{L_1L_3}\right)\Gamma\big(1-\epsilon\big)%
\!\!\!\!\!\!\sum_{(n,\ell_2)\in\Z^{2}\setminus\{\vect{0}\}}\!\!%
\Big(n^2+(2TL_2)^2\ell_2^2\Big)^{\epsilon-1}\nn\\%
&-&\SPEprefacmsb{3}{}\left(\frac{T^2}{L_2L_3}\right)\Gamma\big(1-\epsilon\big)%
\!\!\!\!\!\!\sum_{(n,\ell_1)\in\Z^{2}\setminus\{\vect{0}\}}\!\!%
\Big(n^2+(2TL_1)^2\ell_1^2\Big)^{\epsilon-1}\\
&+&\SPEprefacmsb{2}{3/2}\left(\frac{T^3}{L_1}\right)\Gamma\left(\frac{3}{2}-\epsilon\right)%
\!\!\!\sum_{(n,\ell_2,\ell_3)\in\Z^{3}\setminus\{\vect{0}\}}\!\!%
\Big(n^2+(2TL_2)^2\ell_2^2+(2TL_3)^2\ell_3^2\Big)^{\epsilon-3/2}\nn\\%
&+&\SPEprefacmsb{2}{3/2}\left(\frac{T^3}{L_2}\right)\Gamma\left(\frac{3}{2}-\epsilon\right)%
\!\!\!\sum_{(n,\ell_1,\ell_3)\in\Z^{3}\setminus\{\vect{0}\}}\!\!%
\Big(n^2+(2TL_1)^2\ell_1^2+(2TL_3)^2\ell_3^2\Big)^{\epsilon-3/2}\nn\\%
&+&\SPEprefacmsb{2}{3/2}\left(\frac{T^3}{L_3}\right)\Gamma\left(\frac{3}{2}-\epsilon\right)%
\!\!\!\sum_{(n,\ell_1,\ell_2)\in\Z^{3}\setminus\{\vect{0}\}}\!\!%
\Big(n^2+(2TL_1)^2\ell_1^2+(2TL_2)^2\ell_2^2\Big)^{\epsilon-3/2}\nn\\%
&-&\SPEprefacmsb{1}{2}\Big(T^4\Big)\Gamma\big(2-\epsilon\big)\!\!\!\!\!\!\!\!\!\!\!\!\!%
\sum_{(n,\ell_1,\ell_2,\ell_3)\in\Z^{4}\setminus\{\vect{0}\}}\!\!%
\Big(n^2+(2TL_1)^2\ell_1^2+(2TL_2)^2\ell_2^2+(2TL_3)^2\ell_3^2\Big)^{\epsilon-2}\nn%
,\labeq{FreeEnergyDensityMasslessL1L2L3Pre1}%
\end{eqnarray*}}
\HS{0.12}where the structure of the intermediate divergences, which we mentioned previously, is now obvious. Notice further that as the sum of all the poles in $\epsilon$ identically vanishes, the finite part which is by construction $\lmsb$-dependent will also vanish, leaving the overall result not only finite but also thankfully renormalization scale independent.

Following the massless result of the infinite rectangular tube case, we can again use the Poisson resummation formula~\eqn{PoissonResummation}, together with expressions such as eq.~\eqn{EpsteinZetaTwoDimMassZeroAnaConti} and eq.~\eqn{SplitMatsuContrib2} in order to compute the above two-fold summations while singling out the intermediate stage divergences. In addition, we can use eq.~\eqn{S3Summation1} in order to deal with the three-fold summations, including those which will appear in the decomposition of the four-fold summation. In the end, we are only left with the computation of the last term in the above expression, and to do so we define
{\allowdisplaybreaks
\bea
S_4&\equiv&\sum_{(n,\ell_1,\ell_2,\ell_3)\in\Z^{4}\setminus\{\vect{0}\}}\!\!%
\Big(n^2+(2TL_1)^2\ell_1^2+(2TL_2)^2\ell_2^2+(2TL_3)^2\ell_3^2\Big)^{\epsilon-2}%
,\labeq{S4Summation1}%
\eea}
\HS{0.12}in order to give more details on this last technical step. The above expression can be decomposed and symmetrized following
{\allowdisplaybreaks
\bea
S_4&=&\frac{2}{3}\!\!\sum_{\ell_1=1}^{+\infty}~\sum_{(n,\ell_2,\ell_3)\in\Z^{3}}%
\Big(n^2+(2TL_1)^2\ell_1^2+(2TL_2)^2\ell_2^2+(2TL_3)^2\ell_3^2\Big)^{\epsilon-2}\nn\\%
&+&\frac{2}{3}\!\!\sum_{\ell_2=1}^{+\infty}~\sum_{(n,\ell_1,\ell_3)\in\Z^{3}}%
\Big(n^2+(2TL_1)^2\ell_1^2+(2TL_2)^2\ell_2^2+(2TL_3)^2\ell_3^2\Big)^{\epsilon-2}\nn\\%
&+&\frac{2}{3}\!\!\sum_{\ell_3=1}^{+\infty}~\sum_{(n,\ell_1,\ell_2)\in\Z^{3}}%
\Big(n^2+(2TL_1)^2\ell_1^2+(2TL_2)^2\ell_2^2+(2TL_3)^2\ell_3^2\Big)^{\epsilon-2}\nn\\%
&+&\frac{1}{3}\!\!\sum_{(n,\ell_2,\ell_3)\in\Z^{3}\setminus\{\vect{0}\}}\!\!%
\Big(n^2+(2TL_2)^2\ell_2^2+(2TL_3)^2\ell_3^2\Big)^{\epsilon-2}\nn\\%
&+&\frac{1}{3}\!\!\sum_{(n,\ell_1,\ell_3)\in\Z^{3}\setminus\{\vect{0}\}}\!\!%
\Big(n^2+(2TL_1)^2\ell_1^2+(2TL_3)^2\ell_3^2\Big)^{\epsilon-2}\nn\\%
&+&\frac{1}{3}\!\!\sum_{(n,\ell_1,\ell_2)\in\Z^{3}\setminus\{\vect{0}\}}\!\!%
\Big(n^2+(2TL_1)^2\ell_1^2+(2TL_2)^2\ell_2^2\Big)^{\epsilon-2}.%
\eea}
\HS{0.12}And while the last three terms can be handled via the procedure which we used for the three-fold summations in the previous subsection, the first three terms need to be computed separately. We then define another set of summations, namely $S_{1+3}$, which denotes the first of the above terms. Analytically continuing this set of sums, using for example eq.~\eqn{EpsteinZetaNDimMassiveAnaConti}, leads to the following result
{\allowdisplaybreaks
\bea
S_{1+3}&=&\left(\frac{\pi^{3/2}(TL_1)^{2\epsilon-1}\Gamma\left(\frac{1}{2}-\epsilon\right)\zeta\big(1-2\epsilon\big)}%
{3\times 4^{1-\epsilon}~T^2L_2L_3~\Gamma\big(2-\epsilon\big)}\right)+\left(\frac{\pi^{2-\epsilon}(TL_1)^{\epsilon-\frac{1}{2}}}%
{2^{\frac{1}{2}-\epsilon}\times 3~T^2L_2L_3~\Gamma\big(2-\epsilon\big)}\right)\times\nn\\%
&\times&\sum_{\ell_1=1}^{+\infty}~\sum_{(n,\ell_2,\ell_3)\in\Z^{3}\setminus\{\vect{0}\}}\left[\frac{\Big(n^2+\frac{\ell_2^2}{(2TL_2)^2}%
+\frac{\ell_3^2}{(2TL_3)^2}\Big)^{\frac{1}{4}-\frac{\epsilon}{2}}}{\ell_1^{\frac{1}{2}-\epsilon}}\times\right.\nn\\%
&&\hspa{3.225}\left.\times K_{\frac{1}{2}-\epsilon}\left(4\pi TL_1 \ell_1 \sqrt{n^2+\frac{\ell_2^2}%
{(2TL_2)^2}+\frac{\ell_3^2}{(2TL_3)^2}}\right)\right].%
\eea}

Finally bringing all the results together, and expanding around $\epsilon=0$, we see that all the $\epsilon$- intermediate stage divergences indeed cancel. In addition, the renormalization scale dependence disappears as a consequence of the fact that there is no overall logarithmic UV divergence in the present case. We further perform some of the summations, in order to get closed forms whenever it is possible. Then, relabeling some of the variables, we finally obtain
{\allowdisplaybreaks
\bea
f^{(3)}&=&-\frac{\pi^2 T^4}{90}+\frac{T~\log\Big(8T^3L_1L_2L_3\Big)}{24L_1L_2L_3}%
+\frac{\zeta\left(3\right)T^3(L_1L_2+L_1L_3+L_2L_3)}{4\pi L_1L_2L_3}-\frac{\pi T^2(L_1+L_2+L_3)}{24L_1L_2L_3}\nn\\%
&-&\frac{\zeta\left(3\right)T(L_1^3L_2^3+L_1^3L_3^3+L_2^3L_3^3)}{48\pi L_1^3L_2^3L_3^3}%
+\frac{\pi T(L_1^2L_2+L_1L_2^2+L_1^2L_3+L_1L_3^2+L_2^2L_3+L_2L_3^2)}{144L_1^2L_2^2L_3^2}\nn\\%
&+&\frac{T}{4L_1L_2L_3}\!\sum_{\ell=1}^{+\infty}\Bigg[\log\Big(1-e^{-4\pi TL_1 \ell}\Big)%
+\log\Big(1-e^{-4\pi TL_2 \ell}\Big)+\log\Big(1-e^{-4\pi TL_3 \ell}\Big)\Bigg]\nn\\%
&-&\frac{T^2}{6L_1^2}\!\sum_{\ell=1}^{+\infty}\Bigg[\ell~\text{Li}_{2}\left(e^{-4\pi TL_1\ell}\right)\Bigg]%
-\frac{T}{24\pi L_1^3}\!\sum_{\ell=1}^{+\infty}\text{Li}_{3}\left(e^{-4\pi TL_1\ell}\right)\nn\\%
&-&\frac{T^2}{6L_2^2}\!\sum_{\ell=1}^{+\infty}\Bigg[\ell~\text{Li}_{2}\left(e^{-4\pi TL_2\ell}\right)\Bigg]%
-\frac{T}{24\pi L_2^3}\!\sum_{\ell=1}^{+\infty}\text{Li}_{3}\left(e^{-4\pi TL_2\ell}\right)\nn\\%
&-&\frac{T^2}{6L_3^2}\!\sum_{\ell=1}^{+\infty}\Bigg[\ell~\text{Li}_{2}\left(e^{-4\pi TL_3\ell}\right)\Bigg]%
-\frac{T}{24\pi L_3^3}\!\sum_{\ell=1}^{+\infty}\text{Li}_{3}\left(e^{-4\pi TL_3\ell}\right)\nn\\%
&+&\frac{T^2}{2L_1L_2L_3}\!\!\sum_{n,\ell=1}^{+\infty}\!\!\Bigg[\frac{(L_2\!+\! L_3)~\ell}{n} K_{1}\left(4\pi TL_1 n\ell\right)%
+\frac{(L_1\!+\! L_3)~\ell}{n} K_{1}\left(4\pi TL_2 n\ell\right)\nn\\%
&&\hspa{2.25}+\frac{(L_1\!+\! L_2)~\ell}{n} K_{1}\left(4\pi TL_3 n\ell\right)\Bigg]\nn\\%
&-&\frac{T}{16L_1L_2L_3}\!\!\!\!\!\!\sum_{(n,\ell)\in\Z^{2}\setminus\{\vect{0}\}}\!\!\Bigg[%
\log\left(1\!-e^{-\frac{2\pi L_2}{L_1}\sqrt{\ell^2+(2TL_1)^2n^2}}\right)%
+\log\left(1\!-e^{-\frac{2\pi L_3}{L_1}\sqrt{\ell^2+(2TL_1)^2n^2}}\right)\nn\\%
&&\hspa{3}+\log\left(1\!-e^{-\frac{2\pi L_1}{L_2}\sqrt{\ell^2+(2TL_2)^2n^2}}\right)%
+\log\left(1\!-e^{-\frac{2\pi L_3}{L_2}\sqrt{\ell^2+(2TL_2)^2n^2}}\right)\nn\\%
&&\hspa{3}+\log\left(1\!-e^{-\frac{2\pi L_1}{L_3}\sqrt{\ell^2+(2TL_3)^2n^2}}\right)%
+\log\left(1\!-e^{-\frac{2\pi L_2}{L_3}\sqrt{\ell^2+(2TL_3)^2n^2}}\right)\Bigg]\nn\\
&-&\frac{T}{24L_1^2L_2^2L_3^2}\!\sum_{\ell_1=1}^{+\infty}\sum_{(n,\ell_2)\in\Z^{2}\setminus\{\vect{0}\}}\Bigg[%
L_2L_3^2~\frac{\sqrt{\ell_2^2+(2TL_1)^2n^2}}{\ell_1} K_{1}\left(\frac{2\pi L_2}{L_1}~\ell_1~\sqrt{\ell_2^2+(2TL_1)^2n^2}\right)\nn\\%
&&\hspa{4}+L_3L_2^2~\frac{\sqrt{\ell_2^2+(2TL_1)^2n^2}}{\ell_1} K_{1}\left(\frac{2\pi L_3}{L_1}~\ell_1~\sqrt{\ell_2^2+(2TL_1)^2n^2}\right)\nn\\%
&&\hspa{4}+L_1L_3^2~\frac{\sqrt{\ell_2^2+(2TL_2)^2n^2}}{\ell_1} K_{1}\left(\frac{2\pi L_1}{L_2}~\ell_1~\sqrt{\ell_2^2+(2TL_2)^2n^2}\right)\nn\\%
&&\hspa{4}+L_3L_1^2~\frac{\sqrt{\ell_2^2+(2TL_2)^2n^2}}{\ell_1} K_{1}\left(\frac{2\pi L_3}{L_2}~\ell_1~\sqrt{\ell_2^2+(2TL_2)^2n^2}\right)\nn\\%
&&\hspa{4}+L_1L_2^2~\frac{\sqrt{\ell_2^2+(2TL_3)^2n^2}}{\ell_1} K_{1}\left(\frac{2\pi L_1}{L_3}~\ell_1~\sqrt{\ell_2^2+(2TL_3)^2n^2}\right)\nn\\%
&&\hspa{4}+L_2L_1^2~\frac{\sqrt{\ell_2^2+(2TL_3)^2n^2}}{\ell_1} K_{1}\left(\frac{2\pi L_2}{L_3}~\ell_1~\sqrt{\ell_2^2+(2TL_3)^2n^2}\right)\Bigg]\nn\\%
&+&\frac{T}{24L_1L_2L_3}\!\!\!\!\!\!\sum_{(n,\ell_1,\ell_2)\in\Z^{3}\setminus\{\vect{0}\}}\!\!\Bigg[%
\log\left(1\!-e^{-4\pi TL_1\sqrt{n^2+\frac{\ell_1^2}{(2TL_2)^2}+\frac{\ell_2^2}{(2TL_3)^2}}}\right)\nn\\%
&&\hspa{3.25}+\log\left(1\!-e^{-4\pi TL_2\sqrt{n^2+\frac{\ell_1^2}{(2TL_1)^2}+\frac{\ell_2^2}{(2TL_3)^2}}}\right)\nn\\%
&&\hspa{3.25}+\log\left(1\!-e^{-4\pi TL_3\sqrt{n^2+\frac{\ell_1^2}{(2TL_1)^2}+\frac{\ell_2^2}{(2TL_2)^2}}}\right)\Bigg]%
,\labeq{FinalResultfMasslessL1L2L3}%
\eea}
\HS{0.12}which is the symmetrized and renormalized free energy density of a noninteracting massless neutral scalar field coupled to a heat bath at temperature $T$, and geometrically confined inside a box with sides of respective finite lengths $L_1$, $L_2$, and $L_3$. We refer to eq.~\eqn{M3finalResultSM} for a conveniently much more compact form of this result.

\section{Various proofs}\labsec{AppVariousProofs}

\subsection{Fourier decomposition}\labsec{AppFourierDecomp}

In this section, we first describe the derivation of the form and normalization of the Fourier decomposition of a massless scalar field (adding a mass does not modify the argument), along the direction of a compactified dimension with DBCs. To this end, we apply the method of separation of variables to the scalar field, and consider only, for the sake of argument, the part relevant to one of the DBCs which must also obey the Klein-Gordon equation
{\allowdisplaybreaks
\be
\left(\omega_{\ell}^2+\frac{\partial^2}{\partial\xi^2}\right)\phi_{\ell}(\xi)=0,%
\labeq{KGxi}
\ee}%
where we recall that $\omega_{\ell}\equiv\pi\ell/L$. Compactifying this direction onto a finite length $[0,L]$, and imposing DBCs, implies that
{\allowdisplaybreaks
\be
\phi_{\ell}(0)=\phi_{\ell}(L)=0.%
\labeq{DirichletBoundaryCond}
\ee}
\HS{0.12}We then remember that the general solution to a differential equation such as~\eqn{KGxi} is
{\allowdisplaybreaks
\be
\phi_{\ell}(\xi)=A\sin\left(\omega_{\ell}\xi\right)+B\cos\left(\omega_{\ell}\xi\right).%
\ee}
Thus, by applying the boundary conditions~\eqn{DirichletBoundaryCond} to the above, we obtain
{\allowdisplaybreaks
\be
B\overset{!}{=}0,\quad\omega_{\ell}\overset{!}{=}\frac{\pi\ell}{L},%
\ee}
\HS{0.12}We may then obtain a convenient prefactor by normalizing the field to unity
{\allowdisplaybreaks
\bea
\int_{0}^{L}d\xi~~\phi_{\ell}(\xi)\phi_{\ell}(\xi)&=&\int_{0}^{L}d\xi~~ A^2\sin^2%
\left(\frac{\pi\ell}{L}\xi\right)\overset{!}{=}1,\\%
\Rightarrow A&\overset{!}{=}&\sqrt{\frac{2}{L}}~.%
\eea}
\HS{0.12}Therefore, we obtain the Fourier decomposition along a compactified dimension with DBCs
{\allowdisplaybreaks
\be
\phi_{\ell}\left(\xi\right)=\sqrt{\frac{2}{L}}~\sin\left(\frac{\pi\ell}{L}\xi\right).%
\ee

We now give some useful identities: It should be noted, indeed, that in order to go from eq.~\eqn{ZM1} to eq.~\eqn{ZM1Bis}, the following integral identities, defining the usual Kronecker delta function, are needed
{\allowdisplaybreaks
\bea
\int_{0}^{1/T}\mathop{{{\rm d}}\tau}\!e^{i(\omega_{n_1}+\omega_{n_2})\tau}\equiv\Krondelt{n_1\!+\!n_2}/T\text{,~~}%
\int_{R}\mathop{{{\rm d}^{D-1-c}}\vect{x}}
e^{i(\vect{\omega}_{k_{1}}+\vect{\omega}_{k_{2}})\,\vect{x}}&\equiv& R^{D-1-c} \Krondelt{\vect{n}_{k_1}\!\!+\!\vect{n}_{k_2}}\text{,~~}\\
\inte{z_i}{0}{L_i}\sin\left(\omega_{\ell_{i_1}}z_i\right)%
\sin\left(\omega_{\ell_{i_2}}z_i\right)&\equiv& L_i \Krondelt{\ell_{i_1}\!\!-\!\ell_{i_2}}/2,%
\eea}%
the last equation holding only for all $\ell_i$ strictly positive integers, which is presently the case.

\subsection{Length derivatives}\labsec{AppLengthDerivatives}

We present a short derivation of an important result, i.e. the length derivatives of the internal and free energies, which allows us to compute the pressures directly from the free energy. Consider eq.~\eqn{LegendreTransformDef} which we differentiate following
{\allowdisplaybreaks
\be
dF(T,\,\left\{L_i\right\})+SdT=dE(S,\,\left\{L_i\right\})-TdS.%
\ee}
\HS{0.12}On both sides of the above equation, fixing all variables but one length $L_j$ (preferably the same on both sides) gives us the following important equation
{\allowdisplaybreaks
\be
\partdiffat{F(T,\,\left\{L_i\right\})}{L_j}{T,{\{L_{k\neq j}\}}}dL_j=\partdiffat{E(S,\,\left\{L_i\right\})}{L_j}{S,{\{L_{k\neq j}\}}}dL_j.%
\ee}
\HS{0.12}The above result indicates that both partial differential expressions are equal, namely that
{\allowdisplaybreaks
\be
\partdiffat{F(T,\,\left\{L_i\right\})}{L_j}{T,{\{L_{k\neq j}\}}}=\partdiffat{E(S,\,\left\{L_i\right\})}{L_j}{S,{\{L_{k\neq j}\}}}%
,\labeq{dFdE}%
\ee}
allowing to probe the pressure $P_j$ from either of the two functions $E(S,\,\left\{L_i\right\})$ or $F(T,\,\left\{L_i\right\})$
{\allowdisplaybreaks
\be
P_j=-\frac{L_j}{V}\partdiffat{E(S,\,\left\{L_i\right\})}{L_j}{S,{\{L_{k\neq j}\}}}%
=-\frac{L_j}{V}\partdiffat{F(T,\,\left\{L_i\right\})}{L_j}{T,{\{L_{k\neq j}\}}}.%
\ee}

\bibliography{201029_GeometricConfinement,biblio2}

\end{document}